\documentclass[longauth]{aa}  
\usepackage{graphicx}
\usepackage{subcaption}
\usepackage{xcolor}
\usepackage{txfonts}

\usepackage[hidelinks]{hyperref}
\usepackage[normalem]{ulem}
\usepackage{orcidlink}
\usepackage{booktabs}
\usepackage{verbatim}
\newcounter{pta}
\renewcommand{\thepta}{\Roman{pta}} 

\DeclareRobustCommand{\defpta}[1]{%
   \refstepcounter{pta}%
   \thepta\label{#1}%
}

\newcommand{\refpta}[1]{\ref{#1}}

\newcommand{\logGmu}{\log_{10} G\mu}
\newcommand{\Mc}{\mathcal{M}}

\newcommand{\ndot}{{\dot n}_{\rm 0}}
\newcommand{\alphaM}{\alpha_{\rm \Mc}}
\newcommand{\Mstar}{\Mc_{\rm \star}}
\newcommand{\betaz}{\beta_{\rm z}}
\newcommand{\zo}{z_{\rm 0}}

\makeatletter
\renewcommand*\aa@pageof{, page \thepage{} of \pageref*{LastPage}}
\makeatother

\usepackage[switch]{lineno} 

\def\wm{5}
\begin{document} 

 \title{The second data release from the European Pulsar Timing Array}
   \subtitle{IV. Implications for massive black holes, dark matter, and the early Universe\thanks{{\it Data availability}. The EPTA+InPTA DR2 data used to perform the analysis presented in this paper can be found at:\\ https://zenodo.org/record/8091568\\
https://gitlab.in2p3.fr/epta/epta-dr2}}

\author{
    J.~Antoniadis\orcidlink{0000-0003-4453-776}\inst{\ref{forth}, \ref{mpifr}\ifnum\wm>1, \refpta{epta}\fi}, 
    \ifnum\wm>1 P.~Arumugam\orcidlink{0000-0001-9264-8024}\inst{\ref{IITR}\ifnum\wm>1, \refpta{inpta}\fi}, \fi
    \ifnum\wm>1 S.~Arumugam\orcidlink{0009-0001-3587-6622}\inst{\ref{IITH_El}\ifnum\wm>1, \refpta{inpta}\fi}, \fi
    \ifnum\wm=5 P.~Auclair\orcidlink{0000-0002-4814-1406}\inst{\ref{curl}}, \fi
    S.~Babak\orcidlink{0000-0001-7469-4250}\inst{\ref{apc}\ifnum\wm>1, \refpta{epta}\fi}, 
    \ifnum\wm>1 M.~Bagchi\orcidlink{0000-0001-8640-8186}\inst{\ref{IMSc}, \ref{HBNI}\ifnum\wm>1, \refpta{inpta}\fi}, \fi
    A.-S.~Bak~Nielsen\orcidlink{ 0000-0002-1298-9392}\inst{\ref{mpifr}, \ref{unibi}\ifnum\wm>1, \refpta{epta}\fi}, 
    \ifnum\wm=5 E.~Barausse\orcidlink{0000-0001-6499-6263 }\inst{\ref{sissa}}, \fi
    C.~G.~Bassa\orcidlink{0000-0002-1429-9010}\inst{\ref{astron}\ifnum\wm>1, \refpta{epta}\fi}, 
    \ifnum\wm>1 A.~Bathula\orcidlink{0000-0001-7947-6703} \inst{\ref{IISERM}\ifnum\wm>1, \refpta{inpta}\fi}, \fi
    A.~Berthereau\inst{\ref{lpc2e}, \ref{nancay}\ifnum\wm>1, \refpta{epta}\fi}, 
    M.~Bonetti\orcidlink{0000-0001-7889-6810}\inst{\ref{unimib}, \ref{infn-unimib}, \ref{inaf-brera}\ifnum\wm>1, \refpta{epta}\fi}, 
    E.~Bortolas\inst{\ref{unimib}, \ref{infn-unimib}, \ref{inaf-brera}\ifnum\wm>1, \refpta{epta}\fi}, 
    P.~R.~Brook\orcidlink{0000-0003-3053-6538}\inst{\ref{unibir}\ifnum\wm>1, \refpta{epta}\fi}, 
    M.~Burgay\orcidlink{0000-0002-8265-4344}\inst{\ref{inaf-oac}\ifnum\wm>1, \refpta{epta}\fi}, 
    R.~N.~Caballero\orcidlink{0000-0001-9084-9427}\inst{\ref{HOU}\ifnum\wm>1, \refpta{epta}\fi}, 
    \ifnum\wm=5 C.~Caprini\orcidlink{0000-0001-5393-2205}\inst{\ref{unige}, \ref{CERN}}, \fi
    A.~Chalumeau\orcidlink{0000-0003-2111-1001}\inst{\ref{unimib}\ifnum\wm>1, \refpta{epta}\fi}\ifnum\wm=2\thanks{aurelien.chalumeau@unimib.it}\fi, 
    D.~J.~Champion\orcidlink{0000-0003-1361-7723}\inst{\ref{mpifr}\ifnum\wm>1, \refpta{epta}\fi}, 
    S.~Chanlaridis\orcidlink{0000-0002-9323-9728}\inst{\ref{forth}\ifnum\wm>1, \refpta{epta}\fi}, 
    S.~Chen\orcidlink{0000-0002-3118-5963}\inst{\ref{kiaa}\ifnum\wm>1, \refpta{epta}\fi}\ifnum\wm=3\thanks{sychen@pku.edu.cn}\fi, 
    I.~Cognard\orcidlink{0000-0002-1775-9692}\inst{\ref{lpc2e}, \ref{nancay}\ifnum\wm>1, \refpta{epta}\fi}, 
    \ifnum\wm=5 M.~Crisostomi\orcidlink{0000-0002-7622-4911}\inst{\ref{sissa}}, \fi
    \ifnum\wm>1 S.~Dandapat\orcidlink{0000-0003-4965-9220}\inst{\ref{TIFR}\ifnum\wm>1, \refpta{inpta}\fi}, \fi
    \ifnum\wm>1 D.~Deb\orcidlink{0000-0003-4067-5283}\inst{\ref{IMSc}\ifnum\wm>1, \refpta{inpta}\fi},  \fi
    \ifnum\wm>1 S.~Desai\orcidlink{0000-0002-0466-3288}\inst{\ref{IITH_Ph}\ifnum\wm>1, \refpta{inpta}\fi}, \fi
    G.~Desvignes\orcidlink{0000-0003-3922-4055}\inst{\ref{mpifr}\ifnum\wm>1, \refpta{epta}\fi}, 
    \ifnum\wm>1 N.~Dhanda-Batra \inst{\ref{UoD}\ifnum\wm>1, \refpta{inpta}\fi}, \fi
    \ifnum\wm>1 C.~Dwivedi\orcidlink{0000-0002-8804-650X}\inst{\ref{IIST}\ifnum\wm>1, \refpta{inpta}\fi}, \fi
    M.~Falxa\inst{\ref{apc}, \ref{lpc2e}\ifnum\wm>1, \refpta{epta}\fi}\ifnum\wm=4\thanks{falxa@apc.in2p3.fr}\fi, 
    \ifnum\wm=5 F.~Fastidio\inst{\ref{surrey}, \ref{unimib},}, \fi
    \ifnum\wm=4 I.~Ferranti\orcidlink{0009-0000-1575-2051}\inst{\ref{unimib}, \ref{apc}, \refpta{epta}}, \fi 
    R.~D.~Ferdman\inst{\ref{uea}\ifnum\wm>1, \refpta{epta}\fi}, 
    A.~Franchini\orcidlink{0000-0002-8400-0969}\inst{\ref{unimib}, \ref{infn-unimib}\ifnum\wm>1, \refpta{epta}\fi}, 
    J.~R.~Gair\orcidlink{0000-0002-1671-3668}\inst{\ref{aei}\ifnum\wm>1, \refpta{epta}\fi}, 
    B.~Goncharov\orcidlink{0000-0003-3189-5807}\inst{\ref{gssi}, \ref{lngs}\ifnum\wm>1, \refpta{epta}\fi}
    \ifnum\wm>1 A.~Gopakumar\orcidlink{0000-0003-4274-4369}\inst{\ref{TIFR}\ifnum\wm>1, \refpta{inpta}\fi}, \fi
    E.~Graikou\inst{\ref{mpifr}\ifnum\wm>1, \refpta{epta}\fi}, 
    J.-M.~Grie{\ss}meier\orcidlink{0000-0003-3362-7996}\inst{\ref{lpc2e}, \ref{nancay}\ifnum\wm>1, \refpta{epta}\fi}, 
    \ifnum\wm=5 A.~Gualandris\orcidlink{0000-0002-9420-2679}\inst{\ref{surrey}, \refpta{epta}}, \fi
    L.~Guillemot\orcidlink{0000-0002-9049-8716}\inst{\ref{lpc2e}, \ref{nancay}\ifnum\wm>1, \refpta{epta}\fi}, 
    Y.~J.~Guo\inst{\ref{mpifr}\ifnum\wm>1, \refpta{epta}\fi}\ifnum\wm=3\thanks{yjguo@mpifr-bonn.mpg.de}\fi, 
    \ifnum\wm>1 Y.~Gupta\orcidlink{0000-0001-5765-0619}\inst{\ref{NCRA}\ifnum\wm>1, \refpta{inpta}\fi}, \fi
    \ifnum\wm>1 S.~Hisano\orcidlink{0000-0002-7700-3379}\inst{\ref{KU_J}\ifnum\wm>1, \refpta{inpta}\fi}, \fi
    H.~Hu\orcidlink{0000-0002-3407-8071}\inst{\ref{mpifr}\ifnum\wm>1, \refpta{epta}\fi},  
    F.~Iraci\inst{\ref{unica}\ref{inaf-oac}\ifnum\wm>1, \refpta{epta}\fi}, 
    D.~Izquierdo-Villalba\orcidlink{0000-0002-6143-1491}\inst{\ref{unimib}, \ref{infn-unimib}\ifnum\wm>1, \refpta{epta}\fi}, 
    J.~Jang\orcidlink{0000-0003-4454-0204}\inst{\ref{mpifr}\ifnum\wm>1, \refpta{epta}\fi}\ifnum\wm=1\thanks{jjang@mpifr-bonn.mpg.de}\fi, 
    J.~Jawor\orcidlink{0000-0003-3391-0011}\inst{\ref{mpifr}\ifnum\wm>1, \refpta{epta}\fi}, 
    G.~H.~Janssen\orcidlink{0000-0003-3068-3677}\inst{\ref{astron}, \ref{imapp}\ifnum\wm>1, \refpta{epta}\fi}, 
    A.~Jessner\orcidlink{0000-0001-6152-9504}\inst{\ref{mpifr}\ifnum\wm>1, \refpta{epta}\fi}, 
    \ifnum\wm>1 B.~C.~Joshi\orcidlink{0000-0002-0863-7781}\inst{\ref{NCRA}, \ref{IITR}\ifnum\wm>1, \refpta{inpta}\fi}, \fi
    \ifnum\wm>1 F.~Kareem\orcidlink{0000-0003-2444-838X} \inst{\ref{IISERK}, \ref{CESSI}\ifnum\wm>1, \refpta{inpta}\fi}, \fi
    R.~Karuppusamy\orcidlink{0000-0002-5307-2919}\inst{\ref{mpifr}\ifnum\wm>1, \refpta{epta}\fi}, 
    E.~F.~Keane\orcidlink{0000-0002-4553-655X}\inst{\ref{tcd}\ifnum\wm>1, \refpta{epta}\fi}, 
    M.~J.~Keith\orcidlink{0000-0001-5567-5492}\inst{\ref{jbca}\ifnum\wm>1, \refpta{epta}\fi}\ifnum\wm=2\thanks{michael.keith@manchester.ac.uk}\fi, 
    \ifnum\wm>1 D.~Kharbanda\orcidlink{0000-0001-8863-4152}\inst{\ref{IITH_Ph}\ifnum\wm>1, \refpta{inpta}\fi}, \fi
    \ifnum\wm=5 T.~Khizriev, \inst{\ref{SAI}}, \fi
    \ifnum\wm>1 T.~Kikunaga\orcidlink{0000-0002-5016-3567} \inst{\ref{KU_J}\ifnum\wm>1, \refpta{inpta}\fi}, \fi
    \ifnum\wm>1 N.~Kolhe\orcidlink{0000-0003-3528-9863} \inst{\ref{XCM}\ifnum\wm>1, \refpta{inpta}\fi}, \fi
    M.~Kramer\inst{\ref{mpifr}, \ref{jbca}\ifnum\wm>1, \refpta{epta}\fi}, 
    M.~A.~Krishnakumar\orcidlink{0000-0003-4528-2745}\inst{\ref{mpifr}, \ref{unibi}\ifnum\wm>1, \refpta{epta}\fi\ifnum\wm>1, \refpta{inpta}\fi}, 
    K.~Lackeos\orcidlink{0000-0002-6554-3722}\inst{\ref{mpifr}\ifnum\wm>1, \refpta{epta}\fi}, 
    K.~J.~Lee\inst{\ref{pku}, \ref{naoc}\ifnum\wm>1, \refpta{epta}\fi}, 
    K.~Liu\inst{\ref{mpifr}\ifnum\wm>1, \refpta{epta}\fi}\ifnum\wm=1\thanks{kliu@mpifr-bonn.mpg.de}\fi, 
    Y.~Liu\orcidlink{0000-0001-9986-9360}\inst{\ref{naoc},  \ref{unibi}\ifnum\wm>1, \refpta{epta}\fi}, 
    A.~G.~Lyne\inst{\ref{jbca}\ifnum\wm>1, \refpta{epta}\fi}, 
    J.~W.~McKee\orcidlink{0000-0002-2885-8485}\inst{\ref{milne}, \ref{daim}\ifnum\wm>1, \refpta{epta}\fi}, 
    \ifnum\wm>1 Y.~Maan\inst{\ref{NCRA}\ifnum\wm>1, \refpta{inpta}\fi}, \fi
    R.~A.~Main\inst{\ref{mpifr}\ifnum\wm>1, \refpta{epta}\fi}, 
    \ifnum\wm=4 S.~Manzini\orcidlink{0009-0005-1149-5330}\inst{\ref{unimib}, \ref{apc}\ifnum\wm>1, \refpta{epta}\fi}, \fi
    M.~B.~Mickaliger\orcidlink{0000-0001-6798-5682}\inst{\ref{jbca}\ifnum\wm>1, \refpta{epta}\fi}, 
    \ifnum\wm=5 H.~Middleton\orcidlink{0000-0001-5532-3622}\inst{\ref{unibir}}, \fi
    \ifnum\wm=5 A.~Neronov\inst{\ref{apc}, \ref{EPFL}}, \fi
    I.~C.~Ni\c{t}u\orcidlink{0000-0003-3611-3464}\inst{\ref{jbca}\ifnum\wm>1, \refpta{epta}\fi}, 
    \ifnum\wm>1 K.~Nobleson\orcidlink{0000-0003-2715-4504}\inst{\ref{BITS}\ifnum\wm>1, \refpta{inpta}\fi}, \fi
    \ifnum\wm>1 A.~K.~Paladi\orcidlink{0000-0002-8651-9510}\inst{\ref{IISc}\ifnum\wm>1, \refpta{inpta}\fi}, \fi
    A.~Parthasarathy\orcidlink{0000-0002-4140-5616}\inst{\ref{mpifr}\ifnum\wm>1, \refpta{epta}\fi}\ifnum\wm=2\thanks{aparthas@mpifr-bonn.mpg.de}\fi, 
    B.~B.~P.~Perera\orcidlink{0000-0002-8509-5947}\inst{\ref{arecibo}\ifnum\wm>1, \refpta{epta}\fi}, 
    D.~Perrodin\orcidlink{0000-0002-1806-2483}\inst{\ref{inaf-oac}\ifnum\wm>1, \refpta{epta}\fi}, 
    A.~Petiteau\orcidlink{0000-0002-7371-9695}\inst{\ref{irfu}, \ref{apc}ee\ifnum\wm>1, \refpta{epta}\fi}, 
    N.~K.~Porayko\inst{\ref{unimib}, \ref{mpifr}\ifnum\wm>1, \refpta{epta}\fi}\ifnum\wm=5\thanks{nataliya.porayko@unimib.it}\fi, 
    A.~Possenti\inst{\ref{inaf-oac}\ifnum\wm>1, \refpta{epta}\fi}, 
    \ifnum\wm>1 T.~Prabu\inst{\ref{RRI}\ifnum\wm>1, \refpta{inpta}\fi}, \fi
    \ifnum\wm=5 K.~Postnov\orcidlink{0000-0002-1705-617X}\inst{\ref{SAI},  \ref{KFU}}, \fi
    H.~Quelquejay~Leclere\inst{\ref{apc}\ifnum\wm>1, \refpta{epta}\fi}\ifnum\wm=5\thanks{quelquejay@apc.in2p3.fr}\fi,
    \ifnum\wm>1 P.~Rana\orcidlink{0000-0001-6184-5195}\inst{\ref{TIFR}\ifnum\wm>1, \refpta{inpta}\fi}, \fi
    \ifnum\wm=5 A.~Roper Pol\orcidlink{0000-0003-4979-4430}\inst{\ref{unige}}, \fi
    A.~Samajdar\orcidlink{0000-0002-0857-6018}\inst{\ref{uni-potsdam}\ifnum\wm>1, \refpta{epta}\fi}, 
    S.~A.~Sanidas\inst{\ref{jbca}\ifnum\wm>1, \refpta{epta}\fi}, 
    \ifnum\wm=5 D.~Semikoz\inst{\ref{apc}}, \fi
    A.~Sesana\inst{\ref{unimib}, \ref{infn-unimib}, \ref{inaf-brera}\ifnum\wm>1, \refpta{epta}\fi}\ifnum\wm=5\thanks{alberto.sesana@unimib.it}\fi, 
    G.~Shaifullah\orcidlink{0000-0002-8452-4834}\inst{\ref{unimib}, \ref{infn-unimib}, \ref{inaf-oac}\ifnum\wm>1, \refpta{epta}\fi}\ifnum\wm=1\thanks{golam.shaifullah@unimib.it}\fi, 
    \ifnum\wm>1 J.~Singha\orcidlink{0000-0002-1636-9414}\inst{\ref{IITR}\ifnum\wm>1, \refpta{inpta}\fi}, \fi
    \ifnum\wm=5 C.~Smarra\inst{\ref{sissa}}, \fi
    L.~Speri\orcidlink{0000-0002-5442-7267}\inst{\ref{aei}\ifnum\wm>1, \refpta{epta}\fi}\ifnum\wm=4\thanks{lorenzo.speri@aei.mpg.de}\fi, 
    R.~Spiewak\inst{\ref{jbca}\ifnum\wm>1, \refpta{epta}\fi}, 
    \ifnum\wm>1 A.~Srivastava\orcidlink{0000-0003-3531-7887} \inst{\ref{IITH_Ph}\ifnum\wm>1, \refpta{inpta}\fi}, \fi
    B.~W.~Stappers\inst{\ref{jbca}\ifnum\wm>1, \refpta{epta}\fi}, 
    \ifnum\wm=5 D.~A.~Steer\orcidlink{0000-0002-8781-1273}\inst{\ref{apc}}\fi
    \ifnum\wm>1 M.~Surnis\orcidlink{0000-0002-9507-6985}\inst{\ref{IISERB}\ifnum\wm>1, \refpta{inpta}\fi}, \fi
    S.~C.~Susarla\orcidlink{0000-0003-4332-8201}\inst{\ref{uog}\ifnum\wm>1, \refpta{epta}\fi}, 
    \ifnum\wm>1 A.~Susobhanan\orcidlink{0000-0002-2820-0931}\inst{\ref{CGCA}\ifnum\wm>1, \refpta{inpta}\fi}, \fi
    \ifnum\wm>1 K.~Takahashi\orcidlink{0000-0002-3034-5769}\inst{\ref{KU_J1}, \ref{KU_J2}\ifnum\wm>1, \refpta{inpta}\fi} \fi
    \ifnum\wm>1 P.~Tarafdar\orcidlink{0000-0001-6921-4195}\inst{\ref{IMSc}\ifnum\wm>1, \refpta{inpta}\fi}\fi
    G.~Theureau\orcidlink{0000-0002-3649-276X}\inst{\ref{lpc2e},  \ref{nancay},  \ref{luth}\ifnum\wm>1, \refpta{epta}\fi}, 
    C.~Tiburzi\inst{\ref{inaf-oac}\ifnum\wm>1, \refpta{epta}\fi}, 
    \ifnum\wm=5 R.~J.~Truant\inst{\ref{unimib}}, \fi
    E.~van~der~Wateren\orcidlink{0000-0003-0382-8463}\inst{\ref{astron}, \ref{imapp}\ifnum\wm>1, \refpta{epta}\fi}, 
    \ifnum\wm=5 S.~Valtolina\inst{\ref{aei2}}, \fi
    A.~Vecchio\orcidlink{0000-0002-6254-1617}\inst{\ref{unibir}\ifnum\wm>1, \refpta{epta}\fi}, 
    V.~Venkatraman~Krishnan\orcidlink{0000-0001-9518-9819}\inst{\ref{mpifr}\ifnum\wm>1, \refpta{epta}\fi}, 
    J.~P.~W.~Verbiest\orcidlink{0000-0002-4088-896X}\inst{\ref{FSI}, \ref{unibi}, \ref{mpifr}\ifnum\wm>1, \refpta{epta}\fi}, 
    J.~Wang\orcidlink{0000-0003-1933-6498}\inst{\ref{unibi},  \ref{airub},  \ref{bnuz}\ifnum\wm>1, \refpta{epta}\fi}, 
    L.~Wang\inst{\ref{jbca}\ifnum\wm>1, \refpta{epta}\fi} and 
    Z.~Wu\orcidlink{0000-0002-1381-7859}\inst{\ref{naoc}, \ref{unibi}\ifnum\wm>1, \refpta{epta}\fi}.
    }

\institute{
{Institute of Astrophysics, FORTH, N. Plastira 100, 70013, Heraklion, Greece\label{forth}}\and 
{Max-Planck-Institut f{\"u}r Radioastronomie, Auf dem H{\"u}gel 69, 53121 Bonn, Germany\label{mpifr}}\and
\ifnum\wm>1{Department of Physics, Indian Institute of Technology Roorkee, Roorkee-247667, India\label{IITR}}\and\fi
\ifnum\wm>1{Department of Electrical Engineering, IIT Hyderabad, Kandi, Telangana 502284, India \label{IITH_El}}\and\fi
\ifnum\wm=5{Cosmology, Universe and Relativity at Louvain (CURL), Institute of Mathematics and Physics, University of Louvain, 2 Chemin du Cyclotron, 1348 Louvain-la-Neuve, Belgium \label{curl}}\and\fi
{Universit{\'e} Paris Cit{\'e}, CNRS, Astroparticule et Cosmologie, 75013 Paris, France\label{apc}}\and
\ifnum\wm>1{The Institute of Mathematical Sciences, C. I. T. Campus, Taramani, Chennai 600113, India \label{IMSc}}\and\fi
\ifnum\wm>1{Homi Bhabha National Institute, Training School Complex, Anushakti Nagar, Mumbai 400094, India \label{HBNI}}\and\fi
{Fakult{\"a}t f{\"u}r Physik, Universit{\"a}t Bielefeld, Postfach 100131, 33501 Bielefeld, Germany\label{unibi}}\and
\ifnum\wm=5{Scuola Internazionale Superiore di Studi Avanzati: Via Bonomea 265, I-34136 Trieste, Italy and INFN Sezione di Trieste \label{sissa}}\and\fi
{ASTRON, Netherlands Institute for Radio Astronomy, Oude Hoogeveensedijk 4, 7991 PD, Dwingeloo, The Netherlands\label{astron}}\and
\ifnum\wm>1{Department of Physical Sciences, Indian Institute of Science Education and Research, Mohali, Punjab 140306, India \label{IISERM}}\and\fi
{Laboratoire de Physique et Chimie de l'Environnement et de l'Espace, Universit\'e d'Orl\'eans / CNRS, 45071 Orl\'eans Cedex 02, France \label{lpc2e}}\and
{Observatoire Radioastronomique de Nan\c{c}ay, Observatoire de Paris, Universit\'e PSL, Université d'Orl\'eans, CNRS, 18330 Nan\c{c}ay, France\label{nancay}}\and
{Dipartimento di Fisica ``G. Occhialini", Universit{\'a} degli Studi di Milano-Bicocca, Piazza della Scienza 3, I-20126 Milano, Italy\label{unimib}}\and
{INFN, Sezione di Milano-Bicocca, Piazza della Scienza 3, I-20126 Milano, Italy\label{infn-unimib}}\and
{INAF - Osservatorio Astronomico di Brera, via Brera 20, I-20121 Milano, Italy\label{inaf-brera}}\and
{Institute for Gravitational Wave Astronomy and School of Physics and Astronomy, University of Birmingham, Edgbaston, Birmingham B15 2TT, UK\label{unibir}}\and
{INAF - Osservatorio Astronomico di Cagliari, via della Scienza 5, 09047 Selargius (CA), Italy\label{inaf-oac}}\and
{Hellenic Open University, School of Science and Technology, 26335 Patras, Greece\label{HOU}}\and
\ifnum\wm=5{Université de Gen\`eve, Département de Physique Théorique and Centre for Astroparticle Physics, 24 quai Ernest-Ansermet, CH-1211 Genéve 4, Switzerland\label{unige}}\and\fi
\ifnum\wm=5{CERN, Theoretical Physics Department, 1 Esplanade des Particules, CH-1211 Gen\'{e}ve 23, Switzerland\label{CERN}}\and\fi
{Kavli Institute for Astronomy and Astrophysics, Peking University, Beijing 100871, P. R. China\label{kiaa}}\and
\ifnum\wm>1{Department of Astronomy and Astrophysics, Tata Institute of Fundamental Research, Homi Bhabha Road, Navy Nagar, Colaba, Mumbai 400005, India \label{TIFR}}\and\fi
\ifnum\wm>1{Department of Physics, IIT Hyderabad, Kandi, Telangana 502284, India \label{IITH_Ph}}\and\fi
\ifnum\wm>1{Department of Physics and Astrophysics, University of Delhi, Delhi 110007, India \label{UoD}}\and\fi
\ifnum\wm>1{Department of Earth and Space Sciences, Indian Institute of Space Science and Technology, Valiamala, Thiruvananthapuram, Kerala 695547,India \label{IIST}}\and\fi
\ifnum\wm=5{School of Mathematics and Physics, Faculty of Engineering and Physical Science, University of Surrey, Guildford GU2 7XH, UK\label{surrey}}\and\fi
{School of Physics, Faculty of Science, University of East Anglia, Norwich NR4 7TJ, UK\label{uea}}\and
\ifnum\wm=5{Sternberg Astronomical Institute, Moscow State University, Universitetsky pr., 13, Moscow 119234, Russia\label{SAI}}\and\fi
{Max Planck Institute for Gravitational Physics (Albert Einstein Institute), Am M{\"u}hlenberg 1, 14476 Potsdam, Germany\label{aei}}\and
{Gran Sasso Science Institute (GSSI), I-67100 L'Aquila, Italy \label{gssi}}\and
{INFN, Laboratori Nazionali del Gran Sasso, I-67100 Assergi, Italy \label{lngs}}\and 
\ifnum\wm>1{National Centre for Radio Astrophysics, Pune University Campus, Pune 411007, India \label{NCRA}}\and\fi
\ifnum\wm>1{Kumamoto University, Graduate School of Science and Technology, Kumamoto, 860-8555, Japan \label{KU_J}}\and\fi
{Universit{\'a} di Cagliari, Dipartimento di Fisica, S.P. Monserrato-Sestu Km 0,700 - 09042 Monserrato (CA), Italy\label{unica}}\and
{Department of Astrophysics/IMAPP, Radboud University Nijmegen, P.O. Box 9010, 6500 GL Nijmegen, The Netherlands\label{imapp}}\and
\ifnum\wm>1{Department of Physical Sciences,Indian Institute of Science Education and Research Kolkata, Mohanpur, 741246, India \label{IISERK}}\and\fi
\ifnum\wm>1{Center of Excellence in Space Sciences India, Indian Institute of Science Education and Research Kolkata, 741246, India \label{CESSI}}\and \fi
{School of Physics, Trinity College Dublin, College Green, Dublin 2, D02 PN40, Ireland\label{tcd}}\and
{Jodrell Bank Centre for Astrophysics, Department of Physics and Astronomy, University of Manchester, Manchester M13 9PL, UK\label{jbca}}\and
\ifnum\wm>1{Department of Physics, St. Xavier’s College (Autonomous), Mumbai 400001, India \label{XCM}}\and\fi
\ifnum\wm>1{Department of Astronomy,School of Physics, Peking University, Beijing 100871, P. R. China\label{pku}}\and\fi
\ifnum\wm>1{National Astronomical Observatories, Chinese Academy of Sciences, Beijing 100101, P. R. China\label{naoc}}\and\fi
{E.A. Milne Centre for Astrophysics, University of Hull, Cottingham Road, Kingston-upon-Hull, HU6 7RX, UK\label{milne}}\and
{Centre of Excellence for Data Science, Artificial Intelligence and Modelling (DAIM), University of Hull, Cottingham Road, Kingston-upon-Hull, HU6 7RX, UK\label{daim}}\and
\ifnum\wm=5{Laboratory of Astrophysics, \'Ecole Polytechnique F\'ed\'erale de Lausanne, CH-1015 Lausanne, Switzerland\label{EPFL}}\and\fi
\ifnum\wm>1{Department of Physics, BITS Pilani Hyderabad Campus, Hyderabad 500078, Telangana, India \label{BITS}}\and\fi
\ifnum\wm>1{Joint Astronomy Programme, Indian Institute of Science, Bengaluru, Karnataka, 560012, India \label{IISc}}\and\fi
{Arecibo Observatory, HC3 Box 53995, Arecibo, PR 00612, USA\label{arecibo}}\and
{IRFU, CEA, Université Paris-Saclay, F-91191 Gif-sur-Yvette, France \label{irfu}}\and
\ifnum\wm>1{Raman Research Institute India, Bengaluru, Karnataka, 560080, India \label{RRI}}\and\fi
{Institut f\"{u}r Physik und Astronomie, Universit\"{a}t Potsdam, Haus 28, Karl-Liebknecht-Str. 24/25, 14476, Potsdam, Germany\label{uni-potsdam}}\and
\ifnum\wm=5{Kazan Federal University, 18 Kremlyovskaya, 420008 Kazan, Russia\label{KFU}}\and\fi
\ifnum\wm>1{Department of Physics, IISER Bhopal, Bhopal Bypass Road, Bhauri, Bhopal 462066, Madhya Pradesh, India \label{IISERB}}\and\fi
{Ollscoil na Gaillimhe --- University of Galway, University Road, Galway, H91 TK33, Ireland\label{uog}}\and
\ifnum\wm>1{Center for Gravitation, Cosmology, and Astrophysics, University of Wisconsin-Milwaukee, Milwaukee, WI 53211, USA \label{CGCA}}\and\fi
\ifnum\wm>1{Division of Natural Science, Faculty of Advanced Science and Technology, Kumamoto University, 2-39-1 Kurokami, Kumamoto 860-8555, Japan \label{KU_J1}}\and\fi
\ifnum\wm>1{International Research Organization for Advanced Science and Technology, Kumamoto University, 2-39-1 Kurokami, Kumamoto 860-8555, Japan \label{KU_J2}}\and\fi
{Laboratoire Univers et Th{\'e}ories LUTh, Observatoire de Paris, Universit{\'e} PSL, CNRS, Universit{\'e} de Paris, 92190 Meudon, France\label{luth}}\and
\ifnum\wm=5{Max Planck Institute for Gravitational Physics (Albert Einstein Institute), Leibniz Universit\"at Hannover, Callinstrasse 38, D-30167, Hannover, Germany\label{aei2}}\and\fi
{Florida Space Institute, University of Central Florida, 12354 Research Parkway, Partnership 1 Building, Suite 214, Orlando, 32826-0650, FL, USA\label{FSI}}\and
{Ruhr University Bochum, Faculty of Physics and Astronomy, Astronomical Institute (AIRUB), 44780 Bochum, Germany \label{airub}}\and
{Advanced Institute of Natural Sciences, Beijing Normal University, Zhuhai 519087, China \label{bnuz}}\\
\ifnum\wm>1\\ {\defpta{epta} : The European Pulsar Timing Array}\fi
\ifnum\wm>1\\ {\defpta{inpta} : The Indian Pulsar Timing Array}\fi
}

   \date{Received XXX; accepted YYY}
\titlerunning{GWB Interpretation}
\authorrunning{EPTA+InPTA}

 
  \abstract
{The European Pulsar Timing Array (EPTA) and Indian Pulsar Timing Array (InPTA) collaborations have measured a low-frequency common signal in the combination of their second and first data releases, respectively, with the correlation properties of a gravitational wave background (GWB). Such a signal may have its origin in
a number of physical processes including a cosmic population of inspiralling supermassive black hole binaries (SMBHBs); inflation, phase transitions, cosmic strings, and tensor mode generation by the non-linear evolution of scalar perturbations in the early Universe; and oscillations of the Galactic potential in the presence of ultra-light dark matter (ULDM). At the current stage of emerging evidence, it is impossible to discriminate among the different origins. Therefore, for this paper, we consider each process separately, and investigated the implications of the signal under the hypothesis that it is generated by that specific process. We find that the signal is consistent with a cosmic population of inspiralling SMBHBs, and its relatively high amplitude can be used to place constraints on binary merger timescales and the SMBH-host galaxy scaling relations. If this origin is confirmed, this would be the first direct evidence that SMBHBs merge in nature, adding an important observational piece to the puzzle of structure formation and galaxy evolution. As for early Universe processes, the measurement would place tight constraints on the cosmic string tension and on the level of turbulence developed by first-order phase transitions. Other processes would require non-standard scenarios, such as a blue-tilted inflationary spectrum or an excess in the primordial spectrum of scalar perturbations at large wavenumbers. Finally, a ULDM origin of the detected signal is disfavoured, which leads to direct constraints on the abundance of ULDM in our Galaxy.} 


   \maketitle
%
\section{Introduction}
The recent observation of a common signal with excess power in the nanohertz frequency ranges ~\citep[i.e. a common red signal, as defined in][]{ng12p5gwb+20,gsr+2021,ccg+21} in pulsar timing array  (PTA) datasets, with emerging evidence for quadrupolar correlations\footnote{Readers can refer to~\cite{wm3} for more details on inter-pulsar correlations.} opens a new era in the exploration of the Universe. 
This important milestone has been achieved thanks to the efforts of the European Pulsar Timing Array \citep[EPTA,][]{dcl+16}, the Indian PTA \citep[InPTA][]{bgp+22}, the North American Nanohertz Observatory for Gravitational Waves \citep[NANOGrav,][]{mcl2013}, the Parkes PTA \citep[PPTA,][]{mhb+2013}, and the Chinese PTA \citep[CPTA,][]{2016ASPC..502...19L}.
Although the significance of the signal does not yet reach the 5$\sigma$ mark, which is generally accepted as the `golden rule' for a firm detection claim, the evidence reported by the different collaborations ranges between $2\sigma$ and $4\sigma$ \citep{wm3,ng15yr,pptadr3,cptadr1}, strongly suggesting a genuine gravitational wave (GW) origin. Awaiting decisive confirmation within the International PTA \citep[IPTA,][]{vlh+2016,pdd+2019} framework, with the additional contribution of the  MeerKAT PTA \citep{2023MNRAS.519.3976M}, we are hearing, for the first time, the faint murmur of the GW Universe at frequencies of 1-to-30 nano-Hz, which is ten orders of magnitude lower than the frequencies currently probed by ground-based interferometers \citep{abb2016}. This opens a completely new window onto the Universe, allowing us to look at different phenomena, probe new astrophysical and cosmological sources, and, potentially, new physics as well.

By monitoring an array of millisecond pulsars (MSPs) for decades with a weekly cadence, PTAs are sensitive to GWs in the $10^{-9}$--$10^{-7}$Hz range \citep{fb1990}. At those frequencies, the most anticipated signal to be detected is a stochastic GW background (GWB) produced by the incoherent superposition of waves emitted by adiabatically inspiralling supermassive black hole binaries \citep[SMBHBs,][]{rr1995,jb2003}. 
The signal manifests as a stochastic Gaussian process characterised by a power-law Fourier spectrum of delays-advances to pulse arrival times, with characteristic inter-pulsar correlations of general relativity identified by~\cite{hd1983}.
The statistical properties of the signal are expected to significantly deviate from the typical isotropy, Gaussianity and perhaps even stationarity that is typical of many stochastic signals from the early Universe \citep[e.g.][]{svc08,2012ApJ...761...84R}. In fact, due the shape of the SMBHB mass function and their redshift distribution, the overall signal is often dominated by a few sources, and particularly massive, nearby SMBHBs might result in loud enough signals to be individually resolved as continuous GWs \citep[CGWs,][]{2009MNRAS.394.2255S,Babak:2011mr,2018MNRAS.477..964K} emerging from the GWB. The exact amplitude and spectral shape of the spectrum are intimately related to the cosmological galaxy merger rate and to the dynamical properties of the emitting binaries forming in the aftermath of the merger process \citep{2007PThPh.117..241E,2011MNRAS.411.1467K,2013CQGra..30v4014S,ses2013,2014MNRAS.442...56R}. Therefore, the demonstration of an SMBHB origin of the signal observed by PTAs provides invaluable insights into the formation, evolution, and dynamics of these objects.
Moreover, it brings decisive evidence that SMBHBs merge in nature, thus overcoming the `final parsec problem' \citep{2003ApJ...596..860M}, which is still an open question in our understanding of galaxy and structure formation.

Beyond SMBHBs, a number of processes (potentially) occurring in the early Universe can also produce a stochastic GWB at nanohertz frequencies. 
Tensor modes can be produced as early as during the first tiny fraction of a second after the Big Bang through quantum fluctuations of the gravitational field stretched by the accelerated expansion of inflation \citep{1975JETP...40..409G, 1982PhLB..115..189R,1985SvAL...11..133S,1983PhLB..125..445F, 1984NuPhB.244..541A}. In the literature, these GWs are referred to as `primordial'. In this case, the shape of the power spectrum is defined by the specific model of inflation. Classical tensor mode production invoking the presence of a source term in the GW equation of motion can also take place in the early Universe. There are a plethora of physical processes that can lead to such a 
source term, and trigger the production of GWs. Topological defects, for example decaying cosmic string loops \citep{2000csot.book.....V, 2001PhRvD..64f4008D, 2003PhLB..563....6J, 2004JCAP...03..010D, 2005PhRvD..71f3510D}, particle production during inflation \citep{2011JCAP...06..003S, 2012PhRvD..86j3508B,2013JCAP...11..047C,2012PhRvD..85l3537A}, (magneto-)hydrodynamic turbulence \citep[(M)HD,][]{Kamionkowski:1993fg,
2002PhRvD..66b4030K, 2002PhRvD..66j3505D,2006PhRvD..74f3521C,2007PhRvD..76h3002G,2009JCAP...12..024C}, the collision of bubble walls during a first-order primordial phase transition \citep{1992PhRvD..45.4514K, 1993PhRvD..47.4372K,2008PhRvD..77l4015C,2008JCAP...09..022H,2017PhRvD..95b4009J,2018PhRvD..97l3513C}, sound waves in the aftermath of a first-order phase transition \citep{2014PhRvL.112d1301H,2015PhRvD..92l3009H,2017PhRvD..96j3520H}, as well as scalar perturbations at second order in cosmological perturbation theory \citep{2007PhRvD..76h4019B, 2007PhRvD..75l3518A}, are among the most commonly considered scenarios. GWs decouple from the rest of the matter and radiation immediately after their generation at essentially any temperature in the Universe, so that in the case of clear observational evidence for these types of signals, we can infer nearly unaltered information on the physical processes occurring during or just after the birth of the Universe \citep{2018CQGra..35p3001C}. 

Contrary to the incoherent superposition of GWs from SMBHBs, the stochastic GWB from sources in the early Universe is usually assumed to be statistically homogeneous and isotropic, unpolarised, and Gaussian \citep{Allen:1996vm,Maggiore:1999vm,2018CQGra..35p3001C}. Statistical homogeneity and isotropy are inherited from the same property of the FLRW Universe. The absence of polarisation holds provided no macroscopic source of parity violation is present in the Universe. Gaussianity follows by the central limit theorem in most cases of GWBs generated by processes operating independently in many uncorrelated, sub-horizon regions. This also applies to the irreducible GWB generated during inflation in the simplest scenarios, when the tensor metric perturbation can be quantised as a free field, and hence with Gaussian probability distribution for the amplitude. There are, however, notable exceptions, among which for example rare GWB bursts from cosmic strings cusps and kinks \citep{2000PhRvL..85.3761D,2001PhRvD..64f4008D}, or the GWB from particle production during inflation \citep{Cook:2013xea,Sorbo:2011rz,Anber:2012du}. Therefore, although statistical properties might be useful for discriminating between SMBHBs and several processes in the early Universe, a full assessment of the nature of the GW signal will not be trivial.

Spatially correlated delays of the time of arrivals (TOAs) in an array of MSPs are not a unique imprint of GWs. For example, it is well known \citep[e.g.][]{thk+2016} that such delays can emerge due to the imperfect fitting of the solar system ephemerides (dipolar correlated noise), or due to a miscalibration of the time standard to which the measured TOAs are referred (monopolar correlated noise). Furthermore, individual Fourier harmonics of a common signal in PTA data may include contributions from the oscillations of the gravitational potential associated with the presence of ultralight dark matter ~\citep[ULDM,][]{Smarra_2023}\footnote{In the pulsar timing community, it is customary to refer to the dispersion measure with the acronym DM. Please note that in this paper, DM will exclusively stand for Dark Matter.}, also known as \textit{fuzzy} dark matter (FDM), in the Galactic halo \citep{Khmelnitsky_2014}. The existence of ultralight scalars, generally motivated by string-theoretical frameworks \citep{Green_1987, Svrcek_2006, Arvanitaki_2010}, is also particularly appealing from the astrophysical and cosmological point of view. In fact, several potential issues in the small-scale structure of the Universe, such as the cusp vs core \citep{Flores_1994,Moore_1994, Karukes_2015} or the missing satellite \citep{Klypin_1999, Moore_1999} problems, could be disposed of or, at least, mitigated assuming that dark matter is made of ultralight particles. As predicated by \cite{Khmelnitsky_2014}, the presence of ULDM induces harmonic delays in the arrival times, with a frequency proportional to the ultralight boson mass.

In this paper, we provide a broad overview of the implications of the signal observed in the second data release of the EPTA+InPTA \citep{wm1} for the different physical processes mentioned above. More in-depth analysis of several of these scenarios will be the subject of separate future publications. Unless otherwise stated, we consider each process separately, and we discuss the implications of the detected signal under the hypothesis that it is generated by that specific process. We do not attempt any Bayesian model selection on the signal origin, although a general framework for that is being developed \citep{2021NatAs...5.1268M}. 
The main reason for this choice is that, at this stage of data taking and analysis, the information carried by the signal is not particularly constraining; the evidence of the measurement is still at $\approx3\sigma$, and the amplitude and spectral shape of the signal are not very well measured. With these premises, the result of any model selection is bound to be severely influenced by the priors employed for each of the models under examination. This exercise becomes more meaningful as data get more informative, which we strive to achieve with the analysis of the third release of the combined IPTA data, which is now being assembled.

The paper is organised as follows. In Sec.~\ref{sec:data}, we describe the signal observed by EPTA+InPTA and its main features, including its free spectrum and best-fit parameters. We then proceed with detailing the main implications of the detected signals under the assumption that it is generated by a cosmic population of SMBHBs (Sec.~\ref{sec:SMBHBs}) or by a number of processes occurring in the early Universe (Sec.~\ref{sec:early_universe}). In Sec.~\ref{sec:dark_matter}, we investigate the compatibility of the observed signal with a DM origin and place constraints on ULDM candidates. Finally, in Sec.~\ref{sec:conclusions}, we summarise our main results and discuss future prospects.

\section{The observed signal in the EPTA DR2 dataset}
\label{sec:data}

\begin{figure*}
    \centering
    \includegraphics[width=0.5\textwidth]{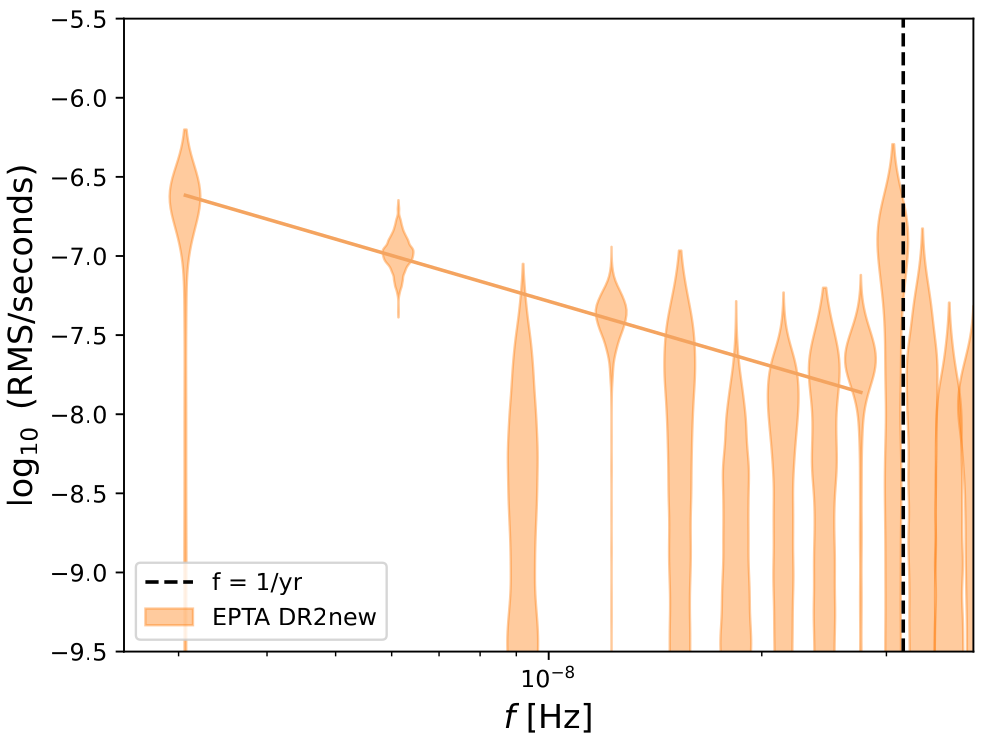}
    \includegraphics[width=0.4\textwidth]{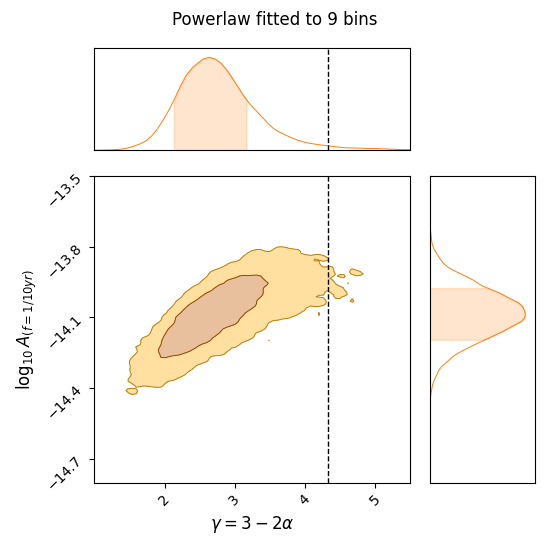}
    \caption{\footnotesize{Properties of the common correlated signal detected in \texttt{DR2new}. Left panel: free spectrum of the RMS induced by the excess correlated signal in each frequency resolution bin (with width defined by the inverse of the data span, $\Delta{f}=T^{-1}$). The straight line is the best power-law fit to the data. Right panel: joint posterior distribution in the $A-\gamma$ plane. \it{Note that we normalise $A$ to a pivotal frequency $f_0=10{\rm yr}^{-1}$.}}}
    \label{fig:EPTADR2}
\end{figure*}

Our investigation is based on the results reported in \citet[][hereinafter PaperIII]{wm3}, which analyses the data of 25  MSPs collected by the EPTA using five of the largest radio telescopes in Europe: the Lovell telescope at the Jodrell Bank Observatory, the Nan\c{c}ay decimetric radio telescope, the Westerbork synthesis radio telescope, the Effelsberg 100\,m radio telescope, and the Sardinia radio telescope. The dataset also includes the Large European Array for Pulsars (LEAP) data, in which individual telescope observations are coherently phased to form an equivalent dish with a diameter of up to 194\,m \citep{bjk+16}. These data are complemented by low-frequency observations of a subset of 10 MSPs performed by the InPTA using the upgraded Giant Metrewave Radio Telescope (uGMRT) and covering about 3.5 years.

The data of each individual pulsar are combined as described in \cite{wm1} and the noise properties of each pulsar are then extracted according to the optimised custom noise models presented in \cite{wm2}. The final result is a dataset of unprecedented sensitivity spanning up to 24.7 years. Four versions of the dataset were analyzed:  
\begin{enumerate}
    \item \texttt{DR2full}. 24.7 years of data taken by the EPTA;
    \item \texttt{DR2new}. 10.3 years of data collected by the EPTA using new-generation wide-band backends;
    \item \texttt{DR2full+}. The same as \texttt{DR2full}, but with the addition of InPTA data; 
    \item \texttt{DR2new+}. The same as \texttt{DR2new}, but with the addition of InPTA data.
\end{enumerate}

The analysis presented in this paper refers to the \texttt{DR2new} dataset only. We do not consider \texttt{DR2full} and \texttt{DR2full+} because evidence of quadrupolar correlation \citep[usually referred to as HD correlation, from][]{hd1983} of the common process is weaker in those datasets, potentially due to the lower quality of early data that were collected with narrowband backends (see discussion in PaperIII). On the other hand, although the analysis of \texttt{DR2new+} produced results in broad agreement with \texttt{DR2new}, that dataset was assembled relatively recently and has not been analysed as thoroughly. For example, the binned free-spectra that we will use in some of the following analyses have only been produced after this work was completed.


Before proceeding with the description of the signal detected in \texttt{DR2new}, here we summarise some notations used in PTA analysis for the benefit of the reader. The perturbation affecting the TOAs, whether produced by GWs or DM, is described in terms of its dimensionless strain $h$. A broad-band stochastic perturbation is defined by its characteristic dimensionless strain $h_c(f)$, often modelled as a power law 
\begin{equation}
    h_c(f)=A\left(\frac{f}{f_0}\right)^{\alpha}.
    \label{eq:hc}
\end{equation}
For example, a population of GW-driven circular SMBHBs produces a spectrum with $\alpha=-2/3$ and amplitude $A\approx10^{-15}$, assuming $f_0=1{\rm yr}^{-1}$. $h_c(f)$ is connected to the differential energy content of the signal per logarithmic frequency through the equation:
\begin{equation}
\Omega(f)=\frac{2\pi^2}{3H_0^2}f^2h_c^2(f),
    \label{eq:omega}
\end{equation}
where $H_0$ is today's Hubble expansion parameter. We note that  $h_c(f)$ and $\Omega(f)$ provide equivalent parametrizations of the spectrum. The former is more popular in the astrophysics domain, whereas the latter is the preferred choice for early Universe and cosmology.

Given $h_c(f)$, the one-sided power spectral density induced by the GW signal in the timing residuals is given by \citep{ltm+2015}: 
\begin{equation}
    S(f)=\frac{h_c^2(f)}{12\pi^2 f^3}=\frac{A^2}{12\pi^2f_0^{2\alpha}}f^{-\gamma},
    \label{eq:sh}
\end{equation}
where $\gamma=3-2\alpha$. PTAs search for HD correlated time delays with such a power spectrum in the data, and measure the parameters $A$ and $\gamma$. For an observation timespan $T$, measurements are discretised in frequency bins $\Delta{f_i}=f_{i+1}-f_i$, where $f_i=i/T$. It is then customary to convert $S(f)$ in RMS residual induced in the TOAs in each frequency bin:
\begin{equation}
    {\rm RMS}_i=\left(\int_{\Delta{f_i}} S(f){\rm d}f\right)^{1/2}\approx\left(S(f_i)\Delta{f_i}\right)^{1/2}=\left(\frac{S(f_i)}{T}\right)^{1/2}. 
    \label{eq:RMS}
\end{equation}

The main properties of the GWB signal observed in \texttt{DR2new} and examined in this paper are shown in Fig.~\ref{fig:EPTADR2}. The length of the dataset is $T=$10.3 years, and excess common correlated power is detected in several frequency bins up to $\approx 30\,$nHz (Fig.~\ref{fig:EPTADR2} left panel). Conversely, some bins are unconstrained, which results in a relatively loose determination of the spectral properties of the observed signal. In the literature, $h_c$ and $S$ in Eqs.~\eqref{eq:hc} and \eqref{eq:sh} are usually anchored to the pivotal frequency $f_0=1{\rm yr}^{-1}$. The data are, however, most informative at the lowest frequencies, while the common power at $1{\rm yr}^{-1}$ is essentially unconstrained. This naturally leads to a strong degeneracy of the $A-\gamma$ 2D posterior, as shown for example in Figure 1 of PaperIII. Therefore, unless otherwise stated, we change the reference frequency to $f_0=10{\rm yr}^{-1}$, where the data are actually constraining, which results in a weaker dependence of $A$ upon $\gamma$, as shown in the right panel of Fig.~\ref{fig:EPTADR2}.    

In the following three sections, we discuss three possible contributions to the signal, probing completely different epochs and scales of our Universe, and the implications for the associated physical processes. 
Namely, the cosmic population of SMBHBs (at redshifts $z \lesssim 1$), the early Universe ($z > 1000$), and DM (within our Galaxy).

\section{Implications I: supermassive black hole binaries} 
\label{sec:SMBHBs}

A cosmic population of SMBHBs is the primary astrophysical candidate to produce a signal in the nanohertz band detectable by PTAs. If we define ${\rm d}^5N/({\rm d}z{\rm d}m_1{\rm d}q{\rm d}e{\rm d}t_r)$ as the cosmic merger rate of SMBHBs as a function of redshift, primary black hole mass, mass ratio, and eccentricity, the general form of the generated GWB as a function of observed frequency $f$ can be written as \citep{2013CQGra..30v4014S}
\begin{multline}
 h_c^2(f) = \int_0^{\infty}{\rm d}z \int_0^{\infty}{\rm d}m_1\int_0^{1}{\rm d}q \frac{{\rm d}^5N}{{\rm d}z{\rm d}m_1{\rm d}q{\rm d}e{\rm d}t_r}\frac{{\rm d}t_r}{{\rm d}{\rm ln}f_{{\rm K},r}}\times\\
 h^2(f_{{\rm K},r})\sum_{n=1}^{\infty}\frac{g[n,e(f_{{\rm K},r})]}{(n/2)^2}\bigg|_{f_{K,r}=f(1+z)/n}.
 \label{eq:hch2}
 \end{multline}

Here, ${\rm d}t_r/{{\rm d}{\rm ln}f_{{\rm K},r}}$ is the time spent by the shrinking binary within a given logarithmic orbital frequency bin, which converts the merger rate into the distribution of rest-frame orbital frequencies of the emitting population. This quantity depends on the physical processes driving the evolution of the SMBHBs including, besides GW emission, interaction with the stellar and gaseous environment surrounding them. As such, it is generally a function of the binary parameters $m_1, q, e$, and extra parameters describing the environment, such as the stellar density in the nucleus of the galaxy host \citep[for more details, see][]{2013CQGra..30v4014S}.
The second line of Eq.~\eqref{eq:hch2} is the strain amplitude produced by each individual eccentric SMBHB binary, cast as the sum of harmonics fulfilling the condition ${f_{K,r}=f(1+z)/n}$. In that expression, $h(f_{K,r})$ is the strain of the equivalent circular binary emitted at twice the orbital frequency of the system, as given in Eq. 4-7 of \cite{2015MNRAS.451.2417R}, and $g(n,e)$ is a combination of Bessel functions \citep[see, e.g.][for details]{2020PhRvD.102j3023B}. 
For a distribution of circular, GW-driven binaries, the only relevant mass parameter is the chirp mass $\Mc = (m_1 m_2)^{3/5} (m_1 + m_2)^{1/5}$, and Eq.~\eqref{eq:hch2} takes the familiar form \citep{svc08}
\begin{equation}
 h_c^2(f) = \int_0^{\infty}{\rm d}z\int_0^{\infty}{\rm d}\Mc\frac{{\rm d}^3N}{{\rm d}z{\rm d}{\cal M}{\rm d}{\rm ln}f_r}h^2(f_r).
\label{eq:hch2_circ}
\end{equation}
This can be recast in terms of the comoving number density of merging binaries ${{\rm d}^2 n}/({{\rm d}z {\rm d}\Mc})$ ~\citep{Phinney:2001}
\begin{equation}
h_{\rm c}^2(f) = \frac{4 G^{5/3}}{3\pi^{1/3}c^2} f^{-4/3} 
               \int {\rm d} \Mc \int {\rm d} z 
               \left(1+z\right)^{-1/3} \Mc^{5/3} \frac{{\rm d}^2 n}{{\rm d}z {\rm d}\Mc}\,,
\label{eq:hcCircularPop}
\end{equation}
which highlights that, in this case, the expected spectrum follows a power law $h_c\propto f^{-2/3}$, and the only free parameter is its overall amplitude. The latter is set by the function ${{\rm d}^2 n}/({{\rm d}z {\rm d}\Mc})$, which contains all the astrophysical knowledge of the cosmic population of merging SMBHBs, and is determined by the cosmological hierarchical assembly of galaxies and their central SMBHs. Conversely, in the most general case described by Eq.~\eqref{eq:hch2} there is also information in the spectral shape of the signal, since coupling with the environment as well as eccentricity affect the function ${\rm d}t_r/{{\rm d}{\rm ln}f_{{\rm K},r}}$, suppressing signal at the lowest frequencies. Moreover, eccentricity distributes the GW power at several higher harmonics of the orbital frequency, slightly modifying the power-law behaviour at high frequencies. In general, the GWB cannot be cast in term of a simple analytical form, although a broken power-law is a sufficient approximation for most situations \citep[see, e.g.][]{2014MNRAS.442...56R,2015ASSP...40..147S,kbh+2017,csd2017}.
 
 The literature investigating the GWB produced by a population of SMBHBs is vast, dating back to the mid-nineties and early 2000s \citep{rr1995,jb2003,wl2003,ses+2004}, and predictions have been made by employing different models and techniques. Models can be broadly classified into two categories: self-consistent theoretical models for SMBH evolution within their galaxies \citep{svc08,2009MNRAS.394.2255S,2012ApJ...761...84R,2015ApJ...799..178K,kbh+2017,2018MNRAS.477.2599B,2020MNRAS.498..537S,2022MNRAS.509.3488I}, and empirical models based on observed properties of galaxy pairs coupled to SMBH-host galaxy relations \citep{ses2013,2015MNRAS.451.2417R,2015MNRAS.447.2772R,2023ApJ...949L..24S}, or on the evolution of the SMBH mass function inferred from observations \citep{2014ApJ...789..156M}. Note that we group both semianalytic models (SAMs) and large cosmological simulations in the first class. The main difference between these two classes is that self-consistent models are constructed to reproduce a large array of observations related to galaxies and the SMBH they host, such as the redshift-dependent galaxy mass function, quasar luminosity function, and so on. Conversely, empirical models are, by construction, consistent with the observations upon which they are based, but are usually not tested against independent constraints. As a consequence, they can generally produce a wider distribution of GWB amplitudes, but consistency with other observations is not necessarily guaranteed.   

In this Section, we investigate the implications of the signal observed in the \texttt{DR2new} dataset for representative models of the two classes. In Sec.~\ref{sec:MBHB_rosado}, we perform a semi-quantitative comparison between the measured signal and predictions of an extended version of the \cite{2015MNRAS.451.2417R}  models (hereinafter RSG15) including binary eccentricity and environmental coupling. In Sec.~\ref{sec:MBHB_inference}, we exploit the framework developed in \cite{2016MNRAS.455L..72M,2017MNRAS.468..404C,csc2019} to draw inference on SMBHB astrophysics from the data, either by assuming astrophysical priors from independent observations, or by using a completely generic model for the SMBHB mass function with minimal assumptions. In Sec.~\ref{sec:MBHB_SAM}, we demonstrate how the measured signal can inform galaxy and SMBH formation models by examining its constraining power on two state-of-the-art SAMs, namely \texttt{L-galaxies} \citep{2015MNRAS.451.2663H} and the model developed by Barausse and collaborators \citep{B12,B+18,B+20}. We discuss caveats and future directions of improvement in Sec.~\ref{sec:considerations}.

\subsection{Qualitative analysis of empirical SMBHB population models}
\label{sec:MBHB_rosado}

To carry out a semi-quantitative comparison between observations and empirical models, we use an extended set of SMBHB populations based on the work of \cite{ses2013} (S13 hereinafter) and RSG15. 

\subsubsection{Description of the models}
\begin{figure}
\includegraphics[width=0.47\textwidth]{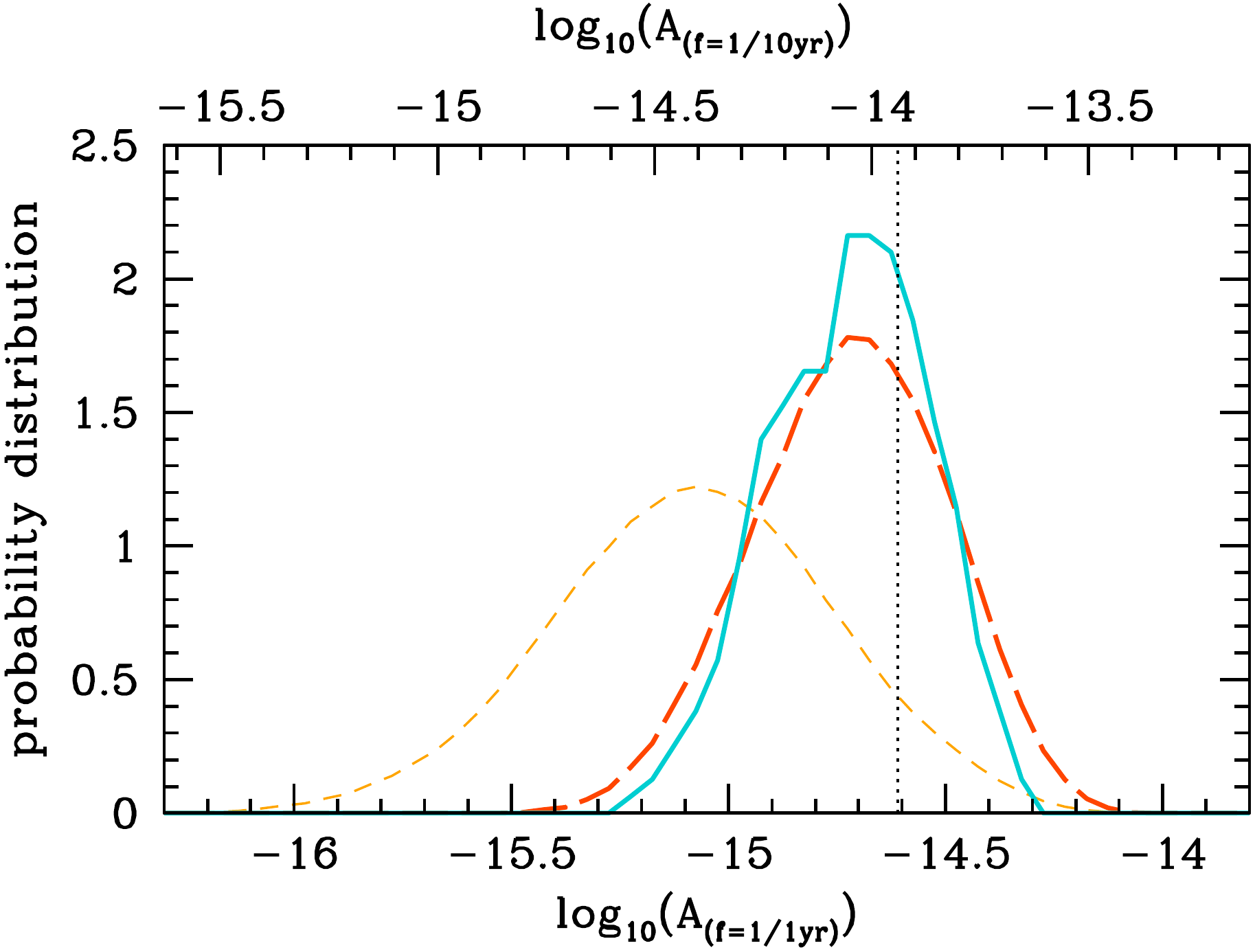}
\caption{\footnotesize{GWB amplitude distributions predicted by the RSG15 models. The thin-dashed yellow line is for the full set of models in RSG15, whereas the thick-dashed orange line is for the subset considered here. The solid blue line is the distribution predicted by the 108 down-selected sample used in the analysis. The vertical line marks the median value of $A$ at $f_0=1{\rm yr}^{-1}$ reported in PaperIII when fixing $\gamma=13/3$ in the search. Note that the lower $x$-axis scale is for A at $f_0=1{\rm yr}^{-1}$, whereas the upper $x$-axis is for A at $f_0=10{\rm yr}^{-1}$ (the normalization used in this paper). Since $\alpha=-2/3$ for circular GW-driven binaries, there is a shift of 0.666 dex between the two.}}
\label{fig:A_Rosado}
\end{figure}

As described in S13, the models are constructed around the number density of merging SMBHs per unit primary mass, mass ratio, and redshift, ${\rm d}^3n/({\rm d}m_1{\rm d}q{\rm d}z)$ obtained by combining different observations of the galaxy mass function and pair fraction, estimated galaxy pair merger timescales, SMBH-host galaxy relations, and prescription for SMBH accretion during mergers (see Section 2 of S13 for full details). Given the large uncertainties in all of the ingredients, the models predict a broad distribution of expected GW amplitudes, as shown in Fig.~\ref{fig:A_Rosado}. 

Guided by the relatively large amplitude of the detected signal and by theoretical and observational advancements in the last decade, we select a sub-sample of those models, as we now justify. First, hydrodynamical simulations of merging galaxies at different scales as well as deep X-ray observations of merging galaxies support accretion activation onto the individual SMBHs prior to merger \citep[e.g.][]{2017MNRAS.469.4437C,2018Natur.563..214K}. Moreover, hydrodynamical simulations of sub-pc scale binaries, have found most of the accretion to occur on the secondary (i.e. less massive) SMBH \citep{2014ApJ...783..134F}. We therefore restrict the analysis to models where SMBHs accrete prior to the merger, either with an equal Eddington ratio or with preferential accretion onto the secondary.\footnote{This is implemented according to a simple scheme whereby accretion is quenched on the primary SMBH until the binary becomes equal mass. If/when this occurs, further accretion is equally distributed among the two (now equal mass) binary components.} Second, observations of overmassive black holes in the centre of large ellipticals \citep{2011Natur.480..215M} has led to an upward revision of the SMBH-galaxy relations. Contrary to S13 and RSG15, here we consider only those revised realations, namely the ones reported by \cite{2013ARA&A..51..511K,2013ApJ...764..184M,2013ApJ...764..151G}. Finally, given the large number of models, to save computing power, we perform an ad hoc down-selection of 108 models that preserves the overall distribution of the expected GWB amplitudes, as shown in Fig.~\ref{fig:A_Rosado}.

As opposed to RSG15, we go beyond the circular-GW driven binary approximation and consider the self-consistent evolution of SMBHBs within their stellar environment.\footnote{Most of the signal comes from SMBHBs hosted in massive, low-redshift galaxies \citep[but see][]{2023ApJ...949L..24S}, which are relatively gas-poor.} This is done by employing the hardening models of \cite{2010ApJ...719..851S} that self-consistently evolve the SMBHB semimajor axis and eccentricity under the combined effect of stellar hardening and GW emission, once a given initial eccentricity $e_0$ at binary formation is given. Those evolutionary tracks allow us to evaluate the term ${\rm d}t/{{\rm d}{\rm ln}f_{{\rm K},r}}$ in Eq.~\eqref{eq:hch2}, and thus to
reconstruct from the density distribution of merging binaries, ${\rm d}^3n/({\rm d}m_1{\rm d}q{\rm d}z)$, the numerical distribution of systems emitting at any time in the whole sky as a function of mass, mass ratio, redshift, orbital frequency and eccentricity, ${\rm d}^5N/({\rm d}m_1{\rm d}q{\rm d}z{\rm d}f{\rm d}e)$. For each model, we consider 10 values of $e_0=0, 0.1, ..., 0.9$ and three different normalizations of the stellar density profile, described as $\rho=C\times \rho_0(r/r_0)^{-1.5}$, with $C=0.1, 1, 10$ \citep[details in][]{2010ApJ...719..851S}. 

In total, we explore 108$\times$10$\times$3$=3240$ models, spanning different eccentricities and densities of the stellar environment. For each model, we draw 100 Monte Carlo samplings of the distribution of the emitting binaries, we discretise the frequency domain in bins of $\Delta{f}=10.3{\rm yr}^{-1}$, and add the signals of binaries falling in the same bin in quadrature. This leads to the binned characteristic strain spectrum $h_c(f)$ that we then convert in $S(f)$ and RMS residuals using Eqs. \eqref{eq:sh} and \eqref{eq:RMS}. The full procedure is described in \cite{2010MNRAS.402.2308A,2020PhRvD.102j3023B}. In this way, we generate a grand total of 324k Monte Carlo realizations of the predicted GW spectrum in the PTA band.

\subsubsection{Comparison with the observed signal}

\begin{figure}
\includegraphics[width=0.5\textwidth]{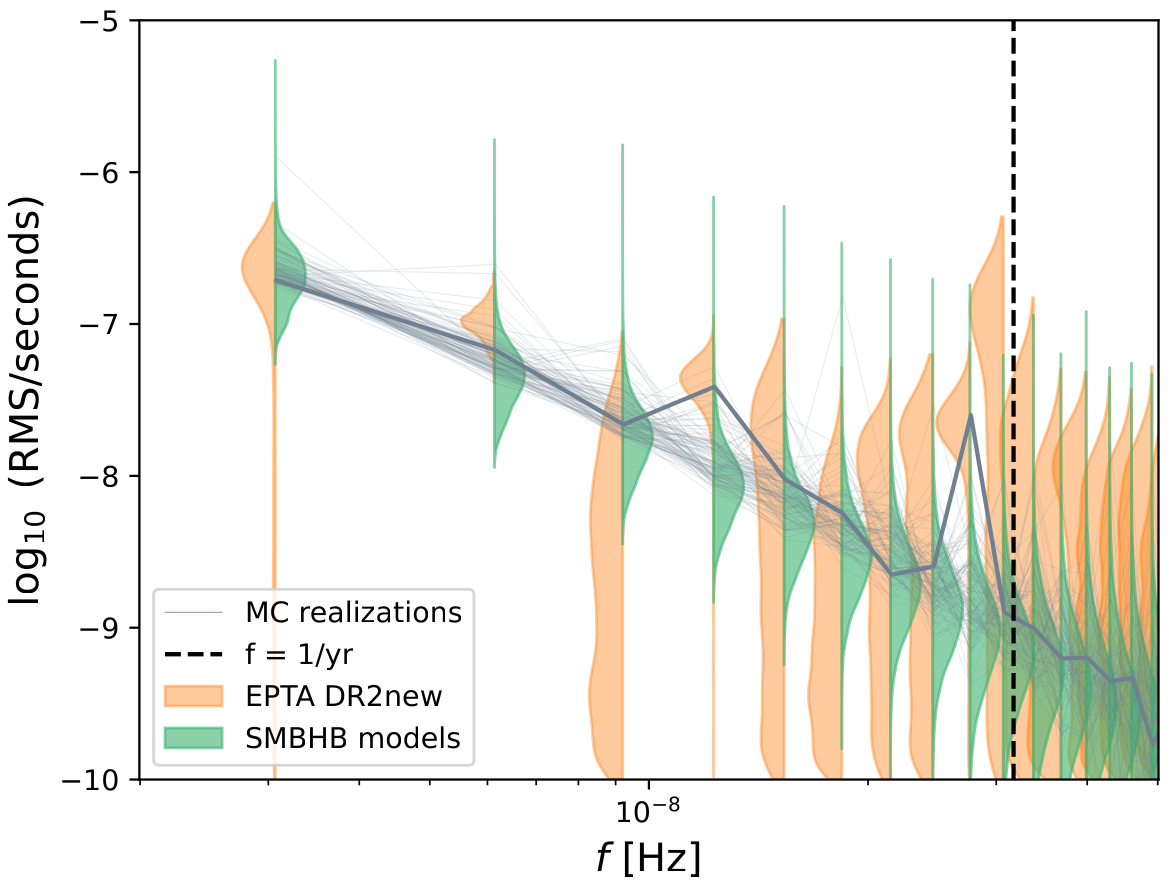}
\caption{\footnotesize{Free spectrum violin plot comparing measured (orange) and expected (green) signals. Overlaid to the violins are the 100 Monte Carlo realizations of one specific model; among those, the thick one represents an example of a SMBHB signal consistent with the excess power measured in the data at all frequencies.}}
\label{fig:violin_allsignal}
\end{figure}

The binned spectrum shown in Fig.~\ref{fig:violin_allsignal} contrasts expectations from the 324k models (green) to the measured correlated signal in \texttt{DR2new} (orange). The two sets of violin plots are in good agreement in the few lowest frequency bins, where measurements are the most constraining. Note that the model prediction distributions are highly non-Gaussian and asymmetric, with long tails extending upwards. This is due to the fact that sparse very massive/nearby binaries can sometimes produce exceptionally loud signals, as illustrated by the 100 individual GWBs overplotted to the violins. In fact, this might explain the extra power measured in the 4th and, most strikingly, in the 9th lowest bins compared to the bulk of the model predictions. We caution that the 9th bin is close to the 1yr$^{-1}$ mark, where PTAs are blind due to fitting for the Earth orbital motion, and leakage from imperfect fitting might affect that measurement. In any case, if this extra power is indeed due to GWs, it can be easily accommodated by theoretical models, as demonstrated by the realization highlighted by the tick grey line. 

\begin{figure}
\includegraphics[width=0.48\textwidth]{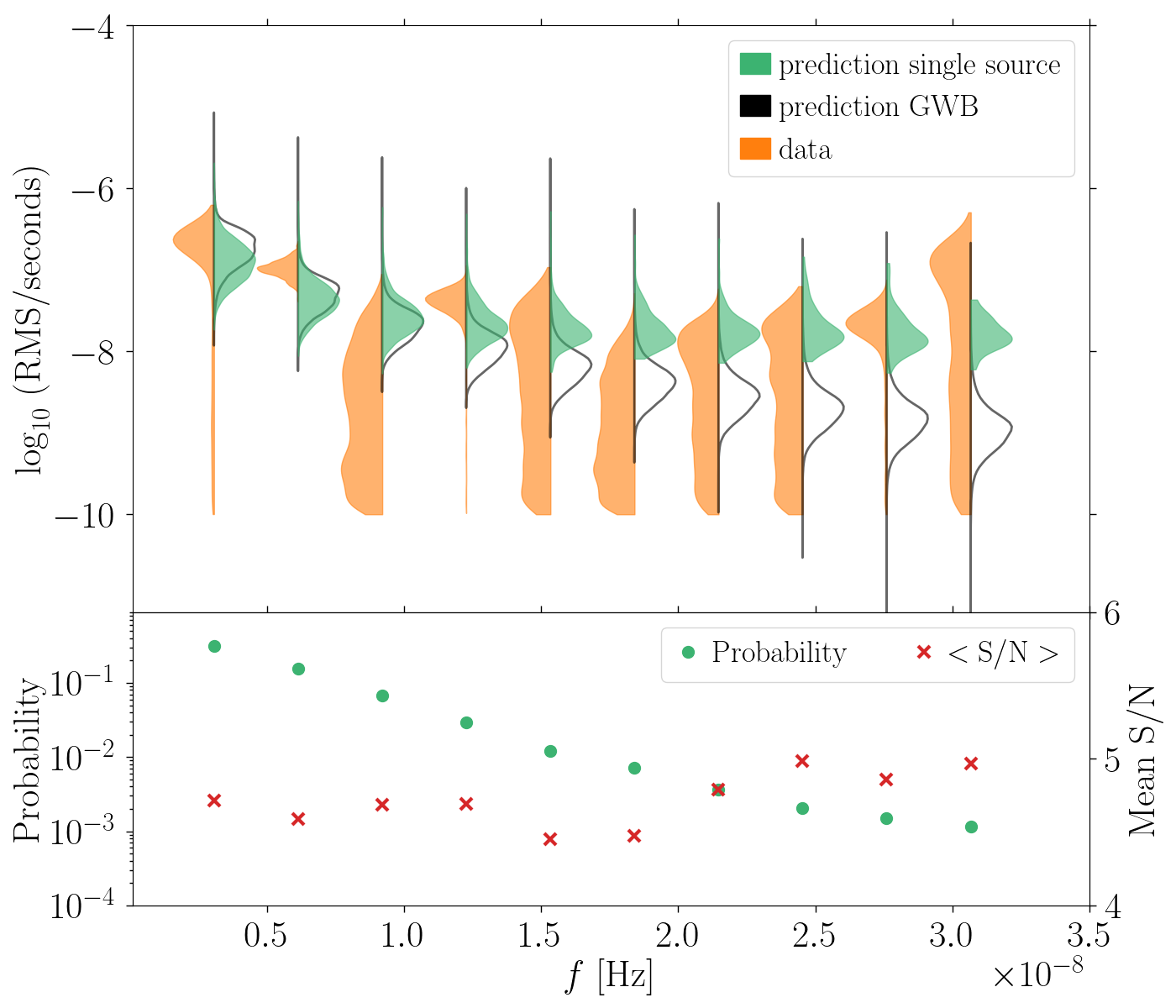}
\caption{\footnotesize{Expected properties of CGWs as a function of frequency. Top panel: free spectrum violin plot comparing the measured signal (orange) to the power distribution of CGWs (green). Empty violins show the full GWB produced by the models for comparison. Bottom panel: the probability of detecting a CGW with S/N$>3$ as a function of frequency (green circles, left $y-$axis scale). The average S/N of CGWs is also shown as red crosses (right $y-$axis scale).}}
\label{fig:violin_single}
\end{figure}

Our Monte Carlo approach to generate the SMBHB population and its associated GW signal also allows us to investigate the occurrence of CGWs in the data, for which evidence in \texttt{DR2new} is found to be inconclusive \citep{wm4}. Since the search performed in that paper was limited to circular binaries, we only carry out this analysis for the 32.4k models with $e_0=0$.\footnote{For binaries starting with $e_0=0$, eccentricities remain well below $0.1$ in the course of stellar hardening-driven evolution, and the GW signals can be approximated as monochromatic.} A full assessment of the detectability of CGWs requires the evaluation of the detection probability of each individual binary for a given false alarm rate, as detailed in RSG15. For the sake of simplicity, and given the qualitative nature of this analysis, we just compute the signal-to-noise ratio (S/N) of each individual binary according to Eq.~(46) of RSG15 (thus also restricting to the Earth term only). When computing the S/N of a source, we model each pulsar noise by using the maximum likelihood values of the single pulsar noise analysis presented in \cite{wm2}, and add the GWB produced by all of the other binaries to the noise spectral density. We arbitrarily set the detectability threshold at S/N$=3$ in the following.

Results are shown in Fig.~\ref{fig:violin_single}, which compares the power distribution of resolvable CGWs to the binned spectra of the overall predicted GW signal and of the \texttt{DR2new} measurements. In line with RSG15, the probability of detecting a CGW is maximum at the lowest frequency, rapidly decaying to less than 0.01 past the 6th bin. Although this seems to suggests that the feature observed at the 9th frequency bin is unlikely to be due to a CGW, it should be noticed that we considered here a threshold of $S/N=3$. The overall GW signal in our data has a total $S/N\approx 3.5-4$ \citep{wm3}, mostly accumulated at the lowest frequency bins. It might still be the case that the feature at the 9th bin is due to an unresolved CGW with S/N$<3$, which would be a more common occurrence in the data.
Note that the average S/N of CGWs slightly increases at higher frequencies, which is primarily due to the frequency dependence of the CGW characteristic strain, $h_c\propto f^{7/6}$. If $p_i$ is the probability of having a CGW of S/N $>3$ in the $i$-th bin, we can compute the probability of detecting {\it at least} one CGW with S/N $>3$ in \texttt{DR2new} according to these models as $p=1-\prod(1-p_i)$, which gives $p=0.49$. It is therefore quite possible that the observed signal is dominated by few bright sources, which might be resolvable in the future with more sensitive datasets. Thus far, searches for CGWs on the current dataset yielded inconclusive evidence \citep{wm4}. This probability is obviously S/N threshold dependent. For example, by increasing this threshold to S/N$>5$, we get $p=0.13$. This is comparable to the 6\% chance found by \cite{2022ApJ...941..119B}. and the slightly larger probability in our models is likely due to the louder overall amplitudes of the signals considered here.
We stress, however, that these findings apply to models where binaries remain essentially circular. The number of resolvable CGWs, tends, in fact, to slightly decrease when the eccentricity increases (Truant et al. in preparation).

\begin{figure}
\includegraphics[width=0.5\textwidth]{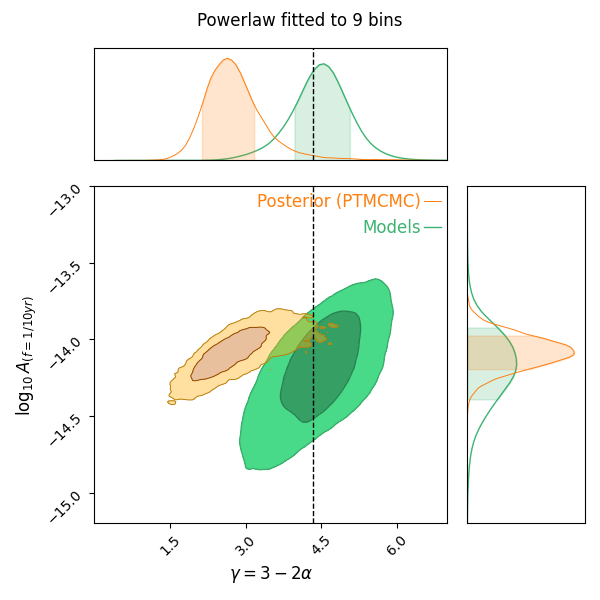}
\caption{\footnotesize{$A-\gamma$ distribution of the measured signal (orange) compared to model predictions (green). 1$\sigma$ and 2$\sigma$ contours are displayed. Shown are also the marginalised $A$ (left) and $\gamma$ (right) distributions, with their 1$\sigma$ credible intervals highlighted as shaded areas.}}
\label{fig:corner_A10yr_gamma}
\end{figure}

Finally, we once again propose the comparison first shown by \cite{msc+2021}, who contrasted the measured 2D $A-\gamma$ posterior to model expectations. For the latter, we just fit the 9 lowest frequency bins of the GWB spectrum of each Monte Carlo realization of the Universe with a straight line in the ${\rm log}A-{\rm log}f$ plane. As described in Sec.~\ref{sec:data}, we normalise the measurement to $f_0=10$yr$^{-1}$, where our data are informative, which alleviates the $A-\gamma$ degeneracy in the posterior. Results are shown in Fig.~\ref{fig:corner_A10yr_gamma}. Although the measured spectrum tends to be shallower than the theoretical one (see also Sec.~\ref{sec:considerations}), the contours overlap at 2$\sigma$ and the marginalised amplitudes are broadly consistent. 


\subsection{Inference on the SMBHB population.}
\label{sec:MBHB_inference}

After checking the general compatibility of the observed signal with expectations from empirical SMBHB population models, we take a more quantitative approach and exploit Bayesian inference to constrain the properties of SMBHBs from the data. We repeat the analysis carried out by \cite{msc+2021} on the common signal detected in the NANOGrav 12.5 year dataset \citep{ng12p5gwb+20}, exploiting the framework laid out in \cite{2016MNRAS.455L..72M,2017MNRAS.468..404C,csc2019}. The SMBHB population of a given model {\bf M} is described by a set of parameters $\{\theta\}$,
we then use Bayesian inference to find the posterior distribution $p(\theta|d, {\bf M})$ for the parameters $\theta$ of model ${\bf M}$ given the observation data $d$:
\begin{equation}
   p(\theta|d,{\bf M})=\frac{p(\theta|{\bf M})p(d|\theta,{\bf M})}{p(d|{\bf M})}. 
\end{equation}
Here, $p(\theta|{\bf M})$ is the prior distribution on the model parameters, $p(d|\theta,{\bf M})$ is the likelihood of model ${\bf M}$ with parameters $\theta$
of producing the data, and $p(d|{\bf M})$ is the evidence. For each set of $\theta$, the likelihood is computed by comparing the signal amplitude produced by ${\bf M}$ at frequencies $f_i=i/(10.3{\rm yr}), i=1,...,9$, to the posterior distribution of the correlated signal measured in \texttt{DR2new} at the same frequencies. We treat each bin as independent, therefore the likelihood takes the form
\begin{equation}
  {\rm log}_{10}p(d|\theta,{\bf M})=\sum_{i=1}^9 {\rm log}_{10} p(A,f_i)|_{A=A_{\bf M}}
\end{equation}
where $p(A,f_i)$ is the probability density of the amplitude $A$ of the correlated signal measured in the $i$-th bin, and it is evaluated at the value $A_{\bf M}$ predicted by the model. 
We estimate the likelihood in each bin using a KDE estimate of the \texttt{DR2new} posteriors, similar to the method used in~\cite{2021NatAs...5.1268M}.

Note that the models we use are deterministic in the sense that they have a 1:1 correspondence between $\theta$ and the predicted $h_c(f)$. In reality, given $\theta$, the ensemble of SMBHBs generating the signal is statistically drawn from the deterministic distribution function, which results in a significant scatter of $h_c(f)$, as demonstrated by the individual spectra shown in Fig.~\ref{fig:violin_allsignal}. We caution that this variance is not captured by these models, and its inclusion in the Bayesian inference pipeline is the subject of ongoing work.

As in~\cite{msc+2021}, we use two models to describe the SMBHB population, which are described in turn below. 

\subsubsection{Agnostic SMBHB population model}


The agnostic model, developed in~\cite{2016MNRAS.455L..72M}, makes minimal assumptions about the underlying population of SMBHBs. Binaries are assumed to be circular, GW-driven and the characteristic strain is computed according to Eq.~\eqref{eq:hcCircularPop} where the source distribution is described by a Schechter function \citep{1976ApJ...203..297S} in $z$ and $\Mc$, 
\begin{equation}
    \frac{{\rm d^2}n}{{\rm d}z{\rm d}\log_{10}\Mc} = 
    \ndot 
    \left[\left(\frac{\Mc}{10^7{\rm M_{\odot}}}\right)^{-\alphaM}
          e^{-\Mc/\Mstar}\right]
    \left[ \left( 1+z \right)^{\betaz} e^{-z/\zo}\right]
    \frac{{\rm d}t_{\rm R}}{{\rm d}z}\,,
    \label{eq:schechter}
\end{equation}
where $t_R$ is the time in the source frame and we assume cosmological parameters from Planck18 \citep{2020A&A...641A...6P}. 
The five model parameters are $\theta = \{ \ndot, \alphaM, \Mstar, \betaz, \zo \}$, where $\ndot$ is the merger rate per unit rest-frame time, co-moving volume, and logarithmic $\Mc$ interval, and the parameter pairs $\{\alpha_M,\Mstar\}$ and $\{\betaz,\zo\}$ control the shape of the $\Mc$ and $z$ distributions, respectively.
The integration limits in $\Mc$ and $z$ are $10^6\leq \Mc/{\rm M_{\odot}} \leq 10^{11}$ and $0\leq z \leq 5$, respectively. The prior ranges of the five model parameters are identical to those used in \citep{msc+2021}.


\subsubsection{Astrophysically-informed SMBHB population model}
\label{sec:smbhb_informed}

The astrophysically-informed model was developed in~\cite{csc2019}. 
This model captures the interaction between the SMBHBs and their environment and allows for eccentric orbits, both of which lead to a characteristic amplitude that does not follow a simple single power law, as in Eq.~\eqref{eq:hch2}. 
The model has $18$ parameters, $16$ of which describe astrophysical observables linking the number of SMBHB mergers to the number of galaxy mergers. The galaxy stellar mass function is modelled as a redshift-dependent Schechter function defined by five parameters: $\{\Phi_{\rm 0}, \Phi_I, M_0, \alpha_{\rm 0}, \alpha_{\rm I}\}$. Both the galaxy pair fraction and merger timescales have power law dependencies on the primary galaxy stellar mass $M$, mass ratio $q$ and redshift function $(1+z)$ and each of them is defined by a set of four parameters: $\{f_{\rm 0}, \alpha_{\rm f}, \beta_{\rm f}, \gamma_{\rm f}\}$ for the pair fraction, and $\{\tau_0, \alpha_{\rm \tau}, \beta_{\rm \tau}, \gamma_{\rm \tau}\}$ for the merger timescale:
\begin{equation}
    f_{\rm pair}=f_0\left(\frac{M}{10^{11}{\rm M_{\odot}}}\right)^{\alpha_f}(1+z)^{\beta_f},\,\,\, {\rm with}\,\,\,\,\,f_0=f'_0\int q^{\gamma_f}{\rm d}q;
\end{equation}
\begin{equation}
    \tau=\tau_0\left(\frac{M}{10^{11}{\rm M_{\odot}}}\right)^{\alpha_\tau}(1+z)^{\beta_\tau}q^{\gamma_\tau}.
\end{equation}
Galaxy pairs are then populated with SMBHs of mass $m$ following a standard black hole to stellar bulge mass relation of the form
\begin{equation}
    m={\cal N}\Biggl\{M_*\left(\frac{M_b}{10^{11}{\rm M_{\odot}}}\right)^{\alpha_*},\epsilon\Biggl\}
\end{equation}
where ${\cal N}\{x,y\}$ is a log normal distribution with mean value $x$ and standard deviation $y$, which adds three further parameters, $\{M_*, \alpha_*, \epsilon\}$, to the model.
The final two parameters describe the eccentricity at SMBHB pairing $\{e_{\rm 0}\}$ and the density of the stellar environment $\{\zeta_{\rm 0}\}$. For the 18 parameters listed above, in the analysis presented here, we use the {\it extended prior} intervals listed in Table I of \cite{csc2019}.


\subsubsection{Results of the inference}

\begin{figure}
\includegraphics[width=0.48\textwidth]{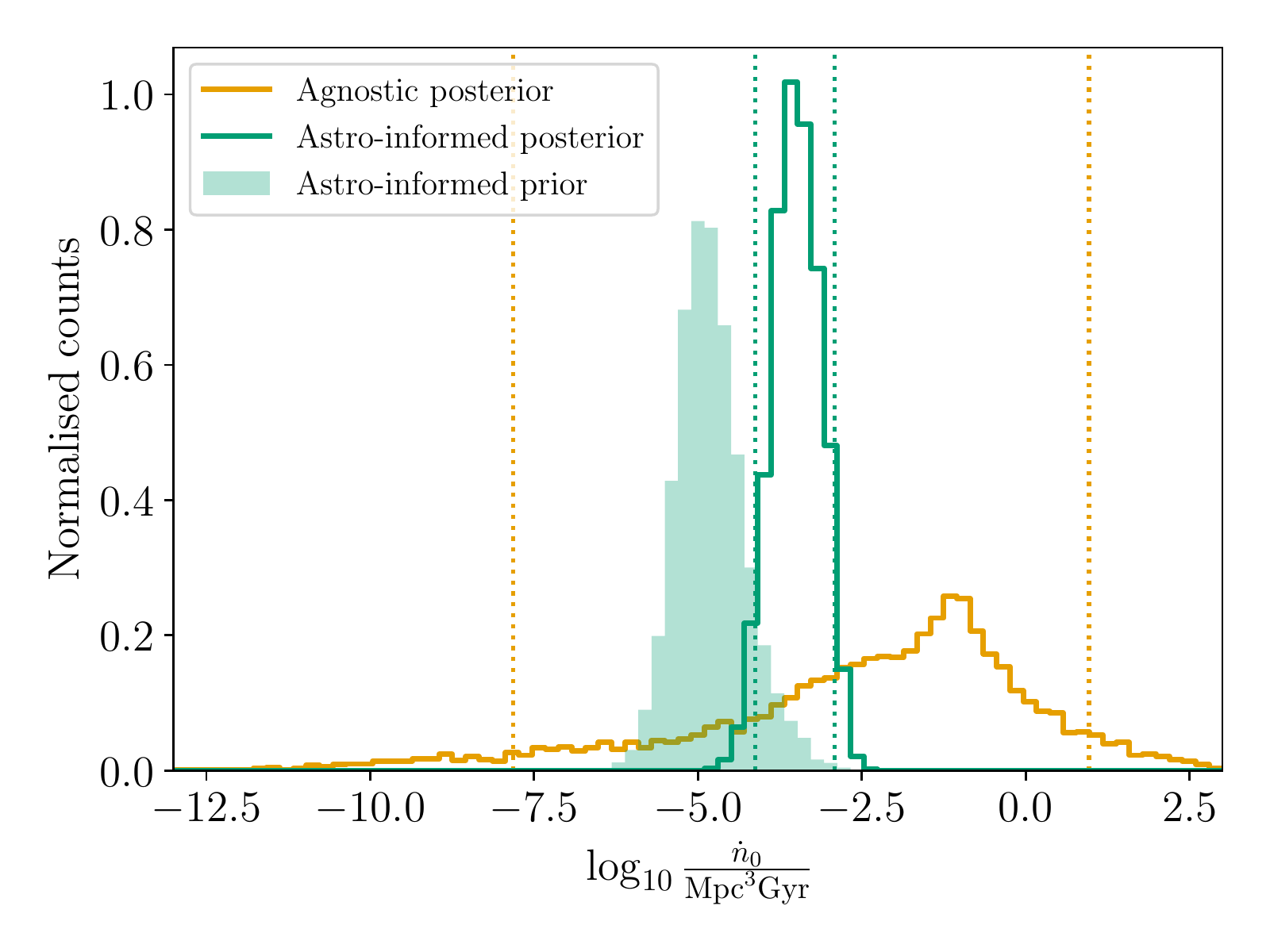}
\caption{\footnotesize{
Marginalised posterior distributions for $\ndot$ using two SMBHB population models. 
The orange and green open histograms show marginalised posteriors for the agnostic and astrophysically-informed models, respectively. 
The filled-green histogram shows the prior for the astrophysically-informed model (the prior for the agnostic model is uniform in the range $[-20,3]$). 
The vertical dotted lines show the $5{\rm th}$ and $95{\rm th}$ percentiles of the posteriors. 
}}
\label{fig:smbhb_numberofmergers}
\end{figure}

\begin{figure}
\includegraphics[width=0.51\textwidth]{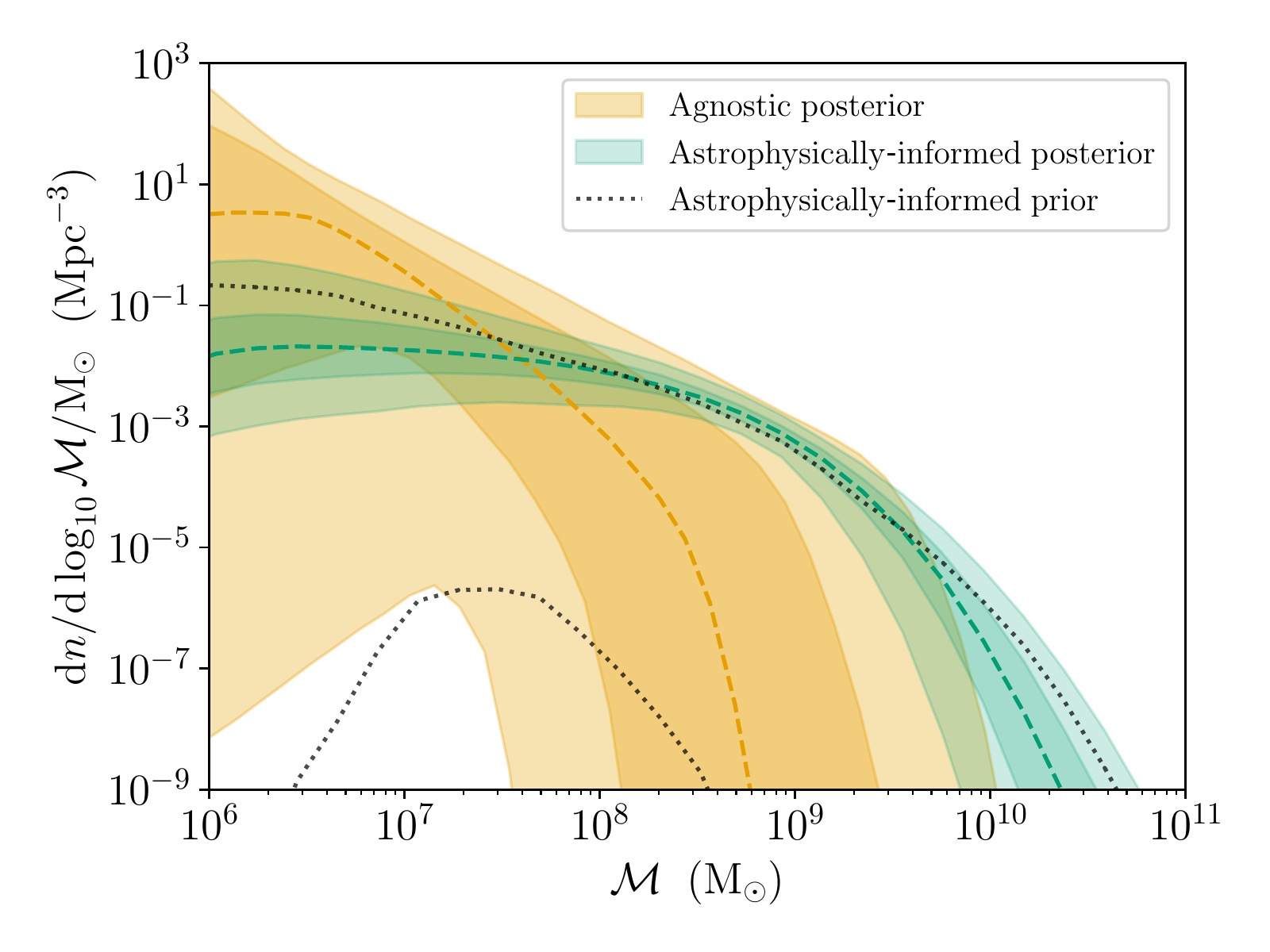}
\caption{\footnotesize{Posterior distribution of the chirp mass function of merging SMBHBs for both the agnostic (orange) and astrophysically informed (green) models. For both models, shaded areas are the central 50\% and 90\% credible regions and the dashed lines show the medias. The black-dotted lines show the central 99\% region for the astrophysical prior.}}
\label{fig:smbhb_dndlogm}
\end{figure}

\begin{figure*}
\includegraphics[width=0.98\textwidth]{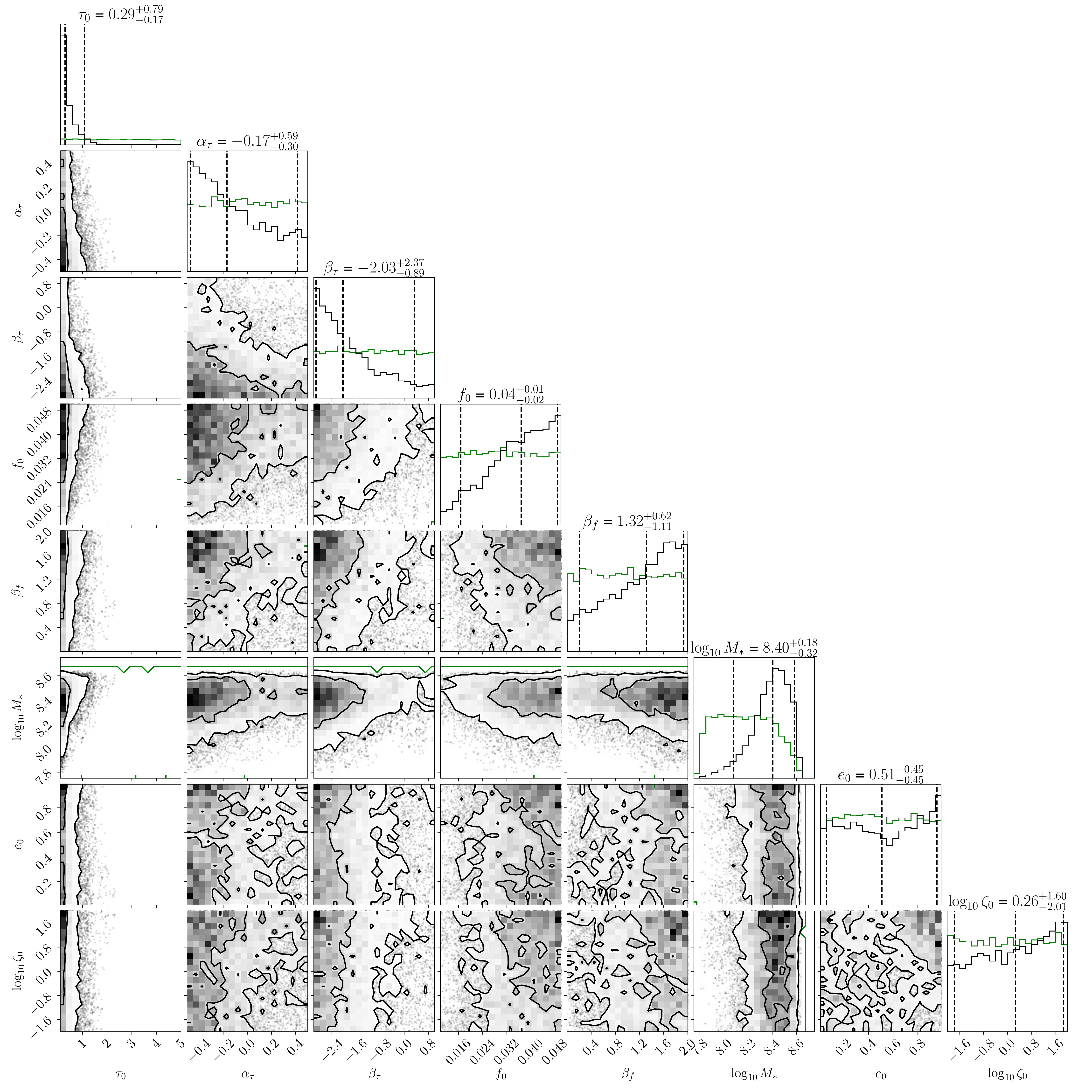}
\caption{\footnotesize{
Posterior distribution of selected parameters for the astrophysically-informed model using nine frequency bins of the free spectrum for the inference. Parameters are described in Sec.~\ref{sec:smbhb_informed}.}}
\label{fig:smbhb_astinformed_selected_corner}
\end{figure*}

The main results of the inference are shown in Figs.~\ref{fig:smbhb_numberofmergers}, ~\ref{fig:smbhb_dndlogm} and~\ref{fig:smbhb_astinformed_selected_corner}. 
Fig.~\ref{fig:smbhb_numberofmergers} shows the marginalised posterior distribution for the normalization of the merger rate density $\ndot$ from the agnostic model compared to an equivalent parameter derived from the astrophysically-informed model.
The constraint on the amplitude of the signal imposes an informative constraint on the normalization of the SMBHB merger density.
Using nine frequency bins, the median and central $90\%$ credible regions for $\log_{10}\ndot / [ {\rm Mpc}^3 {\rm Gyr} ]$ are $-1.95^{+2.91}_{-5.87}$ and $-3.51^{+0.59}_{-0.62}$ for the agnostic and astrophysically-informed models, respectively. 
The measurement essentially constrains the amplitude of the signal, which imposes an informative constraint on $\ndot$.
The astrophysically-informed model clearly shows that the signal favours an $\ndot$ at the upper edge of the astrophysical prior. 
All other parameters of the agnostic model are unconstrained and the posterior is very similar to the prior (see Appendix~\ref{app:smbhb} for full posterior distributions for both models).

%
%
%

Fig.~\ref{fig:smbhb_dndlogm} displays the posterior on the SMBHB chirp mass function for the two models integrated over the redshift range $0<z<1.5$. Although the agnostic model results in a loosely constrained mass function, the measured PTA signal alone places interesting lower limits on the SMBHB binary merger rate in the Universe. For example, we can say at 95\% confidence that for each comoving Gpc$^3$, there have been at least 10$^4$ SMBHB mergers with ${\cal M}\approx 10^7$M$_\odot$ since cosmic noon. When astrophysical priors weights in, the mass function of merging SMBHBs is well constrained by the PTA signal and, as expected from Fig.~\ref{fig:smbhb_numberofmergers}, it lies in the upper range of the astrophysically informed prior. Within this model, the \texttt{DR2new} measured signal implies there have been about 10$^6$ SMBHB mergers for each comoving Gpc$^3$, with ${\cal M}\approx 10^9$M$_\odot$ since $z=1.5$. This points towards a very active merger history of massive galaxies and a very efficient dynamical evolution of the SMBHBs forming in the merger process.



For the astrophysically informed model, \texttt{DR2new} also provides interesting information on several model parameters. This is because the astrophysical prior considerably narrows down the range of signal amplitudes allowed by the model, and the measured signal pushes towards the upper bound of this range. This results in informative constraints on several key parameters, related in particular to the SMBHB merger timescale and the SMBH-bulge mass relation. As shown in Fig.~\ref{fig:smbhb_astinformed_selected_corner}, the SMBH merger timescale $\tau_0$ following galaxy pairing must be shorter than $\approx 1\,{\rm Gyr}$ (90\% confidence), with the data mildly favouring shorter merger times for massive galaxies at low redshifts (i.e. $\alpha_\tau,\beta_\tau<0$). Moreover, the data favour a high normalization of the SMBH-bulge mass relation ${\rm log}_{10}M_*\approx 8.4_{-0.32}^{+0.18}$, compared to a much wider prior range extending all the way down to  ${\rm log}_{10}M_*=7.8$. This is in line with the qualitative analysis of Sec.~\ref{sec:MBHB_rosado}, which showed that the signal is consistent with recent, upward-revised, SMBH-galaxy relations.  There is also a slight preference for a high normalization of the pair fraction $f_0$ with a positive $z$ dependence, $\beta_f>1$. Despite the low $\gamma$ value favoured by the data, indicative of a flatter spectrum compared to the canonical value predicted by circular GW-driven binaries, SMBHB dynamics is largely unconstrained, perhaps with a marginal preference for eccentric binaries evolving in dense environments ($e_0$ and $\zeta_0$ posteriors slightly rising towards the right bound of the prior). 

Altogether, these results point towards efficient orbital decay of SMBHBs in the aftermath of galaxy mergers, providing direct evidence that the `final parsec problem' is solved in nature and that compact sub-parsec SMBHBs must be common in the centre of massive galaxies.

\subsection{Implications for SAMs}
\label{sec:MBHB_SAM}

We now explore the implication of this signal for the joint modelling of the galaxy and SMBH formation and evolution by taking a close look at two state-of-the-art SAMs: the model constructed by Barausse and collaborators~\citep{B12,2014ApJ...794..104S,2015ApJ...806L...8A,K+16,B+18,B+20} 
and \texttt{L-Galaxies} \citep{2015MNRAS.451.2663H,2022MNRAS.509.3488I}.
In this preliminary study, we do not model the dynamical evolution of the binaries and we assume them to be circular and GW driven, thus resulting in a characteristic strain spectrum with $\alpha=-2/3$. 

\subsubsection{SAMs and SMBHB delays}

\begin{figure}
    \centering
    \includegraphics[width=0.5\textwidth]{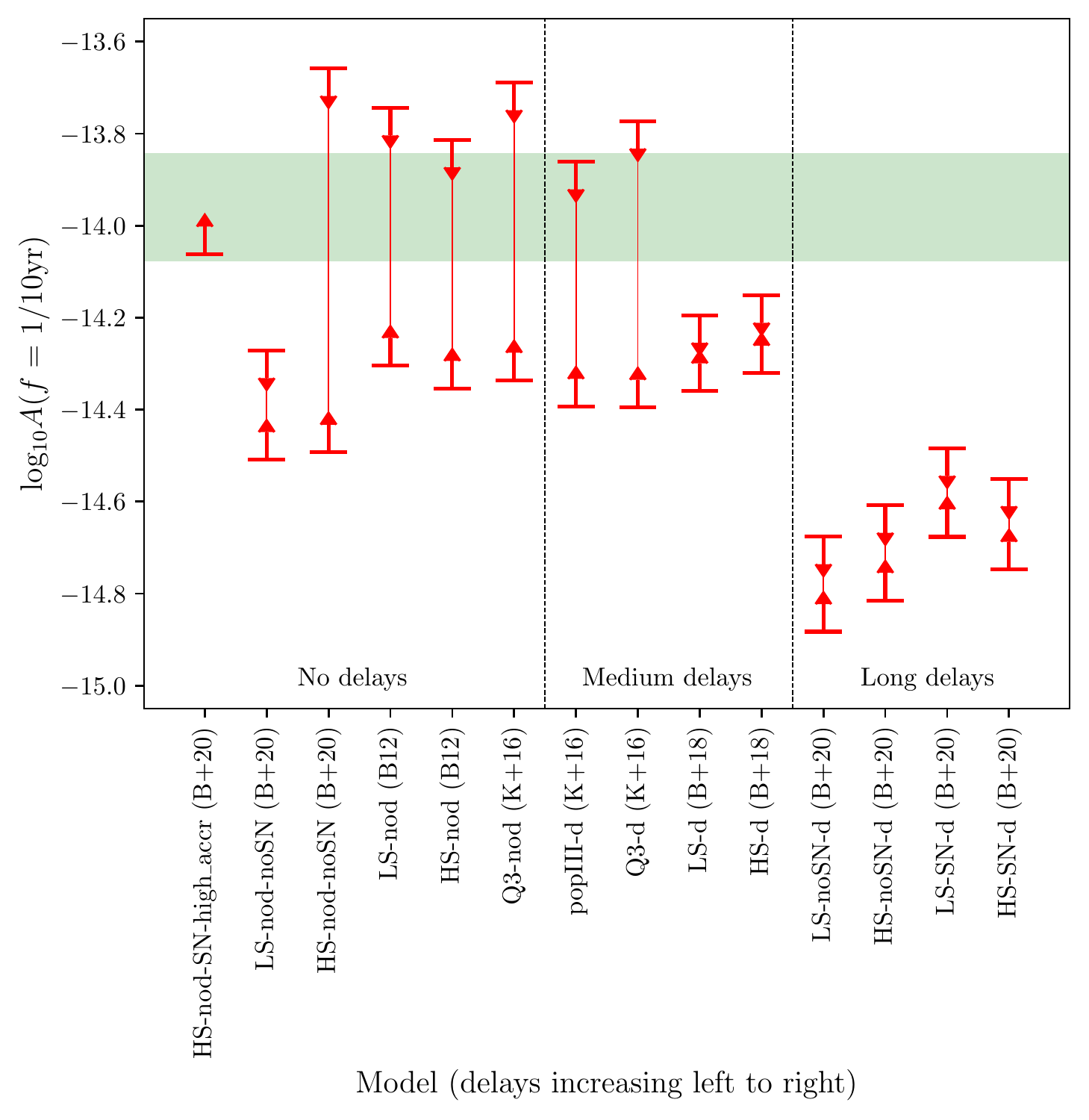}
    \caption{\footnotesize{Predictions for the GWB characteristic strain amplitude at $f=1/10$yr from a range of SAMs published in the literature, assuming quasicircular orbits and no environmental interactions (i.e. $\gamma=13/3$), but different physical prescriptions for the delays (increasing from left to right) between galaxy mergers and black hole mergers. The ``no delays'', ``medium delays'' and ``long delays'' models correspond respectively to the classes of models \textit{(i), (ii)} and \textit{(iii)} defined in the text. The ranges account for the finite resolution of the models. The shaded area is the 
    DR2new 95\% confidence bound.  
    More details about the models are provided in the text.}}
    \label{fig:sam}
\end{figure}

In Fig.~\ref{fig:sam}, we show this comparison for the model of \citet{B12} in its original version (B12) and its subsequent evolutions, which were used to produce the results of \citet{K+16} (K+16), \citet{B+18} (B+18) and \citet{B+20} (B+20).
Besides the specific SAM implementation and (astro)physics, these models mainly differ for the physical prescriptions describing the delays between galaxy and MBH mergers, with 
\textit{(i)}
models ``LS-nod (B12)'', ``HS-nod (B12)'', ``Q3-nod (K+16)'',
``LS-nod-noSN (B+20)'', ``HS-nod-noSN (B+20)''
and ``HS-nod-SN-high-accr (B+20)'' assuming no such delays (except for the delays between the mergers of the halos and those of the corresponding baryonic components)\footnote{Models ``LS-nod-noSN (B+20)'', ``HS-nod-noSN (B+20)''
and ``HS-nod-SN-high-accr (B+20)'' were {\it not} presented in B+20, but are produced
using the model of that paper, setting to zero the delays between  
galaxy and MBH mergers (except for the dynamical friction timescale -- including tidal effects -- between dark matter halos).}; \textit{(ii)} models ``popIII-d (K+16)'', ``Q3-d (K+16)'', ``LS-d (B+18)'', ``HS-d (B+18)''
additionally introducing the effect of stellar hardening, triple MBH interactions and gas-driven migration; and \textit{(iii)} models ``LS-noSN-d (B+20)'', ``LS-SN-d (B+20)'', ``HS-noSN-d (B+20)'' and ``HS-SN-d (B+20)''
accounting for even longer delays (including large contributions from SMBHB separations of hundreds of pc). Note that the labels ``SN'' (and ``noSN'') refer respectively to a putative effect of SN feedback on black hole accretion (and the absence thereof), while ``LS''/``popIII'' and ``HS''/``Q3'' denote respectively 
light and heavy high-redshift seeds for the black hole population.

The predictions are computed by summing the gravitational energy spectra of {\it all} the 
SMBHBs in each model's theoretical population, assuming quasi-circular orbits. As a result, the spectrum has a slope of $\gamma=13/3$ and has no cosmic variance (i.e. we do not account at this stage for the scatter from one realization of the SMBHB population 
to another). 
The range shown for each model represents the uncertainty due to the correction for the finite resolution of the SAM's merger tree. In more detail, the lower end of the range represents a model's prediction at the finite resolution, while the upper end is the extrapolation -- performed as described in Figure 4 of  \citet{K+16} --
to infinite resolution. The lower arrow (pointing up) should therefore be interpreted as a lower limit, while the upper arrow (pointing down) should be understood as an upper bound (due to the uncertainty of the extrapolation procedure). 
The extrapolation
has not been performed for the model HS-nod-SN-high-accr (B+20), for which we report only the (more robust) prediction at finite resolution. The latter already agrees with the measured amplitude, as a result of a higher accretion rate (by a factor $\sim 4$) for SMBHs. 

\begin{figure}
    \centering
    \includegraphics[width=0.5\textwidth]{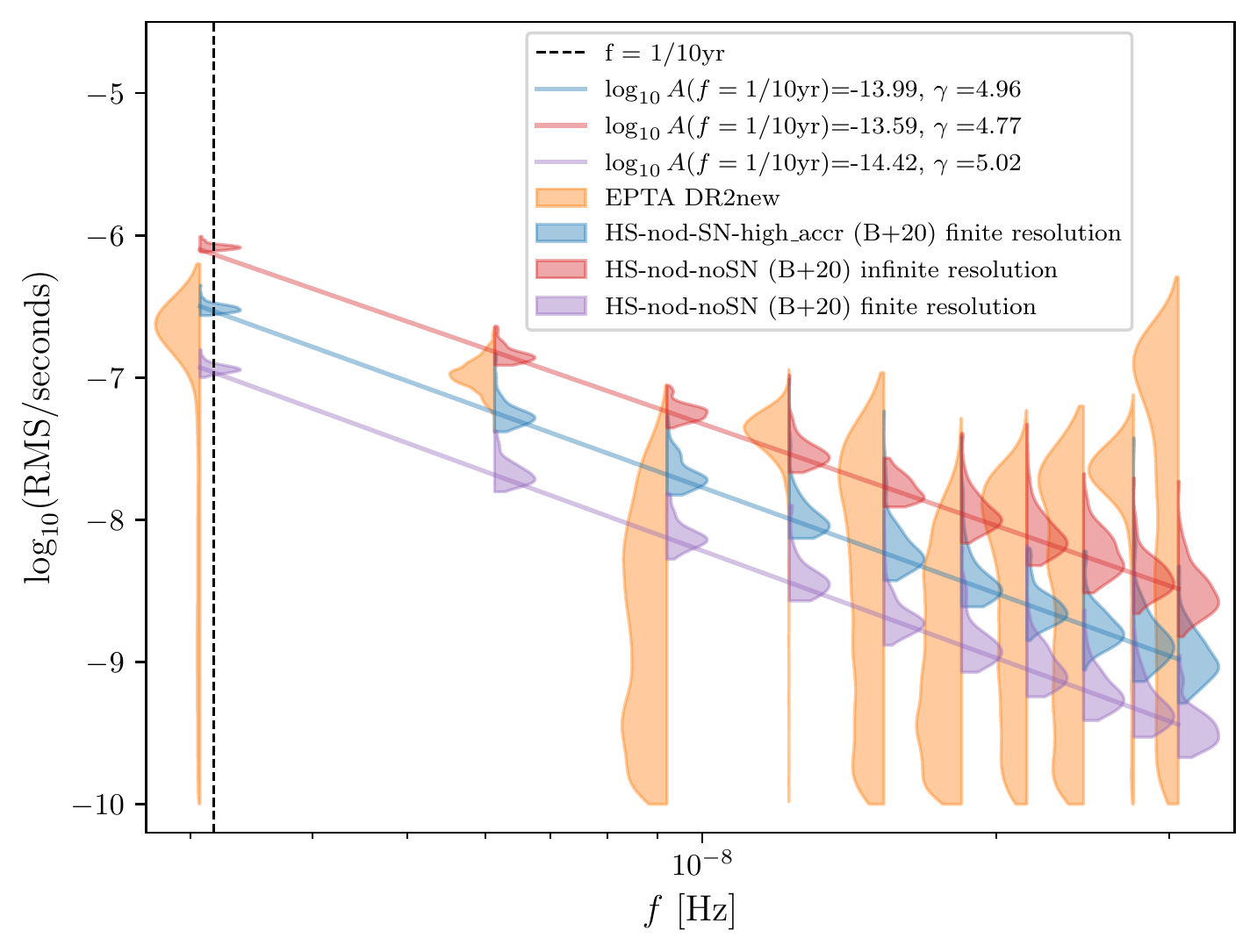}
    \caption{\footnotesize{Binned spectrum of the predicted GWB amplitude for models ``HS-nod-SN-high-accr (B+20)'' and ``HS-nod-noSN 
    (B+20)''. The distribution of the predictions represents the scatter among different realizations of the SMBHB population (``cosmic variance''). Also shown are power-law fits to the predictions.}}
    \label{fig:sam2}
\end{figure}

\begin{figure}
    \centering
    \includegraphics[width=0.47\textwidth]{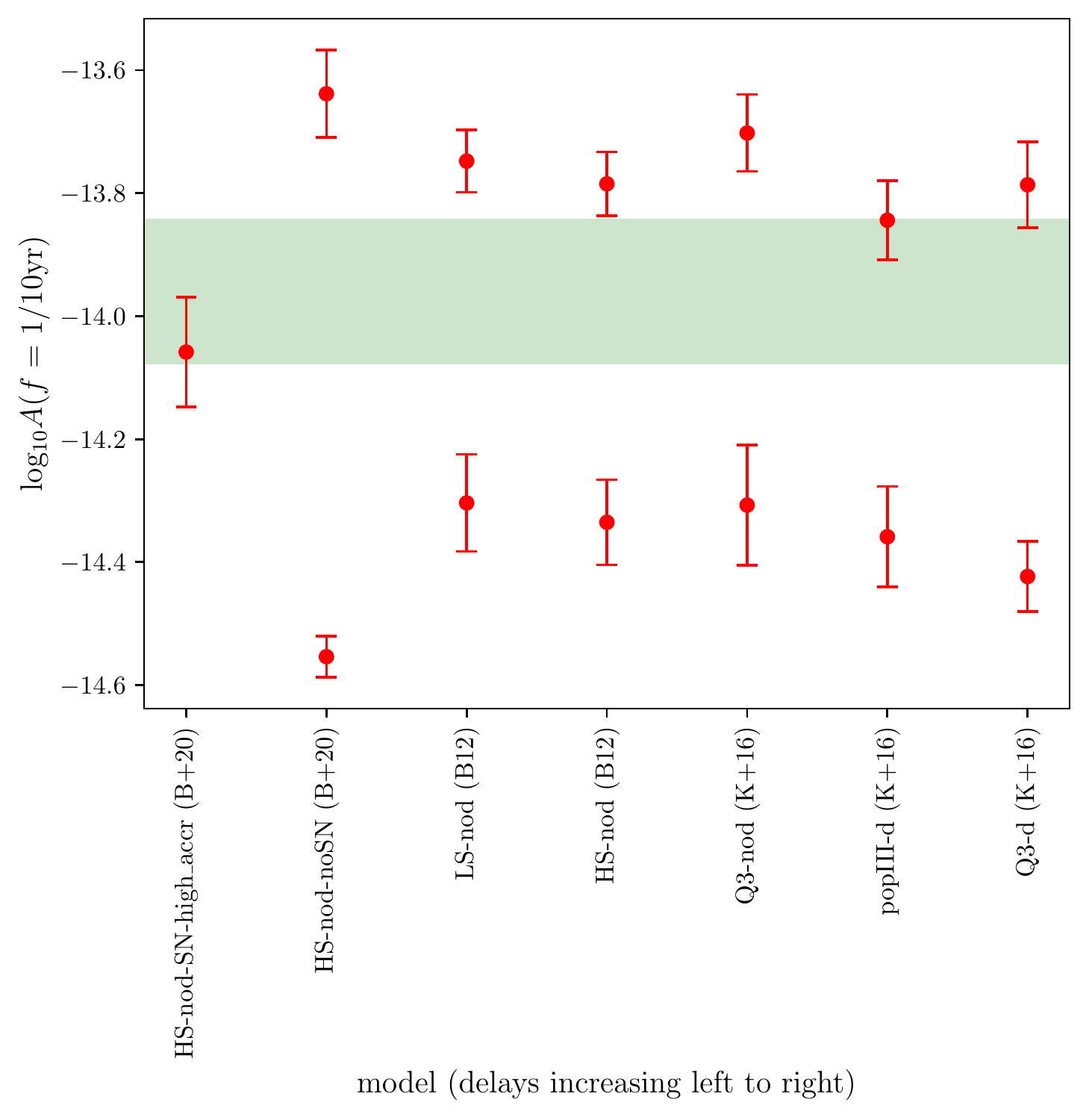}
    \caption{\footnotesize{Predictions for $A(f=1/10{\rm yr})$
    in various SAMs, obtained by fitting the spectrum in the first 9 frequency
    bins with $\gamma=13/3$ for multiple realizations of the SMBHB population. The error bars
    represent the 95\% confidence interval for the predictions, and account for the
    scatter due to cosmic variance. For each model (except for the
    boosted accretion model HS-nod-SN-high-accr (B+20)), the higher prediction is the
    extrapolation to infinite SAM resolution, while the lower one is the finite-resolution
    prediction. The shaded area is the 95\% confidence interval for the measurement of $A(f=1/10{\rm yr})$, fixing $\gamma=13/3$. For HS-nod-SN-high-accr (B+20) we only show the  result uncorrected for resolution.}}
    \label{fig:sam3}
\end{figure}

For  two of the models in better qualitative agreement with the data (i.e. ``HS-nod-SN-high-accr (B+20)'' and ``HS-nod-noSN (B+20)''), we compare the predicted signal with the measured \texttt{DR2new} free spectrum 
in Fig.~\ref{fig:sam2}.
Unlike in the case of Fig.~\ref{fig:sam}, these predictions were obtained for multiple specific {\it realizations} of the SMBHB population, following~\citet{svc08}\footnote{Note however that following \citet{svc08}, and different from RSG15, we average over sky position and binary inclination.}. The 
probability distribution function plotted in each bin represents the scatter of these multiple realizations, and should therefore be  interpreted as a ``cosmic variance''. Similarly, in Fig.~\ref{fig:sam3} we show the theoretical forecasts for $A(f=1/10{\rm yr})$ from a subset of the models presented above (namely those in qualitative agreement with the data in Fig.~\ref{fig:sam}). These forecasts are obtained by fitting the model predictions (from multiple realizations of the SMBHB population)
in the first 9 frequency bins with a power law with $\gamma=13/3$.
The error bars represent the 95\% confidence regions (accounting for the scatter due to cosmic variance), while the shaded area indicates the  95\% confidence region of the posterior for $A(f=1/10{\rm yr})$ (assuming again $\gamma=13/3$).

Overall, this qualitative comparison, while somewhat dependent on the details of the SAM implementation, suggests that {\it (i)} large delays arising from the dynamics of black hole pairs at large $\sim100$ pc separations are disfavoured, {\it (ii)} SMBHB mergers proceed efficiently after their host galaxies have coalesced. Moreover, these results seem to suggest that {\it (iii)} accretion onto SMBHs proceeds efficiently, possibly resulting in a larger local SMBH mass function at high masses. 

\subsubsection{\texttt{L-Galaxies}}

\begin{figure}
    \centering
    \includegraphics[width=0.48\textwidth]{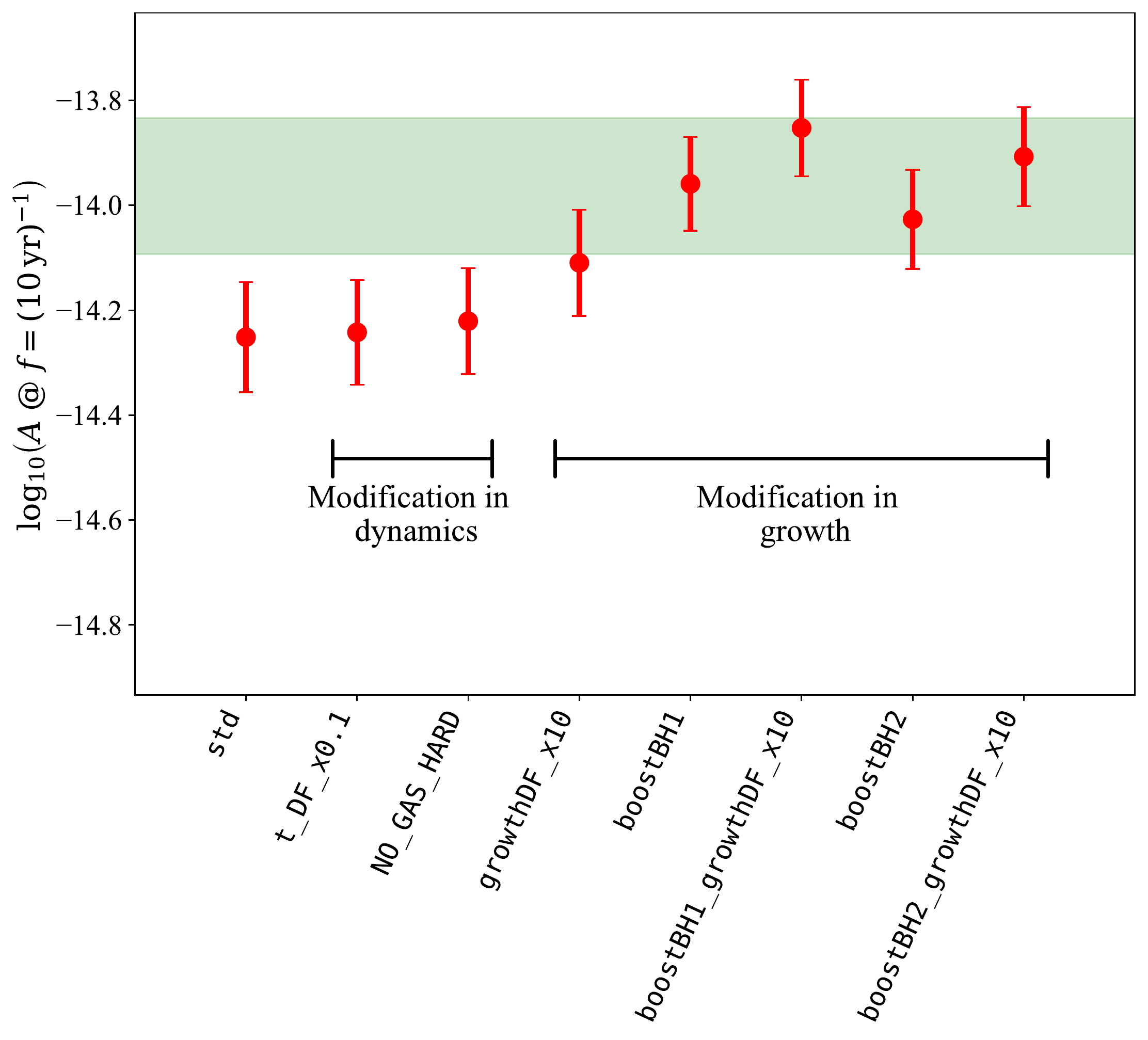}
    \caption{\footnotesize{Predictions for the GWB characteristic strain amplitude at $f=10/$yr$^{-1}$ from a range of \texttt{L-Galaxies} semi-analytical model versions, assuming that $h_c(f)\,{\propto}\, f^{-2/3}$. The error bars are computed taking into account the cosmic variance. To this end, we divided the \texttt{Millennium} box into 125 sub-boxes and we compute the GWB in each one. The standard deviation provided by the 125 GWBs corresponds to the extension of our error bars.}}
    \label{fig:LGalaxies_SAM}
\end{figure}

Next, we explore the implications that the EPTA results have for
\texttt{L-Galaxies}  \citep{2015MNRAS.451.2663H,2022MNRAS.509.3488I}, a sophisticated SAM constructed on the dark matter merger trees extracted from the \texttt{Millennium} simulation suite \citep{2005Natur.435..629S,2009MNRAS.398.1150B}. On top of the galaxy physics, \texttt{L-Galaxies} features a comprehensive modelling for the assembly of SMBHs, including gas accretion triggered by galactic mergers and disc instabilities and dynamical evolution of SMBHBs within the galaxy merger process. The latter accounts for dynamical friction (DF), stellar and gas hardening and, eventually, GW emission. All of these processes are governed by a set of parameters that are tuned to reproduce a vast array of observations including, among others, the galaxy mass function and morphological distribution, the quasar luminosity function and the SMBH-host galaxy relations. 


\cite{2022MNRAS.509.3488I} found that the standard \texttt{L-Galaxies} tuning results in a GWB with ${\rm log_{10}}A=-14.9$ at $f_0=1{\rm yr}^{-1}$, lower than that measured in \texttt{DR2new}. Here we perform a systematic investigation of how the stochastic GWB at nanohertz frequencies depends upon the parameters governing the physics of SMBHs and SMBHBs in the SAM. To this aim, we run \texttt{L-Galaxies} in the following configurations:
\begin{itemize}
\item \texttt{std}: standard configuration \citep{2022MNRAS.509.3488I};
\item \texttt{t\_DF\_x0.1}: SMBH dynamical friction (DF) time reduced by a factor of ten; 
\item \texttt{NO\_GAS\_HARD}: only stellar hardening;
\item \texttt{growthDF\_x10}: accretion boosted by ten in the DF phase;
\item \texttt{boostBH1}: gas accretion doubled after galaxy mergers and disc instabilities;
\item \texttt{boostBH2} gas accretion doubled after galaxy mergers and tripled after disc instabilities.
\item \texttt{boostBH1\_growthDF\_x10}: adding accretion boost in the DF phase to model \texttt{boostBH1};
\item \texttt{boostBH2\_growthDF\_x10}: adding accretion boost in the DF phase to model \texttt{boostBH2};
\end{itemize}


Results are shown in Fig.~\ref{fig:LGalaxies_SAM}. Changes to the dynamics of SMBHs appear to have a minor effect on the amplitude of the GWB. While shortening the DF time (\texttt{t\_DF\_x0.1}) allows more SMBHBs to merge within the Hubble time, the most massive ones, which are responsible for the bulk of the GW signal, already have short DF timescales, and the overall GW signal is only mildly increased. Turning off gas hardening results in longer-lived SMBHBs that tend to merge at lower redshifts, resulting in louder GW signals. The effect is, however, negligible.  
Conversely, the tuning of gas accretion can significantly change the masses of the SMBHBs, resulting in a larger impact on the GWB. Model \texttt{growthDF\_x10}  leaves the general population of SMBHs untouched, only boosting the growth of those in the dynamical friction phase. This alone increases the level of the GWB by a factor $\approx1.5$ compared to model \texttt{std}. Finally, the models \texttt{boostBH1} and \texttt{boostBH2} increase the gas accretion onto the whole population of SMBHs, making the GWB a factor of 2.5 louder with respect to the baseline model. Boosting accretion onto the whole population {\it and} in the hardening phase further amplifies the expected GWB, reaching the upper bound of the measured value (models \texttt{boostBH1\_growthDF\_x10} and \texttt{boostBH2\_growthDF\_x10} in the figure). Although these models can accommodate the GWB signal measured in PTAs, the boosted accretion and subsequently larger SMBH masses can make it harder to reproduce the observed SMBH mass and quasar luminosity functions \citep{2022MNRAS.509.3488I}. Additionally, more work is required to find a model that can reproduce all observational constraints in the light of the PTA GW signal (Izquierdo-Villalba et al. in preparation).

\subsection{Further considerations on the measured spectrum: eccentricity and statistical biases.}
\label{sec:considerations}

The analyses presented so far give strong indications that the signal is compatible with a cosmic population of SMBHBs swiftly coalescing in the aftermath of galaxy mergers. The relatively flat slope of the measured spectrum might be indicative of strong environmental coupling and high eccentricities, although inference from the data is inconclusive in this respect (see Fig.~\ref{fig:smbhb_astinformed_selected_corner}). 

The eccentricity of SMBHBs is of particular relevance as it might carry important information on the dynamical processes driving binary evolution at sub-pc scales \citep[see e.g.][]{2012JPhCS.363a2035R}. While gas-driven dynamics is expected to saturate the binary eccentricity at a value $e\approx 0.4-0.6$ \citep{2011MNRAS.415.3033R,2021ApJ...914L..21D}, stellar hardening is known to statistically increase eccentricity without any saturation point \citep{1996NewA....1...35Q}, potentially leading to extremely eccentric systems \citep{2010ApJ...719..851S}. A large binary eccentricity has two important implications for the interpretation of the current data: it flattens the low-frequency spectrum and it speeds up the SMBHB merger process, as inferred by the small $\tau_0$ derived in Sec.~\ref{sec:MBHB_inference}. 

\begin{figure}[ht]
    \centering
    \includegraphics[width=0.52\textwidth]{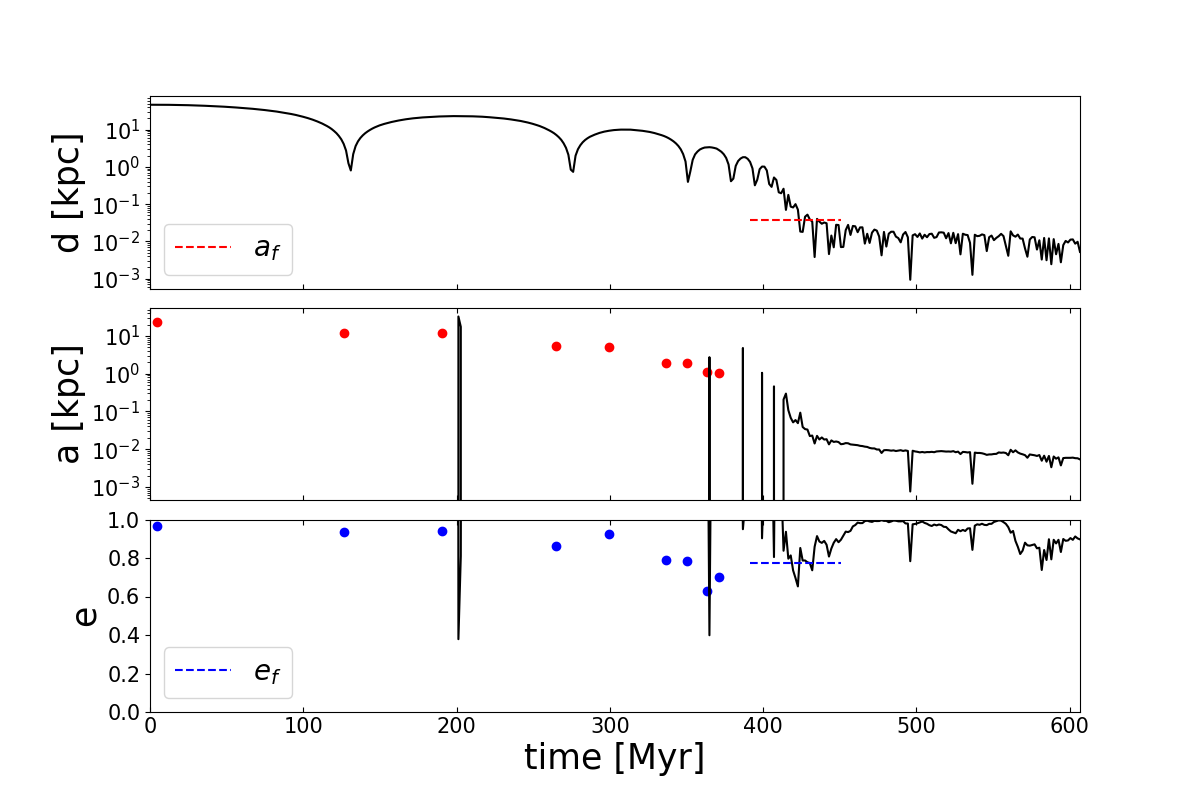}
   \caption{\footnotesize{Orbital parameters (distance between the SMBHs, semi-major axis and eccentricity) of a SMBHB formed in a representative $N$-body simulation of a galactic merger with parameters drawn from progenitors of likely PTA sources in the {\tt IllustrisTNG100-1} cosmological simulation. 
   Mergers are selected from the merger trees of the 100 most massive galaxies at $z=0$, based on galaxy mass ratio (major mergers with mass ratio $1:4$ or higher) and redshift ($z\leq2$). 
   The dashed lines indicate the critical separation $a_f$ and the corresponding eccentricity $e_f$ at the time in the evolution marking approximately the end of the SMBH inspiral due to DF and the beginning of the hardening phase. Dots represent $a$ and $e$ computed from the apoastron-periastron separation of the two SMBHs {\it before} pairing in a bound binary.}}
    \label{fig:ecc}
\end{figure}

\begin{figure}[ht]
    \centering
    \includegraphics[width=0.45\textwidth]{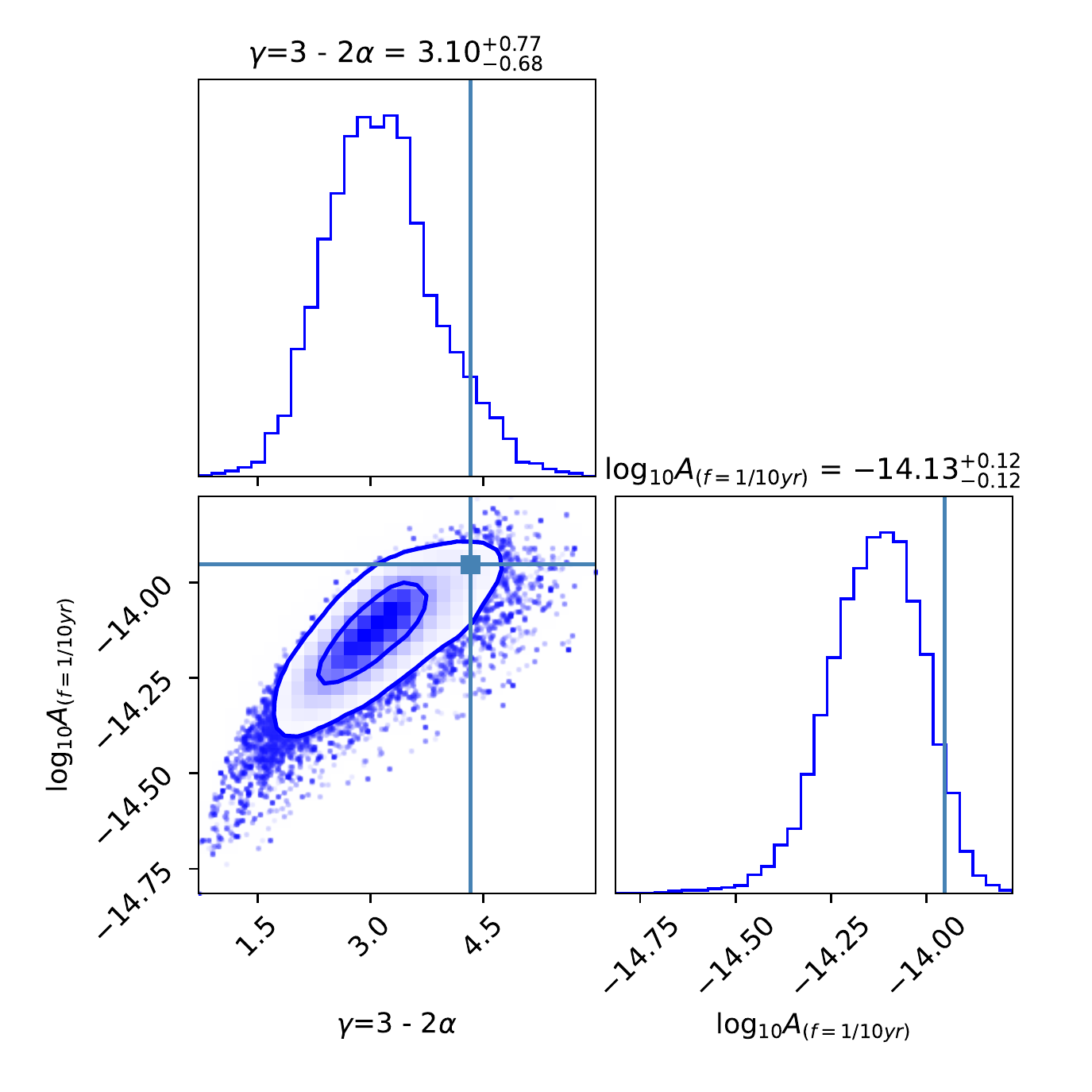}
    \includegraphics[width=0.5\textwidth]{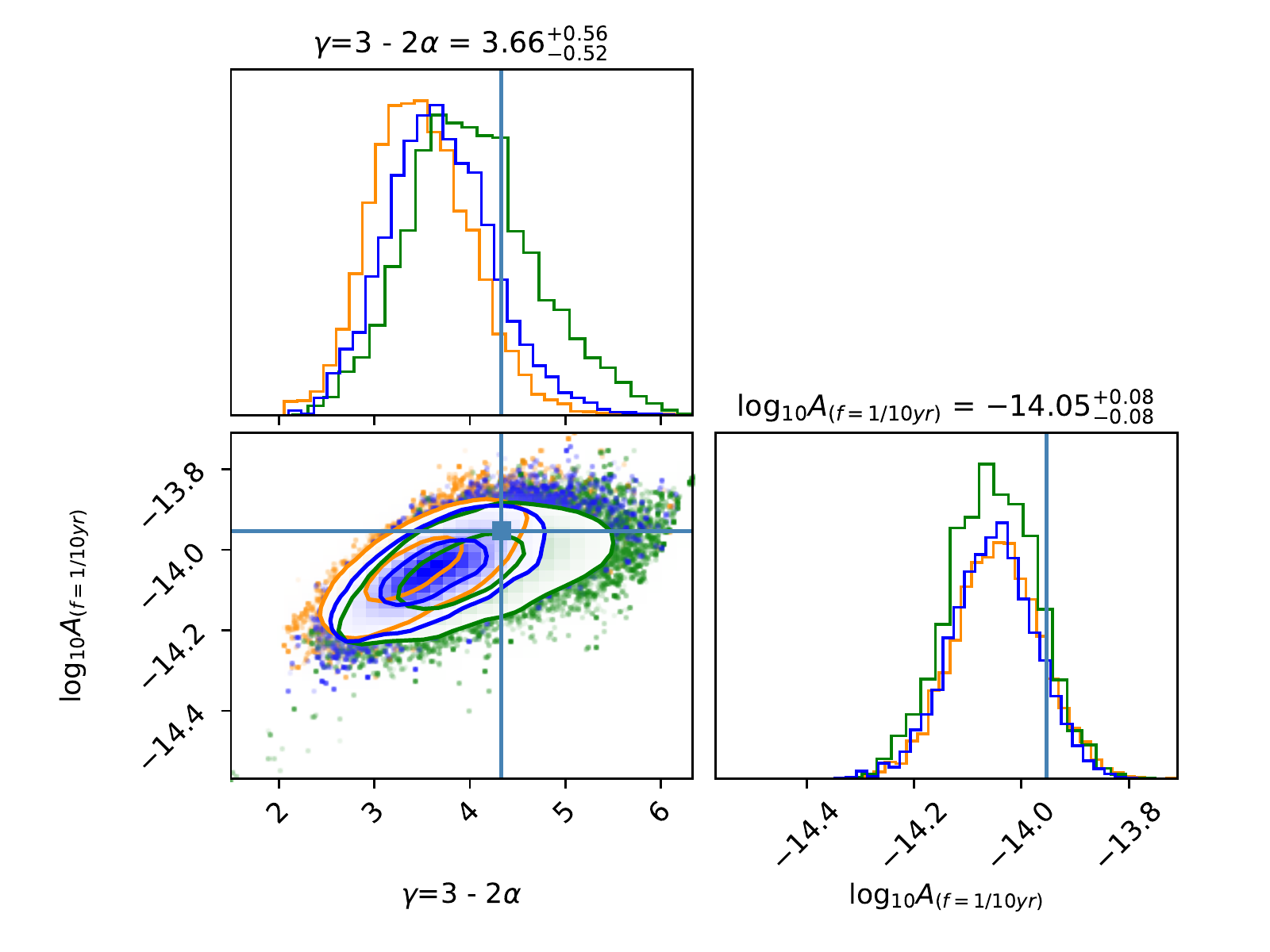}
    \caption{\footnotesize{Posterior distributions of the recovered GWB from injections on synthetic data mimicking \texttt{DR2new}. Top panel: statistical offset in an ideal dataset. The square marks the injected value and the blue contours are 1$\sigma$ and 2$\sigma$ of the recovered posterior. Bottom panel: effect of high-frequency noise mismodeling on the recovery. The orange, blue and green contours are respectively obtained when EFAC$=0.8, 1, 1.2$ are used for the recovery (injected EFAC$=1$).}}
    \label{fig:recovery_bias}
\end{figure}

Low-redshift massive galaxies are generally gas-poor, and stellar-driven hardening represents the main mechanism driving the evolution of the binaries comprising the bulk of the PTA GW signal. Modelling the whole dynamical evolution, from the first galaxy encounter to black hole pairing and final coalescence, is theoretically and numerically challenging and has been the subject of many studies \citep[e.g.][]{2011ApJ...732L..26P,2012ApJ...749..147K,2016ApJ...828...73K,2021MNRAS.502.4794N, 2022MNRAS.511.4753G}. In particular, the binary eccentricity is very sensitive to the initial orbits of the merging galaxies \citep{2022MNRAS.511.4753G}.
In ongoing work (Fastidio et al. in prep), we are connecting the sub-pc dynamics of SMBHBs to the large-scale parameters of the galactic encounters extracted from the \texttt{IllustrisTNG100} simulation \citep{2018MNRAS.473.4077P}. Preliminary results show that mergers of massive galaxies occur preferentially on nearly radial orbits, potentially resulting in highly eccentric binaries. 
Fig. \ref{fig:ecc} shows the orbital parameters of a SMBHB formed in a high-accuracy $N$-body simulation of a representative galactic merger with properties taken from a merger tree in {\tt IllustrisTNG100}. Merger trees are selected to represent likely PTA sources at low redshifts. The galactic merger is followed from early times through the inspiral, pairing and hardening of the SMBHs via a hybrid numerical scheme able to model the evolution self-consistently from kpc to mpc scales \citep{2014ComAC...1....1D}. 
Despite the intrinsic stochasticity of the processes driving binary formation and hardening \citep{2020MNRAS.497..739N}, binaries tend to form with a large eccentricity, which then further grows due to encounters with background stars, as also found by scattering experiments \citep{2010ApJ...719..851S}. 
Although more work is needed to determine the distribution of expected binary eccentricities and current EPTA data are not yet strongly informative, this pilot study shows the potential of using these measurements in the near future to constrain the physics of galaxy and SMBH mergers.

When drawing conclusions on the physical properties of the sources of the GW signal, it is useful to bear in mind not only that the constraints on the spectrum are relatively loose (see Fig.~\ref{fig:EPTADR2}), but also that the measured parameters can be subject to statistical and systematic biases. To address this, we are conducting an extensive campaign of simulations by injecting and recovering different types of signals in synthetic PTAs mimicking the properties of the EPTA \texttt{DR2new} dataset \citep{2024A&A...683A.201V}. We generated individual noises for 25 pulsars using the maximum likelihood values of the measured white noise and drew the red noise and dispersion measure parameters from the posterior distribution of the customised noise models \citep{wm2}. We simulated TOAs from multi-frequency observations and added a GWB spectrum from an astrophysical population of circular SMBHBs producing a nominal GWB with $A_{1{\rm yr}^{-1}}=2.5\times 10^{-15}$, consistent with the \texttt{DR2new} measurement at $\gamma=13/3$. We performed 100 experiments by changing the specific noise realization and the sampling of the injected SMBHB population. The analysis was carried using the \texttt{ENTERPRISE} software package \citep{2020zndo...4059815E}.

Two illustrative examples of injection-recovery mismatch are shown in Fig.~\ref{fig:recovery_bias}. The top panel shows one of the GWB recoveries. Although the injection did not present particular features (e.g. loud CGWs), for this specific noise realization, the recovered signal has a shallow slope with a median value of $\gamma=3.10$. Similar cases have been observed when injecting a pure $\gamma=13/3$ power law with the \texttt{createGWB} function of \texttt{libstempo} \citep{2020ascl.soft02017V}. This shows that even with an ideal setup when simultaneously fitting multiple parameters (102 in this case) in a complex problem, the stochastic properties of the noise can easily bias the recovered signal, especially if the S/N is low (S/N$\approx3.5$ for \texttt{DR2new}). Multiple injections with the realistic GWB model and \texttt{createGWB} show systematic biases of recovery of the realistic GWB signal, when modelled with an ideal power law. This is explored in detail in a follow-up work. In the bottom panel of Fig.~\ref{fig:recovery_bias}, we show how the presence of some extra high-frequency noise unaccounted for in the MSP noise model can also influence the recovery of the parameters. The setup is the same as in the left panel, but we simulate high-frequency noise mismatch by setting different values of EFAC$=0.8, 1, 1.2$ in the recovery. Although the posterior of the signal amplitude is hardly affected, $\gamma$ can shift significantly depending on whether the high-frequency noise is slightly over- or under-estimated. While these are only two examples, they highlight the complexity of PTA measurements and invite us to be cautious when drawing conclusions that might strongly be influenced by potential biases in the recovered signal.


\section{Implications II: physics of the early Universe}
\label{sec:early_universe}

Although a GWB generated by an ensemble of the putative SMBHBs is the most plausible source of the observed common-red noise process in pulsar data, more exotic explanations are possible, such as signals originated in the early Universe. 
The various possible types of cosmological backgrounds of GWs associated with early Universe physics are reviewed in \cite{2018CQGra..35p3001C} and are found to be stochastic. Similarly to the traditional case invoking SMBHBs, the angular spatial correlation for those scenarios follows the HD curve. However, the spectral shapes of the predicted GW spectra are generally different, which can help to disentangle between different types of backgrounds. 
In this work, we focus on four possible scenarios:
\begin{enumerate}
\item an inflationary GWB from the amplification of quantum fluctuations of the gravitational field, 
\item a GWB from a network of cosmic string loops,
\item a GWB from vortical (M)HD turbulence at the QCD energy scale, 
\item a scalar-induced GWB arising from inflationary scalar perturbations at the 2nd order in perturbation theory.
\end{enumerate}
Given the low significance of the detected signal and the limited number of probed frequency bins due to the short timespan of the data, one cannot currently perform a reliable model selection. Therefore, throughout the section, we consider these scenarios separately and assume that each of them can fully explain the detected signal independently. Analysis invoking more complex models with simultaneous fits for multiple scenarios as well as opportunities to disentangle between those~\citep[e.g.][]{ppta+22, 2022ApJ...938..115K} will be considered in a number of future works.

\subsection{\label{sec:inf}Implications on a stochastic background of primordial (inflationary) gravitational waves} 

Here we address the GWB possibly generated during inflation  
\citep{1975JETP...40..409G, 1982PhLB..115..189R,1985SvAL...11..133S,1983PhLB..125..445F, 1984NuPhB.244..541A}. 
In the standard inflationary scenario, 
tensor quantum vacuum fluctuations of the metric are amplified by the accelerated expansion, leading to a GWB as they subsequently re-enter the horizon during the radiation (or matter) era.
The cosmic microwave background (CMB) and Big Bang Nucleosynthesis (BBN) provide precise measurements of the radiation energy density, from which one can derive weak upper bounds on the amplitude of such a GWB ~\citep[see e.g.][and references therein]{2018CQGra..35p3001C}.  
Furthermore, tensor metric perturbations lead to CMB temperature anisotropies and polarisation at large angular scales \citep{1967ApJ...147...73S,1985SvAL...11..133S,1996AnPhy.246...49K,1994PhRvD..50.3713A}. 
Since the anisotropies and polarisation detected so far are  
due to scalar perturbations, 
it is 
possible to constrain the energy density of a GW background by placing an upper limit on the tensor-to-scalar ratio $r$ at CMB scales: recent upper bounds are given e.g.~in \cite{2022PhRvD.105h3524T,2023JCAP...04..062G}. 
Another parameter to consider is the tensor spectral index $n_T$ of the tensor perturbations. 
In the context of slow-roll single-field inflation, these two parameters are linked via the consistency relation $r=-8 n_T$ .
By fixing the consistency relation, \cite{2022PhRvD.105h3524T} finds 
$r < 0.032$ at 95\% CL, 
while by relaxing it, \cite{2020A&A...641A..10P} finds $r < 0.076$ and $-0.55<n_T<2.54$ at 95\% CL. 

Within the slow-roll consistency relation, the GWB spectral slope is therefore slightly red-tilted, causing this signal to be out of reach of most current and planned GW detectors such as PTAs, LISA, aLIGO, aVirgo or the Einstein Telescope. 
On the other hand, it is fair to consider that the spectral slope could vary over the large frequency span ranging from CMB scales to those probed by GW detectors \citep{2016PhRvX...6a1035L}.
Inflationary scenarios breaking the consistency relation at small scales might therefore produce a detectable GWB, if they lead to a blue-tilted spectrum. 
In this case, PTAs, LISA and ground-based devices can place upper limits on $n_T$ ~\citep[see e.g.][]{2017PhRvL.118l1101A}. 
It is interesting to investigate which kinds of processes could give rise to a blue-tilted GW background while 
still obeying CMB constraints at large scales.
One possibility is 
the presence, after inflation, of a stiff component in the Universe, with an equation of state $w > 1/3$ \citep{2008PhRvD..78d3531B,1998PhRvD..58h3504G,2008PhRvD..77f3504B}. 
The enhancement of the tensor spectra can also be produced during inflation thanks to processes such as, for example, (i) particle production during inflation (see e.g.~\cite{2011JCAP...06..003S, 2012PhRvD..86j3508B,2013JCAP...11..047C,2012PhRvD..85l3537A,2016JCAP...12..026B}) (ii) enhancement of tensor perturbations for example by spectator fields, or space-dependent inflation (see e.g.~\cite{2007PhRvD..76f1302B,2013PhRvD..88j3518B,2015JCAP...04..011B,2015PTEP.2015d3E01F})  (iii) modified gravity theories such as $f(R)$ or Horndeski gravity  (\cite{1974IJTP...10..363H,2010RvMP...82..451S})  and iv) enhanced scalar perturbations at small scales and/or primordial black holes,  which are treated in Sec.~\ref{sec:secondorder}. 

\subsubsection{Analysis}

Similarly to what was done in \cite{2016PhRvX...6a1035L}, we constrain the key parameters defining the GWB,
$r$ and 
$n_T$, 
while being agnostic on the underlying process generating the blue-tilted spectrum.
If we assume that the common quadrupolar red noise signal present in EPTA data is of 
inflationary origin, these two parameters can be estimated using the \texttt{DR2new} dataset.
Note that the spectral index $n_T$ is expected to vary with the frequency scale considered ~\citep[see e.g.][]{2021PhLB..81536137G,2023MNRAS.520.1757G,2022PhRvD.106f3512A}. 
However, for simplicity, here we consider a constant $n_T$ all the way from CMB scales to those corresponding to the (narrow) EPTA frequency band. It is then possible to approximate the fractional characteristic GW energy density using \citep{2016PhRvX...6a1035L,2018CQGra..35p3001C}
\begin{eqnarray}
   & \lefteqn{\Omega_{\rm GW} (f)=\frac{3}{128}\Omega_{\rm rad}\,r\,\mathcal{P}^*_{\mathcal{R}}\left( \frac{f}{f_*} \right)^{n_T} \left[\frac{1}{2} \left( \frac{f_{\rm eq}}{f} \right)^{2}+\frac{16}{9}\right]}& \nonumber\\ %
    &&\approx 1.5 \times 10^{-16} \left( \frac{r}{0.032} \right) \left( \frac{f}{f_*} \right)^{n_T}, \label{eq: inf spectrum}
\end{eqnarray}
where the second line is valid in the PTA frequency band, and has been obtained by setting $h^2\Omega_{\rm rad}=2.47\cdot 10^{-5}$ with $h=0.67$, the amplitude of the scalar spectrum $\mathcal{P}^*_{\mathcal{R}}=2\cdot 10^{-9}$, and
$f_* \approx 7.7 \times 10^{-17}$ Hz related to the CMB pivot scale $k_*=0.05/{\rm Mpc}$ \citep{2014A&A...571A..16P}.
$f_{\rm eq}$ denotes the frequency entering
the horizon at matter-radiation equality.

\begin{figure}
    \centering
    \includegraphics[width=0.46\textwidth]{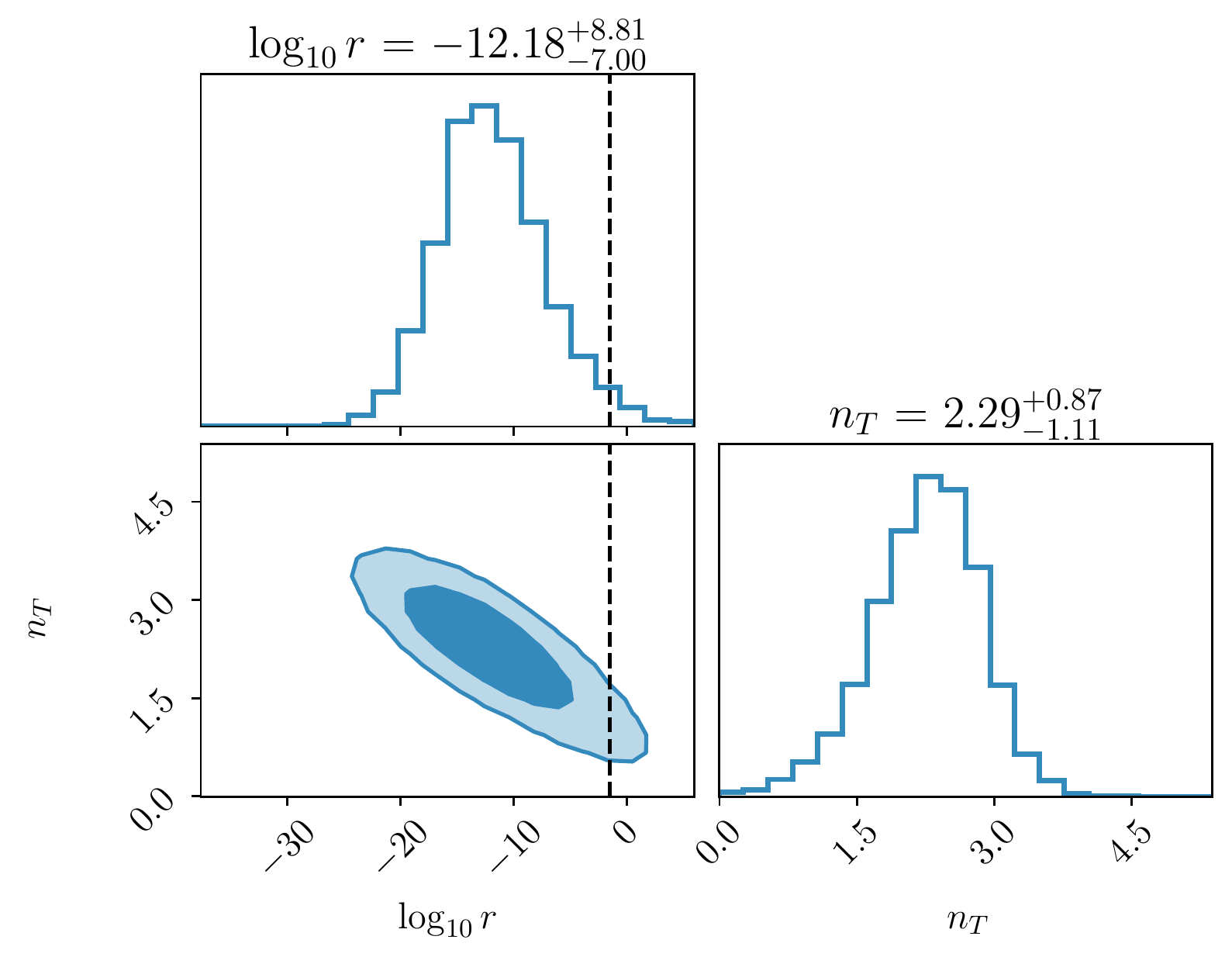}
    \caption{\footnotesize{2D posteriors of the tensor-to-scalar ratio (in $\log_{10}$) and the fractional energy density spectral index $n_T$ in the PTA frequency range. The 68\% and 95\% credible regions are displayed. The black dashed line represents the tensor-to-scalar ratio upper bound found in \cite{2022PhRvD.105h3524T} assuming single-field slow-roll inflation.}}
    \label{fig:inflation_post}
\end{figure}

We then use the nine lowest frequency posteriors of the RMS free spectrum shown in Fig.~\ref{fig:EPTADR2} \citep[see][for details on the method]{2021NatAs...5.1268M, Taylor_NFS, Leclere:2023ryt} to fit the inflationary spectrum of Eq.~\ref{eq: inf spectrum} and obtain posteriors on $\log_{10} r$ and $n_T$. Results are reported in  Fig.~\ref{fig:inflation_post}. 
Note that, since $\gamma=5-n_T$, the correlation between the amplitude and spectral index of the signal is compatible with  Fig.~\ref{fig:corner_A10yr_gamma}.
The 90\% credible (symmetric) intervals are $\log_{10} r = -12.18^{+8.81}_{-7.00}$ and $n_T = 2.29^{+0.87}_{-1.11}$. 
The obtained value of $n_T$ corresponds to a PSD spectral index of $\gamma \simeq 2.7$, as in Fig.~\ref{fig:EPTADR2}. 
The excessively small value of $r$ is a consequence of the simplistic parameterisation of Eq.~\ref{eq: inf spectrum}, which assumes a constant $n_T$ at all scales. 
The fractional energy density spectrum obtained from the maximum a posteriori parameter values is plotted in Fig.~\ref{fig:spectrum_plot}.

We have so far considered a primordial background to be the only source of GWs in our data. We now recall that the most plausible and loud source of a GW background at these frequencies remains that of a SMBHB background. It is therefore likely that any signature for a cosmological background needs to be considered in parallel with a SMBHB background, or in this case more accurately termed `foreground'. \cite{2022ApJ...938..115K} have explored the likelihood of detecting a cosmological background in the presence of a SMBHB foreground using simulations, and found that the shallower the slope of the cosmological background (for example $\gamma=4$ as opposed to $\gamma=5$), the harder it is to detect (and the longer it takes, possibly more than 20 years). According to these simulations, this does not bode well for an even shallower slope, similar to the one detected in \texttt{DR2new} with a possible $\gamma < 3$. 

Here we explore a superposition of these two backgrounds in the \texttt{DR2new} dataset. Considering a two-component GWB for the common red noise model, we place constraints on $\log_{10} r$ for given values of $n_T$ spanning the range $[-1, 3]$. In this case, our null hypothesis is a GWB from a population of GW-driven circular SMBHBs parameterised only by the PSD amplitude $\log_{10} A$ of Eq.~\eqref{eq:sh} (we fix $\gamma = 13/3$). We run several analyses with a fixed $n_T$ for the inflationary background, sampling over $(\log_{10} r, \log_{10} A)$. For each of the $n_T$ values, we obtain a distribution for $\log_{10} r$ and take the 95\% quantile as an upper bound. As found in \cite{2016PhRvX...6a1035L}, $n_T$ and the $\log_{10} r$ upper bounds are related with good precision by a linear relation: 
\begin{equation}
    n_T = a \log_{10} \left( \frac{r}{0.032} \right) + b.
\end{equation}
Our analysis gives $a=-0.16$ and $b=0.70$, which is comparable to the forecast values given in  \cite{2016PhRvX...6a1035L}
(note that they normalise $r$ to 0.11). 

\subsubsection{Discussion}

From the analysis of the \texttt{DR2new} dataset above, we have obtained credible intervals for the tensor-to-scalar ratio $r$ and the spectral index $n_T$. This was performed assuming that reheating is instantaneous, and that inflation is followed directly by the radiation-dominated era, for which the equation of state parameter of the Universe is $w=1/3$. Under this assumption, one finds that the best-fit value for the tensor spectral index is $n_T=5-\gamma\simeq 2.3$, which is directly linked to the best-fit PSD spectral index $\gamma \simeq 2.7$. This high value of $n_T$ is not consistent with slow roll inflation. 
However, if inflation is followed by a stage in which $w\neq 1/3$, the relation between the PSD spectral index $\gamma$ and the primordial tensor spectral index $n_T$ changes to \citep{2016ApJ...821...13A,2018CQGra..35p3001C}
\begin{equation}
\gamma=5-n_T+\frac{2(1-3w)}{3w+1},
\label{eq:stiff}
    \end{equation}
again with $\gamma \simeq 2.7$. 
If a stiff fluid component ($w>1/3$) were to dominate the Universe for a finite amount of time after inflation, the last term in Eq.~\ref{eq:stiff} would be bounded between 0 and $-2$. 
Hence, $n_T \gtrsim 0.3$, meaning that even 
allowing for the presence of a stiff component after inflation,
it does not seem possible to explain the common red noise in the context of slow roll inflation ($n_T \simeq 0$) for the best fit value $\gamma \simeq 2.7$. 
However, by broadening the range of possible values to $\gamma \geq 3$, $n_T \simeq 0$ does become compatible with the common red noise. 

\subsection{Implications on a background of cosmic strings}

Cosmic strings are line-like topological defects that may form after a symmetry-breaking phase transition in the early Universe \citep{Kibble:1976sj,Hindmarsh:1994re}; they are generic predictions of most Grand Unification Theories scenarios~\citep{Jeannerot:2003qv}.
These 1D objects are characterised by the string tension $G\mu$ (or equivalently their energy per unit length) which is related to the energy scale of the phase transition.

Overall, cosmic strings combine relativistic velocities and large energy densities, making them natural sources of GWs.
These GWs may take the form of bursts from cusps, kinks and kink-kink collisions on the loops~\citep{Damour:2001bk}, and have been searched for in the LIGO/Virgo/KAGRA  (LVK) detectors~\citep{LIGOScientific:2017ikf,LIGOScientific:2021nrg}.
Cusps are points on the string which, in the Nambu-Goto approximation, propagate at the speed of light, and the string doubles back on itself. On the other hand, kinks are discontinuities in the tangent vector of a string and are formed at each intercommutation of strings. 
The future space-based detector LISA will be sensitive to cosmic string bursts for tensions as low as $G\mu \sim 10^{-11}$~\citep{Auclair:2023brk}.
Most notably, in the event that LISA detects GW bursts from cosmic strings, they will likely repeat multiple times a year due to the periodic nature of the cosmic string loops~\citep{Auclair:2023mhe}.

The incoherent sum of these GW bursts also produces a stochastic GWB (SGWB), which has been looked for by LVK~\citep{LIGOScientific:2017ikf,LIGOScientific:2021nrg}.
LISA is expected to be able to detect a SGWB from cosmic strings with tensions as low as $G\mu \sim 10^{-17}$~\citep{Auclair:2019wcv}.
The strings background has already been looked for in EPTA~\citep{Sanidas:2012ee,Leclere:2023ryt}, in NANOGrav~\citep{Blasi:2020mfx,Ellis:2020ena} and PPTA~\citep{Bian:2022tju,Chen:2022azo}.
These earlier analyses consistently found a good agreement between the common uncorrelated red signal (CURN) present in the data and a SGWB from cosmic strings with tensions $G\mu \sim 10^{-10}-10^{-11}$.

\subsubsection{Description of the models}

Since the SGWB is sourced by sub-horizon cosmic string loops, the central quantity in our analysis is the loop density distribution, $\textbf{n}(\ell,t)$, where $\ell$ is the invariant length of the loop (defined by its total energy divided by $\mu$) and $t$ is cosmic time. 
We focus on the most up-to-date loop distributions, calibrated on large scales with Nambu-Goto simulations \citep{Lorenz:2010sm,Blanco-Pillado:2013qja}. 
Loops are formed from the intercommutation of super-Horizon infinite strings, with a maximal size $\ell\approx 0.1 t$ at time $t$. 
(Note that the simulations assume an intercommutation probability of $1$, as is the case for field theory strings.) 
The two models --- which we refer to as BOS~\citep{Blanco-Pillado:2015ana} and LRS~\citep{Lorenz:2010sm}, after the names of their authors --- differ in their treatment of small loops, particularly on scales at which gravitational radiation (not included in numerical simulations) can have important effects.  
Indeed, compared to the BOS model, the LRS model has an additional population of small loops, which leaves an imprint in the SGWB spectrum \citep{Auclair:2020oww}. 
The explicit expressions for the two loop distributions are given in \cite{LIGOScientific:2017ikf}, and both have been considered by the LVK \citep{LIGOScientific:2021nrg} and LISA \citep{Auclair:2019wcv} collaborations.
The third observing run of LVK placed constraints on $G\mu$, based on the non-detection of a SGWB.
These are $G\mu \lesssim 10^{-8}$ for BOS and $G\mu \lesssim 10^{-15}$ for LRS.
At the frequency of ground-based detectors, the SGWB signal is produced by loops formed during the radiation era.
At low PTA frequencies, the SGWB signal is dominated by larger loops, namely those formed at recent times, in transition from the radiation to matter era and also in the matter era \citep{Ringeval:2017eww,Auclair:2020oww,Leclere:2023ryt}. 

Other than $G\mu$, a further parameter appears in $\textbf{n}(\ell,t)$, namely $\Gamma$, which describes the rate at which loops lose energy through gravitational radiation: $\dot{\ell} = - \Gamma G\mu$.
If $\Gamma$ is large, fewer loops are present in the distribution since loops decay more rapidly. 
On the other hand, GWs are emitted more intensively: the final effect is a combination of these two, which impacts the SGWB of the BOS and LRS models in subtly different ways \citep{Auclair:2020oww}. 
The value of $\Gamma$ depends on the properties of loops, and in particular on how many cusps, kinks, and kink-kink they contain as \citep{Damour:2001bk,Siemens:2006yp,Ringeval:2017eww} 
\begin{equation}
    \label{eq: total Gamma}
    \Gamma = N_c\Gamma^{c} + N_k\Gamma^{k} + \frac{N_k^2}{4} \Gamma^{kk},
\end{equation}
where $N_{c,k}$ is the number of cusps/kink events per oscillation period of the loop,  $N_k^2/4$ the number of kink-kink collisions, and
\begin{equation}
    \label{different gamma values for each events}
    \Gamma^c = \frac{3(\pi g_1^c)^2}{2^{1/3}g_2^{2/3}}, \: \Gamma^k = \frac{3(\pi g_1^k)^2}{2^{2/3}g_2^{1/3}}, \: \Gamma^{kk} = 2(\pi g_1^{kk})^2 \;,
\end{equation}
where $g_2 = \sqrt{3}/4, g_1^c\approx 0.85, g_1^k\approx 0.29, g_1^{kk}\approx 0.1$.

In this paper, we consider 2 cases. For the first model (i) 
$N_c = 2$ and $N_k=0$, for which $\Gamma=57$~\citep{Leclere:2023ryt}, a value close to that observed in numerical simulations of individual loops~\citep{Vachaspati:1984gt,Blanco-Pillado:2017oxo}. Therefore, the only free parameter for this model is $G\mu$. 
As for the second model (ii), we fix $N_c=1$ and allow $N_k$ to vary between 1 and 200 (with a uniform prior) so as to account for  theoretical uncertainties on the initial number of kinks at loop creation, and on the efficiency of the gravitational backreaction that should smooth out the loops \citep[see the LVK analysis,][for a similar approach]{LIGOScientific:2021nrg}. 
Therefore, this is a two-parameter model $(G\mu, N_k)$.

The fractional energy density of the SGWB per logarithmic interval of frequency is given by
\begin{multline}
    \label{eq: full SGWB energy density}
    \Omega_{\rm{GW}}(t_0, f) = \frac{16 \pi (G\mu)^2}{3H_0^2} \sum_b \frac{N_b \Gamma^{(b)}}{\zeta(q_b)} \\ 
    \times \sum_{n=1}^{+\infty} \int \frac{n^{1-q_b} \mathrm{d}\,{z}}{(1+z)^5 H(z)}
\textbf{n}\left[\frac{2n}{(1+z)f}, t(z)\right],
\end{multline}
where $H_0$ is the Hubble constant, $H(z)$ is the Hubble parameter for which we assume standard $\Lambda$CDM cosmology with the Planck-2018 fiducial parameters \citep{Planck:2018vyg}, and $t_0$ is the cosmic time today. The sum is over the cusp, kink and kink-kink contributions (labelled with the index $b$) for which $q_b=4/3$, $5/3$  and $2$ respectively.

\subsubsection{Analysis results}

Our analysis follows the one presented in \cite{Leclere:2023ryt}, which was based on six pulsars with the best timing precision from the EPTA early Data Release 2 \citep{ccg+21}.
We now analyse the \texttt{DR2new} dataset with $25$ pulsars, using the (computationally heavy) hierarchical likelihood data analysis method described in PaperIII. We sample the SGWB parameters $(G\mu)$ or $(G\mu,N_k)$ as well as the individual pulsar noise parameters.

\begin{figure}
    \centering
    \includegraphics[width=0.48\textwidth]{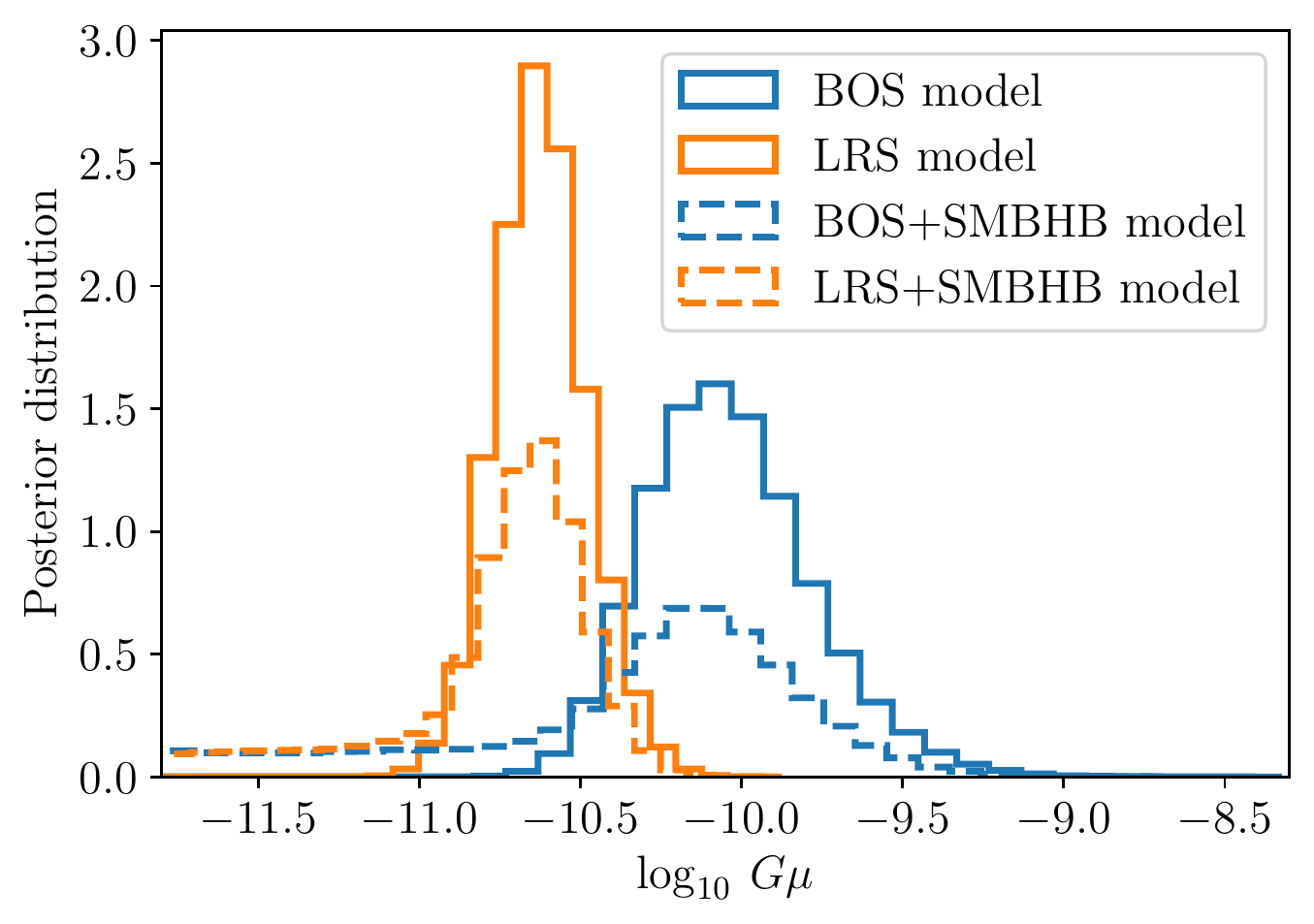}
    \caption{\footnotesize{Comparison of the string tension posteriors for the two string models (BOS and LRS) in case \textit{(i)}, $N_c=2$ and $N_k =0$ ($\Gamma = 57$). Solid lines assume only a cosmic string background, dashed lines assume both a population of GW-driven circular SMBHBs and cosmic strings.}} 
    \label{fig:cs_gmu_post}
\end{figure}

\begin{figure*}
    \centering
    \includegraphics[width=0.75\textwidth]{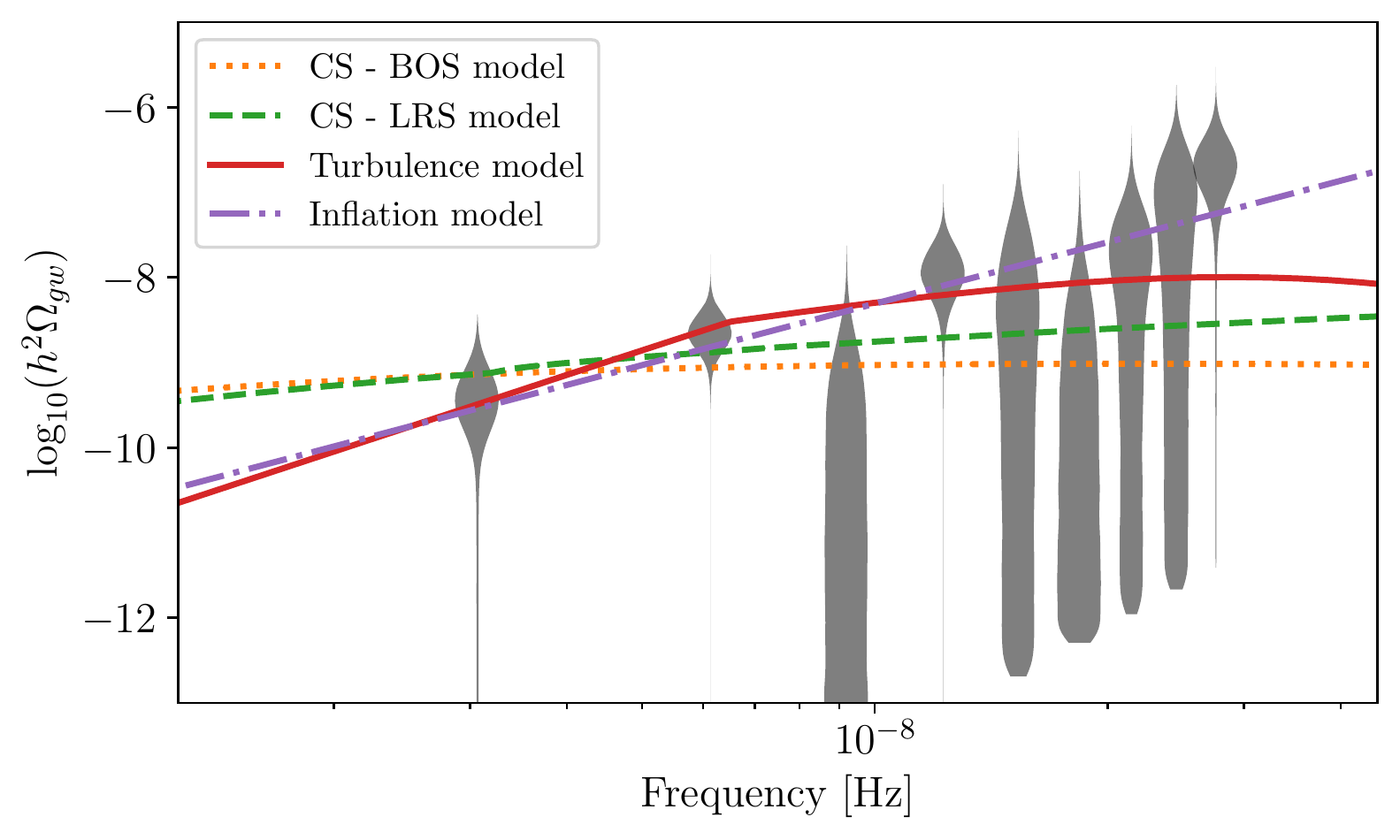}
    \caption{\footnotesize{SGWB spectra (in terms of $\log_{10} h^2\Omega_{gw}$) for four different early Universe SGWB models considered in this paper. BOS/LRS correspond to a cosmic string background with $N_c=2$ and $N_k =0$ ($\Gamma = 57$), and $\logGmu = -10.1$/$-10.6$. The GWB from turbulence is plotted in solid line for $\lambda_* \mathcal{H}_* = 1$, $\Omega_*=0.3$, and $T_* = 140$ MeV.  The inflationary spectra is shown for $\log_{10} r = -13.1$ and $n_T = 2.4$ (maximum a posteriori value). Power spectrum of the 2nd-order GWB from the scalar curvature perturbations described by the powerlaw model with $A_\zeta^{\text{10yr}}=-2.9$ and $n_s=2.1$ is shown with brown puncture-dot line. The nine first Fourier bins posteriors of the common signal are represented by the grey violin areas.}}
    \label{fig:spectrum_plot}
\end{figure*}

Solid lines in Fig.~\ref{fig:cs_gmu_post} show the posterior distribution on $\log_{10}(G\mu)$ in case \textit{(i)}, for which $N_k=0$ and $\Gamma=57$. The string tension 90\% credible (symmetric) interval is $\logGmu = -10.07^{+0.47}_{-0.36}$ (resp.~$-10.63^{+0.24}_{-0.22}$) for the BOS (resp.~LRS) model.
The corresponding SGWB spectra are shown in Fig.~\ref{fig:spectrum_plot} where, for each model, we set $G\mu$ to their median values.
We also consider the two-component SGWB model made of the sum of a background from cosmic strings and one from a population of GW-driven circular SMBHBs characterised by the PSD of Eq.~\eqref{eq:sh}. 
The posteriors of the two background parameters $(\log_{10} A, \logGmu)$ are highly correlated, since both provide a possible explanation for the detected signal. As a result, the posterior on $\logGmu$ no longer has compact support, but a tail to lower values (see the dashed lines in Fig.~\ref{fig:spectrum_plot}). We therefore extract the 95-quantile of the string tension posterior to obtain an upper bound of $\logGmu < -9.77$ (resp.~$-10.44)$ for the BOS (resp.~LRS) models.  

The \texttt{DR2new} dataset exhibits a shallower slope for the PSD of the common red signal than \texttt{DR2full}. While cosmic strings were a good fit to the common red signal of 6 pulsars of \texttt{DR2full} \citep{Leclere:2023ryt}, this is no longer true for \texttt{DR2new}. This is because the predicted SGWB PSD is generally steeper than the measured correlated red signal in the data, as can be seen in Fig.~\ref{fig:spectrum_plot}. 

For case \textit{(ii)}, we obtain very similar results to those discussed in \cite{Leclere:2023ryt}. Namely, we obtain quasi non-informative posteriors for $N_k$, showing that the data can be equally explained by a population of kinky loops with $N_k \gtrsim 120$. 
In other words, we cannot extract any upper bound on the number of kinks, since this quantity is degenerate with $G\mu$. 

\subsection{Implications on background from turbulence around the QCD energy scale}



Turbulence can arise in the early Universe in the aftermath of a first-order phase transition \citep{Witten:1984rs,Kamionkowski:1993fg}, or can be driven by
pre-existing primordial magnetic fields \citep{Quashnock:1988vs,Brandenburg:1996fc}.
If the (magneto-)hydrodynamic turbulence were present around the QCD epoch, when the Universe had a temperature of $T_*\sim 100$ MeV, it would generate a GWB 
in the PTA band.
The characteristic scale of the turbulence, determining the characteristic GW frequency, is in fact 
related to the (comoving) Hubble radius at that epoch $\lambda_*\simeq  \mathcal{O}(\mathcal{H}_*^{-1})$, where 
\begin{equation}
    {\cal H}_* \simeq 10^{-8} \frac{T_*}
    {100 {\rm \, MeV}} \biggl(\frac{g_*}{10}\biggr)^{1\over 6} \, {\rm \, Hz},
\end{equation}
and $g_*$ denotes the number of relativistic degrees of freedom. 
If a large lepton asymmetry and/or primordial magnetic fields were present in the early Universe, the QCD phase transition might have been of first order \citep{2009JCAP...11..025S,2018PhRvL.121t1302W,2020arXiv200900036M,2022PhRvD.105l3533M,2021PhRvL.126a2701V,2023PhRvD.107a4021C}. 
In this case, one would expect additional sources of GWs, from the collision of broken phase bubbles and the subsequent development of sound waves in the primordial fluid \citep{1992PhRvD..45.4514K, 1993PhRvD..47.4372K,2008PhRvD..77l4015C,2008JCAP...09..022H,2017PhRvD..95b4009J,2018PhRvD..97l3513C,2014PhRvL.112d1301H,2015PhRvD..92l3009H,2017PhRvD..96j3520H}. 
This was analysed for PTAs, e.g.~in \cite{2021NatAs...5.1268M,NANOGrav:2021flc,Xue:2021gyq}.
In what follows, we focus on the GWB generated by decaying (M)HD turbulence. 

\subsubsection{Description of the model}

The presence of bulk velocity and magnetic fields produce anisotropic stresses, which in turn act as a source of GWs \citep{Kamionkowski:1993fg,
2002PhRvD..66b4030K, 2002PhRvD..66j3505D,2006PhRvD..74f3521C,2007PhRvD..76h3002G,2009JCAP...12..024C}.
This has been recently studied via numerical simulations in \cite{RoperPol:2018sap,RoperPol:2019wvy,Brandenburg:2021tmp,RoperPol:2022iel}.
In particular, \cite{RoperPol:2022iel} show that the envelope of the GWB produced by decaying
MHD turbulence can be estimated analytically, assuming that the
anisotropic stresses from the velocity and magnetic fields
vary more slowly than the dynamical production of GWs.
This was also validated by numerical simulations of purely kinetic turbulence in \cite{Auclair:2022jod}.
This assumption leads to the following GWB signal:
\begin{equation}
    \Omega_{\rm GW} (f) = 3 \, {\cal A} \, \Omega_*^2 \, \bigl(\lambda_* {\cal H}_*\bigr)^2 F_{\rm GW, 0} \,
    S_{\rm turb} (\lambda_* f),
    \label{eq:OMGWturb}
\end{equation}
where $\Omega_*$ is the ratio of the (M)HD turbulent energy density to the radiation one, and $\lambda_* {\cal H}_*$ is the ratio of the characteristic length scale of the turbulence, $\lambda_*$, to the comoving Hubble horizon ${\cal H}_*^{-1}$ at the QCD epoch.
The parameter ${\cal A} \simeq 1.75 \times 10^{-3}$ is the efficiency of GW production,\footnote{This estimate is conservative since it only considers the decaying stage of turbulence.
Numerical simulations find larger values when including a stage of turbulence production \citep{RoperPol:2019wvy,Kahniashvili:2020jgm,RoperPol:2021xnd}} estimated in \cite{RoperPol:2022iel}.
The function $F_{{\rm GW}, 0}$ is the fractional radiation energy density at the epoch of GW generation to its value at the present time.
It depends on the temperature scale $T_*$ via the number of degrees of freedom $g_*$,
\begin{equation}
    F_{\rm GW, 0} \simeq 8 \times 10^{-5} \, \biggl(\frac{10}{g_*}
    \biggr)^{1/3}.
\end{equation}
The spectral shape of the GWB signal, 
$S_{\rm turb} (f)$, is
\begin{align}
    S_{\rm turb} (\lambda_* f) 
    = &\, {\cal B} \, 
    \bigl(\lambda_*
    f\bigr)^3 \,
    p_\Pi (\lambda_* f) \, \nonumber \\ \times &\,
    \left\{ \begin{array}{ll}
        \ln^2 \bigl[1 + {\cal H}_* \, \delta t_{\rm fin}/(2 \pi) \bigr], & {\rm \ if \ } f < 1/\delta t_{\rm fin},  \\
        \ln^2 \bigl[1 + \lambda_* {\cal H}_*/(2 \pi \lambda_* f)
        \bigr], & {\rm \ if \ } f \geq 1/\delta t_{\rm fin},
    \end{array}\right.
    \label{eq:Sturb}
\end{align}
where ${\cal B} \simeq 50 \,
    \bigl(\lambda_* {\cal H}_* \bigr)^{-2}$ is a normalising factor,
and $\delta t_{\rm fin}$ denotes the effective  duration of the turbulence. 
The latter can be estimated, from the numerical simulations performed in \cite{RoperPol:2022iel},
to be $\delta t_{\rm fin} \simeq 2 \lambda_*/\sqrt{1.5\, \Omega_*}$.
The function $p_\Pi (\lambda_* f)$ in Eq.~\eqref{eq:Sturb} denotes  the spectrum of the anisotropic stresses. 
For solenoidal fields (e.g. a primordial magnetic field or 
vortical bulk fluid motion) characterised by a typical correlation scale of the order of the turbulence scale $\lambda_*$, it is constant
for $f < 1/\lambda_*$. Furthermore, it decays as  $f^{-11/3}$ for $f \gtrsim 1/\lambda_*$, if the turbulence is of the Kolmogorov type, as we assume here.
Hence, the resulting spectral shape of the GWB in Eq.~\eqref{eq:OMGWturb} presents
three power laws: $f^3$ at frequencies below the inverse effective duration of the turbulence $f < 1/\delta t_{\rm fin}$, $f$ at
intermediate frequencies $1/\delta t_{\rm fin} < f < 1/\lambda_*$, and $f^{-8/3}$ at large frequencies
$f > 1/\lambda_*$.

The GWB produced from vortical (M)HD turbulence is therefore determined by three parameters:
the temperature scale $T_*$,
the turbulence strength $\Omega_*$, and the turbulence characteristic length scale $\lambda_* {\cal H}_*$.
By causality, $\lambda_* {\cal H}_*$ is bound to be smaller than one.
In general, also $\Omega_* \lesssim 1$, otherwise turbulence would
change the dynamics of the Universe.
However, note that the template described above has been validated in principle only for non-relativistic plasma motions, for which $\Omega_* \lesssim {\cal O} (0.1)$.

\subsubsection{Analysis results}

\begin{figure}
    \centering
    \includegraphics[width=0.45\textwidth]{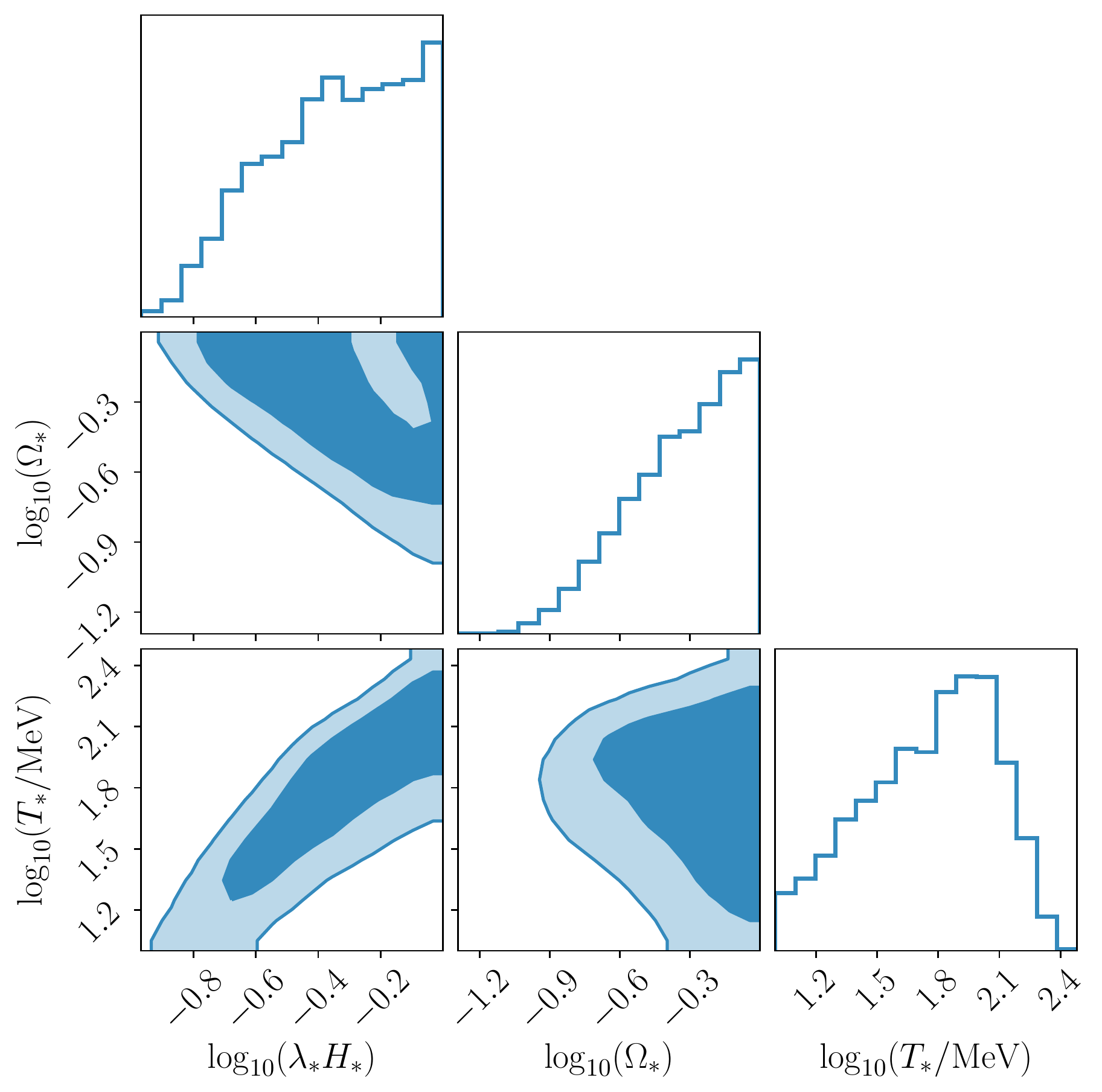}
    \caption{\footnotesize{2D posteriors for the parameters of the background from turbulence around the QCD energy scale obtained using a free spectrum fit on \texttt{DR2new} data. The 68\% and 95\% credible regions are displayed.}}
    \label{fig:qcd_post}
\end{figure}

As in \autoref{sec:inf}, here we use the fast free spectrum analysis method on \texttt{DR2new} data to constrain the model, considering the nine first Fourier bins of the RMS spectrum of \autoref{fig:EPTADR2}. We use $\log_{10}$-uniform priors for the model parameters, choosing $\log_{10} (\lambda_* \mathcal{H}_*)\in[-3, 0]$, $\log_{10} \Omega_*\in[-2, 0]$, and $\log_{10} (T_* / 1 \textup{MeV}) \in [1, 3]$. The 2D posteriors obtained are shown in \autoref{fig:qcd_post}.

For values of $\Omega_*$ below 0.1, the model can only explain the level of correlated noise at the lowest frequency bin if the amplitude of the spectrum is sufficiently high. This can be achieved only if $\lambda_*\mathcal{H}_*$ is close to 1 and the peak frequency lies within the PTA frequency range, implying $T_* \sim 60$ MeV.
However, at frequencies around the peak, the signal corresponds to a
power spectral density for the residuals steeper than $\gamma \sim 4$,
which cannot fit the data well.
For this reason, values of $\Omega_* \lesssim 0.1$ are disfavoured.

For larger values of $\Omega_*$, the $f^3$ part of the 
spectrum at frequencies below $\delta t_{\rm fin}^{-1} \propto \sqrt{\Omega_*}/(\lambda_* \mathcal{H}_*) \times \mathcal{H}_*(T_*)$ can enter the PTA band with a sufficiently high amplitude. Furthermore, the distance between the break at $\delta t_{\rm fin}^{-1}$ and the spectral
peak at $1/\lambda_*$ becomes minimal in the limit $\Omega_* \sim 1$.
Both of these characteristics lead to a better fit to the data. This is recovered in the posteriors of \autoref{fig:qcd_post}, together with the degeneracy between $\lambda_* \mathcal{H}_*$ and $\Omega_*$ from the signal amplitude (see \autoref{eq:OMGWturb}), and the degeneracy between $\lambda_* \mathcal{H}_*$ and $T_*$ from the break at $1/\delta t_{\rm fin}$ (note that the dependence of the latter on $\sqrt{\Omega_*}$ is subdominant).

The model therefore provides a good fit to the data in the limit of large $\Omega_*$, close to the upper bound of the prior. The extension of the dataset to longer observation time will be crucial for further constraining this model at low frequencies.

\subsection{Implications on the 2nd-order GWB produced by primordial curvature perturbations}
\label{sec:secondorder}

\begin{figure*}[ht]
    \centering
    \includegraphics[width=0.4\textwidth]{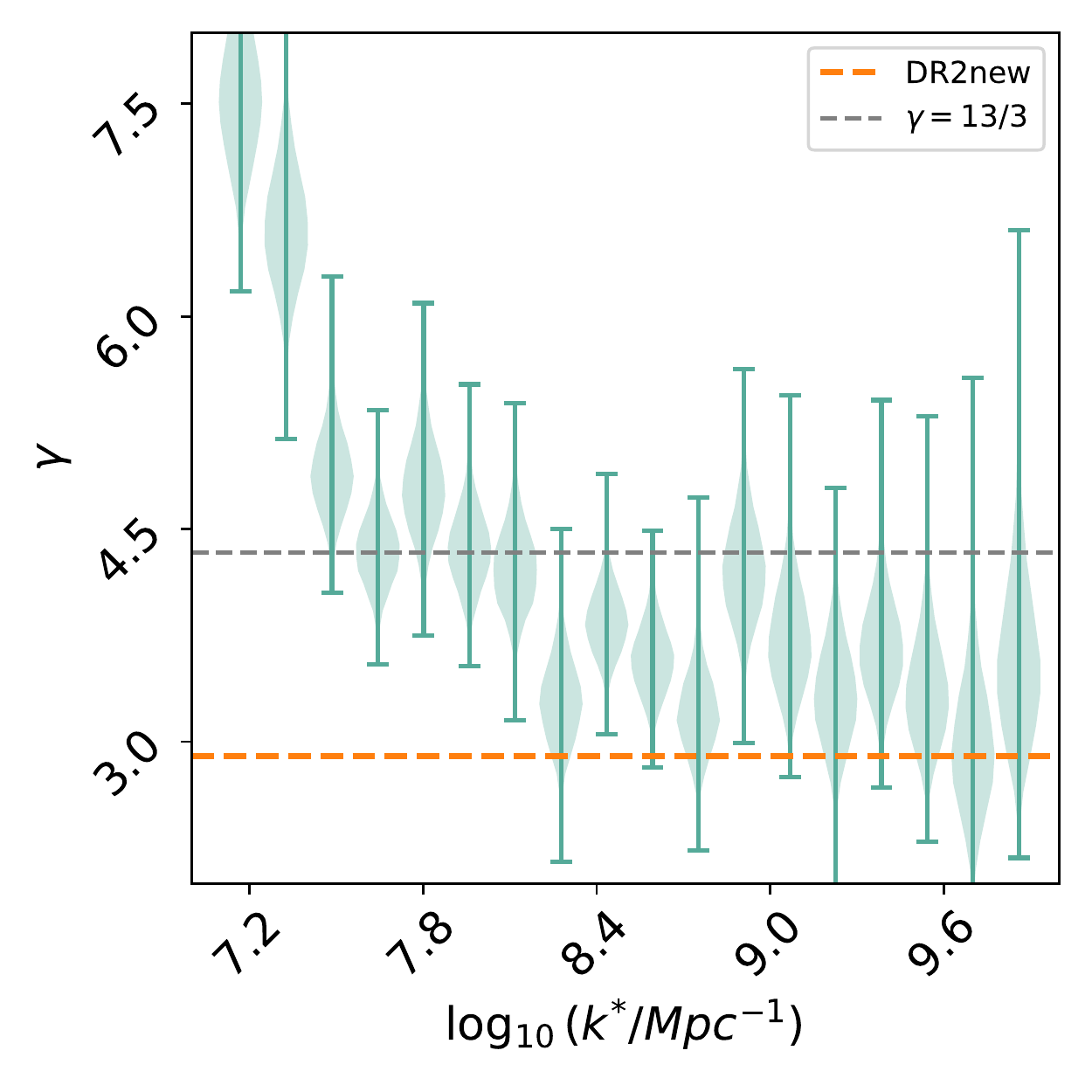}
    \includegraphics[width=0.4\textwidth]{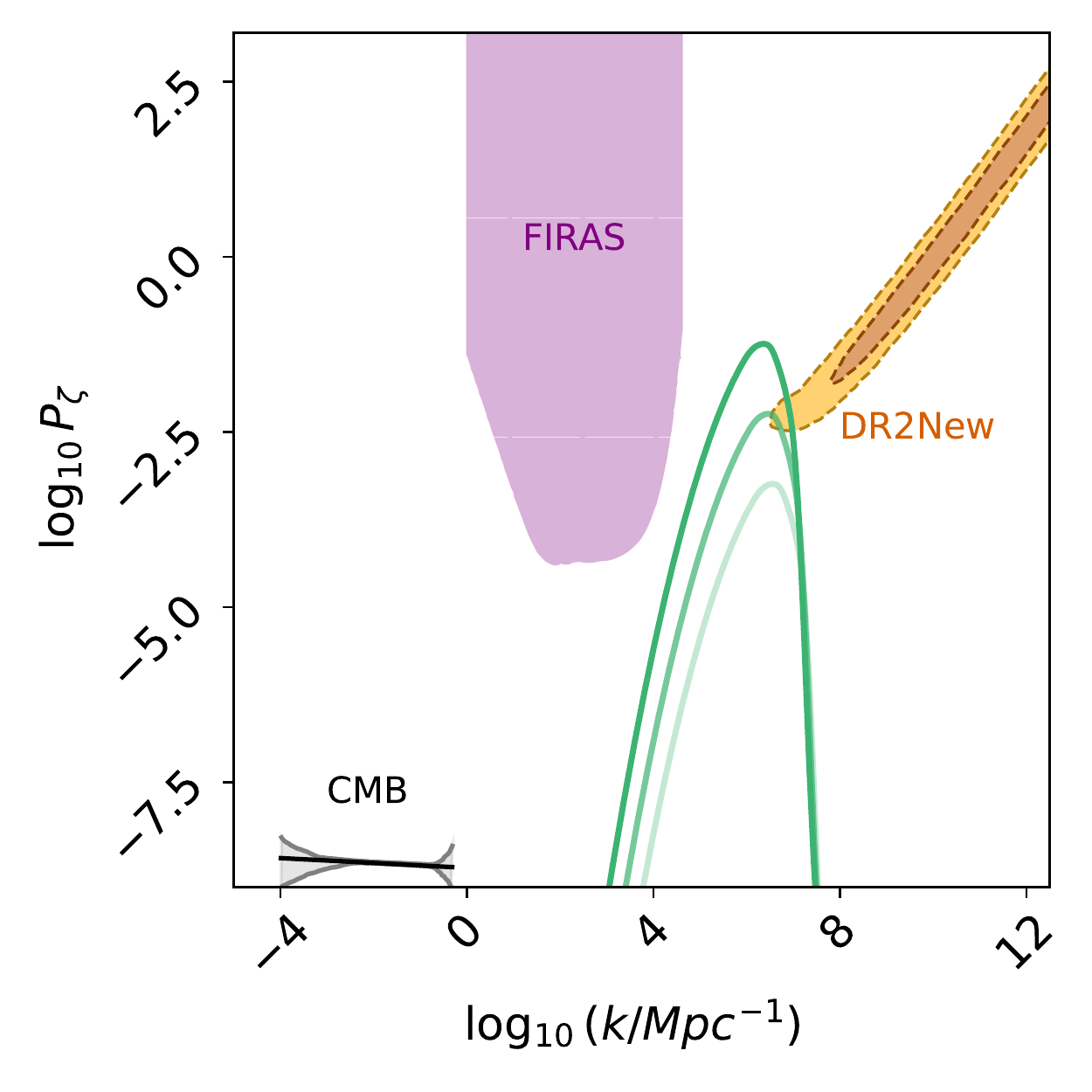}
    \caption{\footnotesize{Results for the monochromatic curvature perturbations described by Eq.~\ref{eq:monochromatic_sp}. Left panel: recovered slopes $\gamma$ of a simple power-law model as a function of characteristic scale $k^{*}$ of the injected GWB generated by the monochromatic curvature perturbations. The horizontal lines show the theoretical value of $\gamma$ from a population of circular, GW-driven SMBHBs (grey) and the one obtained in PaperIII (orange). Right panel: 1$\sigma$ and 2$\sigma$ contours of the posterior distributions on the amplitude $A_{\zeta}$ and characteristic scale of fluctuations $k^{*}$ for \texttt{DR2new} (orange colour). The posterior distribution is overlaid with the current constraints on the primordial power spectrum using Planck data (CMB). The grey colour depicts the 2-$\sigma$-confidence intervals. The purple shaded area represents the bounds from spectral distortions \citep{2012ApJ...758...76C}. For comparison in green we place the prediction of the primordial spectrum of scalar perturbations in the two-field model of inflation described in \cite{2020JCAP...08..001B} for a range of the model parameters. All three models result in PBH mass functions peaked at $\sim35$~$M_\sun$ with the brightest line corresponding to the dark matter fraction of PBHs of $\sim0.01$.}} 
    \label{fig:summ_cmb_delta}
\end{figure*}

\begin{figure*}[ht]
    \centering
    \includegraphics[width=0.4\textwidth]{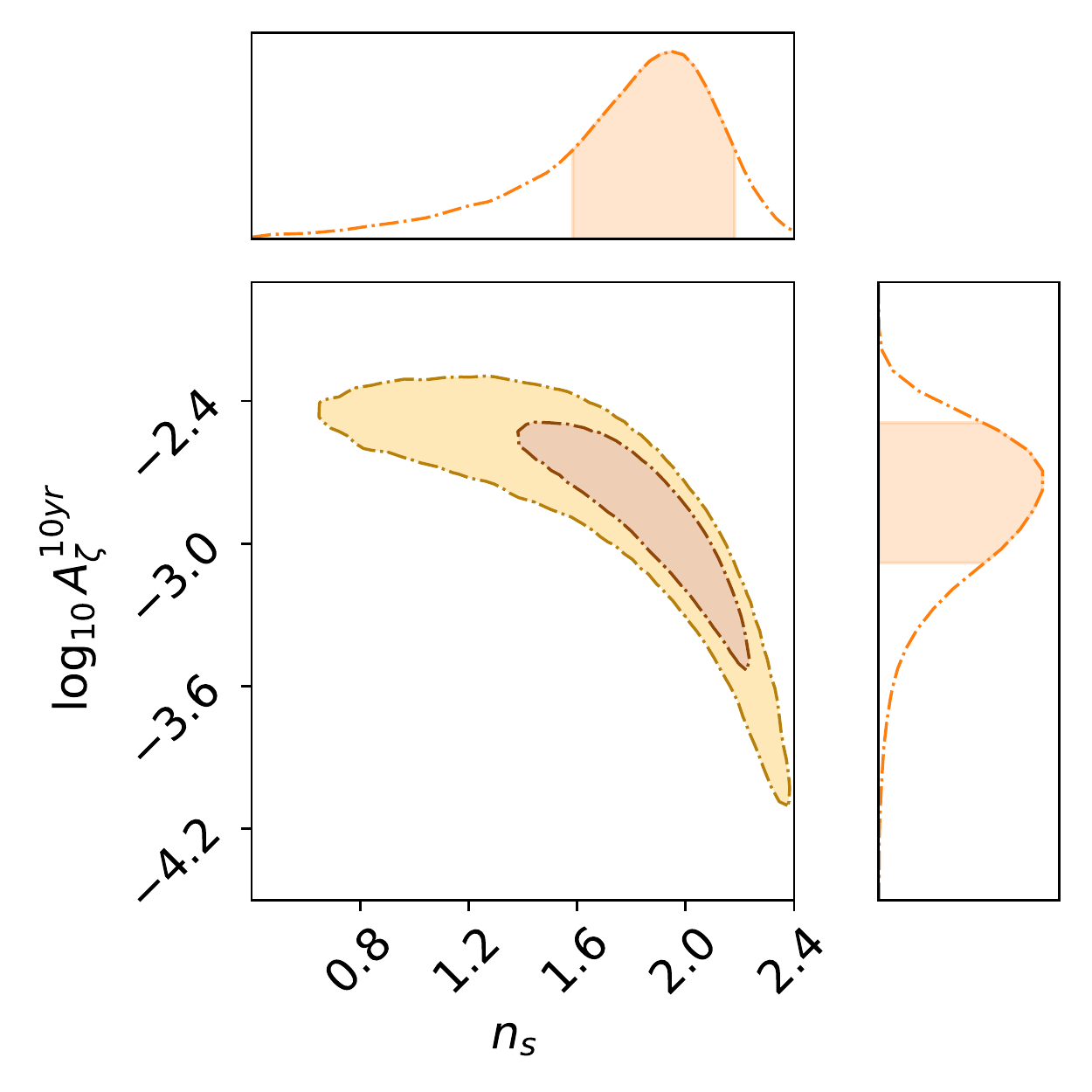}
    \includegraphics[width=0.4\textwidth]{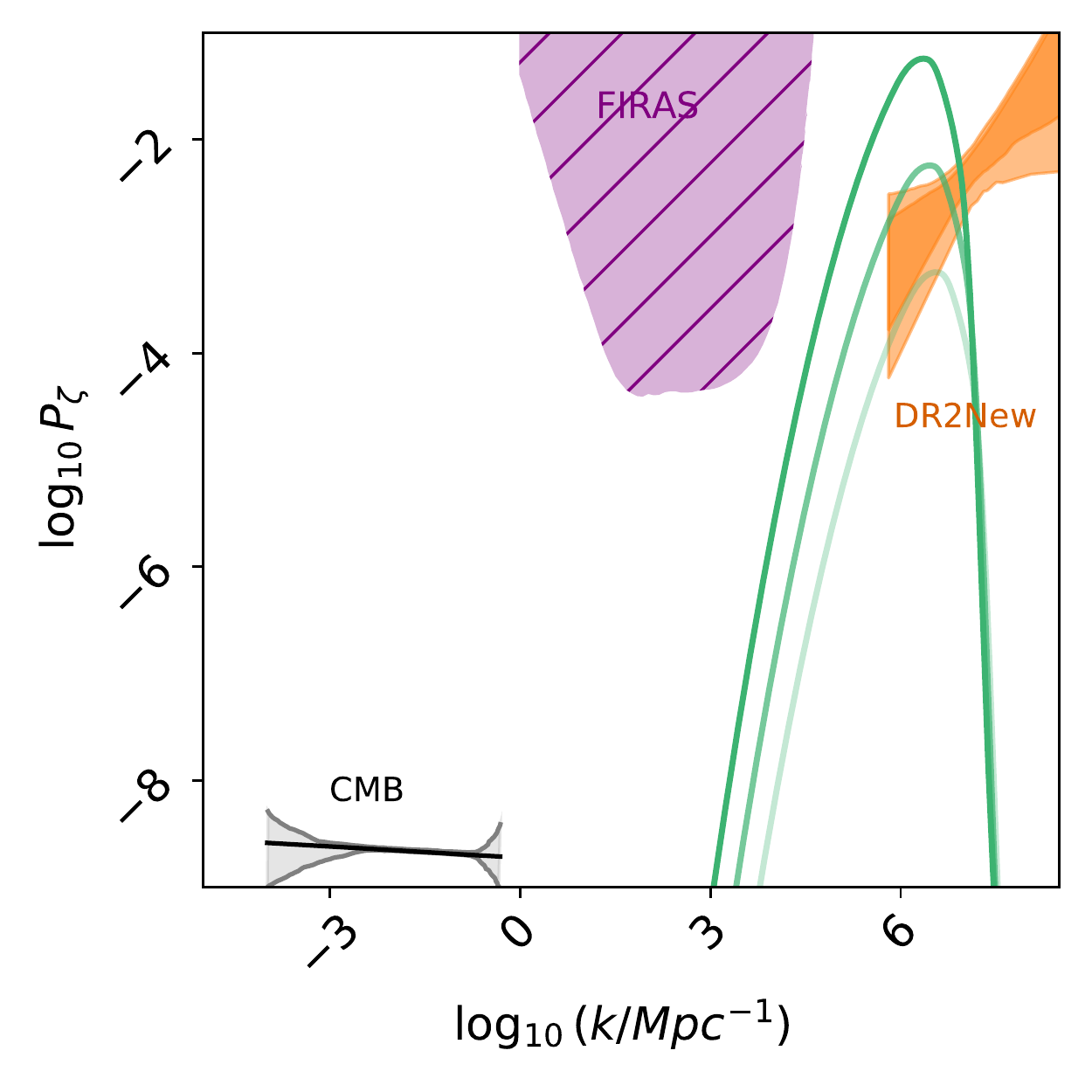}
    \caption{\footnotesize{Results for the power-law model of the curvature perturbations described by Eq.~\eqref{eq:powerlaw_sp}. Left panel: 1$\sigma$ and 2$\sigma$ contours of the posterior distributions on the amplitude $A_{\zeta}$ and the slope of the power spectrum $n_s$ obtained by the analysis of \texttt{DR2New}. Right panel: 1$\sigma$ and 2$\sigma$ contours of the power spectra inferred from the \texttt{DR2New} analysis by picking 1000 random samples from the posteriors overlaid with the current constraints on the primordial power spectrum using the latest Planck data. The grey colour depicts the 2$\sigma$-confidence intervals. The green lines and purple shaded areas are the same as in Fig.~\ref{fig:summ_cmb_delta}.}}
    \label{fig:summ_cmb_powerlaw}
\end{figure*}

It is well-known that scalar, vector and tensor modes of the perturbed metric do not mix at linear order of the Einstein equations \citep{1946ZhETF..16..587L, baumann_2022}. However, scalar curvature perturbations will source propagating tensorial modes (GWs) at the 2nd order in perturbation theory \citep{1967PThPh..37..831T, 1993PhRvD..47.1311M, 1998PhRvD..58d3504M, 2004PhRvD..69j4011N, 2005PhRvD..71d3508C, 2007PhRvD..75l3518A, 2007PhRvD..76h4019B}. Such scalar curvature perturbations and associated primordial density fluctuations inevitably exist in the Universe and can be directly constrained by observations of the CMB. The latest Planck data \citep{2020A&A...641A..10P} suggests that the power spectrum of the curvature perturbations is nearly scale-invariant with the amplitude $A_{\zeta}\sim2\times 10^{-9}$, which implies a marginal energy-density of the generated GWB. Specifically, when projected to the PTA sensitivity band, the fractional contribution of the energy density in the associated GWs becomes $\Omega_{\text{gw}}\sim10^{-24}$, which is practically non-detectable by current experiments. On the other hand, some models of inflation \cite[see, for example,][and references therein]{2018JCAP...07..007D,
2019JCAP...06..028B,2020JCAP...08..001B, 2023EPJC...83...82Y} make it possible to produce a sharp increase in the power spectrum of primordial curvature perturbations over many orders of magnitude  at small scales.

While the CMB is only capable of directly sampling large cosmological scales with $k\sim10^{-3}-10^{-1}$~Mpc$^{-1}$, small scales stay largely uncovered. PTAs provide a unique opportunity to complement the CMB measurements by indirectly probing the scalar curvature perturbations in a scale range $k\sim10^{6}-10^{8}$~Mpc$^{-1}$ through the second-order generated GWB, and to place bounds on the steepest possible growth of the power spectrum as well as corresponding models of inflation \citep{2009PhRvL.102p1101S, 2011PhRvD..83h3521B, 2020PhRvL.124y1101C, 2023arXiv230207901D, 2023Univ....9..157Z}. 

In this work, we consider two models of the primordial curvature power spectrum: i) monochromatic and ii) powerlaw. For the former, the primordial spectrum is modelled as:
\begin{equation}
P_{\zeta}=A_{\zeta} \delta (\text{log}k/k^{*})\,
\label{eq:monochromatic_sp}
\end{equation}
where $A_{\zeta}$ is a dimensionless amplitude and $k^{*}$ is a wavenumber at which the monochromatic power spectrum has a Dirac-delta peak. For the second case:
\begin{equation}
P_{\zeta}=A^{10\text{yr}}_{\zeta}\left(\frac{k}{k_{\text{10yr}}}\right)^{(n_s-1)}\,
\label{eq:powerlaw_sp}
\end{equation}
where $n_s$ characterises the slope and $k^\text{10yr}$ is a normalizing scale $k_\text{10yr}=2\pi/(10yr\times c)$, so that $A^{10\text{yr}}_{\zeta}$ corresponds to dimensionless amplitude at ten years.  

In the first scenario, a  semi-analytical solution for the induced spectrum of GWB exists and is given by \citep{2018PhRvD..97l3532K, 2018JCAP...09..012E}:
\begin{multline}
\Omega_{\text{GW}}\left(f=\frac{kc}{2\pi}\right)  = \frac{3 A_{\zeta}^{2}}{64} \left( \frac{4 - \tilde{k}^{2}}{4} \right)^{2} \tilde{k}^{2} \left( 3 \tilde{k}^{2} - 2\right) \Theta(2 - \tilde{k}) \times\\
\times \left( \pi^{2} ( 3 \tilde{k}^{2} - 2)^{2} \Theta (2 \sqrt{3} - 3 \tilde{k}) + \left( 4 + ( 3 \tilde{k}^{2} - 2)^{2} \log \left|1 - \frac{4}{3 \tilde{k}^{2}} \right|\right)^{2} \right)\,
\label{eq:gwb_monochromatic}
\end{multline}
where $\tilde{k}=k/k^{*}$ and $\Theta$ is the Heaviside theta function. In spite of being nonphysical, the $\delta$-function peak approximately describes the maximum of the produced GWB in the inflationary model with the steepest possible $k^4$ growth of a spectral peak in the single-field inflation at small scales \cite[see Figure 7 in][]{2019JCAP...06..028B}.

In the second case of a more general (and more realistic) power-law spectrum, the result can only be obtained numerically \citep{2018PhRvD..97l3532K}:
\begin{equation}
\Omega_{\text{GW}}\left(f=\frac{kc}{2\pi}\right) = Q(n_s)(A_{\zeta}^{\text{10yr}})^2\left(\frac{k}{k_{\text{10yr}}}\right)^{2(n_s-1)}\,
\label{eq:gwb_powerlaw}
\end{equation}
where $Q(n_s)$ is the scaling factor which can be evaluated in a range of $n_s$ using interpolation points from Table 1 of \cite{2018PhRvD..97l3532K}. 

After its production, the GWB is damped due to quantum interactions with the particles of the primordial plasma at the radiation-dominated epoch, and redshifted inversely proportionally to the scale factor (as it also occurs to radiation) starting from the epoch of matter-radiation equality \citep{2018JCAP...05..035S}. The present value of the fractional energy density is then:
\begin{equation}
\Omega^{0}_{\text{GW}} = 2\Omega_{r}^{0}\left(\frac{g_{*}(T)}{g_{*}(T_{\text{eq}})}\right)\left(\frac{g_{*s}(T)}{g_{*s}(T_{\text{eq}})}\right)^{-\frac{4}{3}}\Omega_{\text{GW}}\,
\end{equation}
where $T$ is the temperature of the Universe at the moment when structures of a typical size $1/k$ re-enter the horizon\footnote{We conservatively set the temperature at the epoch of production to 17.35 K.}, $T_{\text{eq}}$ is the temperature of the Universe at the epoch of matter-radiation equality, $g_{*}$ and $g_{*s}$ are relativistic degrees of freedom and degrees of freedom in entropy, respectively. The final expression for the auto-power spectral density of the timing residuals is:
\begin{equation}
S(f)=\frac{H_0^2}{8\pi^4}\frac{\Omega_\text{GW}^0(f)}{f^5},
\end{equation}
where $H_0$ is the Hubble constant at the present epoch.

The outlined formalism was applied to the \texttt{DR2new} version of the latest EPTA dataset. The number of frequency components which was used for the Fourier representation of the signal was fixed to 9. We have chosen broad uninformative priors for the parameters: uniform in $[-6, 3]$ for $\log_{10}A_{\zeta}$ and $\log_{10}A_{\zeta}^\text{10yr}$, uniform in $[4, 12]$ for $\log_{10}(k^{*}/\text{Mpc}^{-1})$, and uniform in $[0.4, 2.6]$ for $n_s$. Boundaries for the latter are constrained by the limitations of the numerical approximation of the power law model. For this analysis, we assumed that the common red noise process detected in the latest EPTA dataset can be fully explained by the 2nd-order scalar-induced GWs. 

Results for monochromatic and power law models are shown in Figures~\ref{fig:summ_cmb_delta} and \ref{fig:summ_cmb_powerlaw}, respectively. The 2D posterior distribution of the model parameters of the monochromatic model is depicted with orange contours on the right panel of Fig.~\ref{fig:summ_cmb_delta}. One may notice that the regions of the highest probability are strongly elongated due to a strong positive correlation between $A_{\zeta}$ and $k^{*}$; these parameters are essentially degenerate. Therefore, \texttt{DR2new} can only provide lower limits on the characteristic scale and amplitude of the monochromatic model: $\log_{10}(k^{*, 0.05}/\text{Mpc}^{-1})=7.6$ and $\log_{10}A^{0.05}_{\zeta}=-1.7$ meaning that a whole range of models predicting Dirac-delta power spectrum can equally good describe the signal of \texttt{DR2new}. 
This behaviour is explained in the left panel of Fig.~\ref{fig:summ_cmb_delta}, for which we have simulated GWB signal generated by the monochromatic primordial scalar perturbations, Eq.~\eqref{eq:gwb_monochromatic}, and attempted to recover a more general power law model of the form $S(f)\sim f^{-\gamma}$ used in PaperIII to model an arbitrary common red noise process. After a rapid decrease, the recovered slope stabilises at $\gamma\sim 3.5$ in the limit of large $k^{*}$, and becomes consistent with both values obtained with \texttt{DR2new} and the theoretically predicted 13/3 for the background from circular, GW-driven SMBHBs. This degeneracy can raise important issues when one tries to disentangle one signal from another. 
For the power-law case, the 2D posteriors are shown in the left panel of Fig.~\ref{fig:summ_cmb_powerlaw} with the following means and 1-$\sigma$ uncertainties: $\log_{10}A_{\zeta}=-2.94^{+0.42}_{-0.46}$ and $n_s=2.11^{+0.25}_{-0.32}$. 

On the right panels of Figs.~\ref{fig:summ_cmb_delta} and \ref{fig:summ_cmb_powerlaw}, we also overplot the inferred primordial power spectrum with the one obtained from the Planck data. The orange areas on the right panel of Fig.~\ref{fig:summ_cmb_powerlaw} are 1$\sigma$ and 2$\sigma$ contours of the power-law model described by Eq.~\ref{eq:powerlaw_sp} reconstructed using 1000 random draws from the posteriors. To explain the observed signatures of the \texttt{DR2new} in terms of the second-order GWB from the primordial scalar perturbations, an excess in the primordial spectrum at low scales should be invoked without violating the CMB inflationary parameters. Such excess has been proposed in many papers, e.g. in the aforementioned works by \cite{1994PhRvD..50.7173I, 2017PDU....18....6G, 2018JCAP...07..007D, 2018PhRvL.121h1306C, 2018JCAP...07..032B, 2019JCAP...06..028B,2020JCAP...03..002M, 2020JCAP...08..001B,2023EPJC...83...82Y}. Notably, this power excess would lead to a copious production of primordial black holes (PBHs) at the radiation-dominated stage, which is sometimes taken by the authors as the motivation to introduce them as cold dark matter candidates \cite[][and reference therein]{2016PhRvD..94h3504C}. 
The PBH formation from cosmological perturbations has been extensively explored \cite{2018CQGra..35f3001S}. On the radiation-dominated stage, the PBH mass is related to the mass inside the horizon at the time of the perturbation entering, $M\sim M_H\approx m_{Pl}^2t\sim 8 M_\odot (100 {\rm MeV}/T)^2(40/g_*)^{1/2}$, where $m_{Pl}$ is the Planckian mass, $T$ is the temperature. 
In terms of the mode comoving wavenumber at the moment of the horizon crossing, $k=aH$, the part of the horizon mass collapsing into a PBH reads 
\begin{equation}
    M(k)\approx 30 M_\odot \left(\frac{10.75}{g_*}\right)^{-1/6}\left(\frac{3\times 10^5\rm Mpc^{-1}}{k}\right)^{2}\,.
\end{equation}
For example, in the two-field inflationary model by \cite{2020JCAP...08..001B}, a peak around $k\sim 2\times 10^6$~Mpc$^{-1}$ could explain the power $A_\zeta\sim 10^{-2}$ and simultaneously lead to the production of PBHs with masses peaked at $\sim 35 M_\odot$ \citep[see also the analysis of the NANOGrav results in ][]{2021PhRvL.126e1303V}. Interestingly, such a peak seems to be observed in the chirp mass distribution of LVK merging binary black holes \citep{2023PhRvX..13a1048A}. A more general list of primordial power spectra, as well as a careful retranslation of them to the PBH abundance and their mass function, will be considered in a follow-up paper (Porayko et al., in prep.).

\section{Implications III: dark matter}
\label{sec:dark_matter}

Unlike spatially and temporally-correlated stochastic processes discussed in Secs.~\ref{sec:SMBHBs} and~\ref{sec:early_universe}, in this Section, we explore a possible deterministic contribution to the EPTA signal from ultralight scalar-field dark matter (ULDM).
For comparison, the morphology of a putative ULDM signal, predicted by~\cite{Khmelnitsky_2014}, is similar to a CGW from a SMBHB, that is, it is prominent only in one frequency bin.
Given that the signal observed with \texttt{DR2new} is mostly apparent in the first two fundamental $T^{-1}$ frequency bins, it is of interest to consider possible contributions from physical processes with narrowband spectra.
Therefore, the analysis presented here complements the CGW interpretation of the signal by~\cite{wm4}.
Additionally, the ULDM search with~\texttt{DR2full} is performed in~\cite{Smarra_2023}.

Dark matter currently constitutes approximately 26\% of the energy density of the Universe, as confirmed by, e.g., galactic rotation curves \citep{Rubin_1970, Rubin_1980, de_Salas_2019}, baryonic acoustic oscillations and cosmic microwave background measurements \citep{Bennett_2013, Planck:2018vyg} as well as galaxy surveys \citep{Escudero_2015}. 
The standard cold dark matter (CDM) picture, whose leading candidates are the Weakly Interacting Massive Particles (WIMPs) \citep{Arcadi_2018} and the QCD axion \citep{Di_Luzio_2020}, successfully grasps the large-scale structure of the Universe. However, it presents some well-known issues, when it comes to explaining observations at scales smaller than $\mathcal{O}$(kpc). Among these, the \textit{cusp-core problem} \citep{Flores_1994,Moore_1994,Karukes_2015} concerns the inconsistency between the observation of a flat density profile in the centre of galaxies and the power-law-like behavior predicated by CDM, while the mismatch between the simulated and observed number of dwarf galaxies in the proximity of our Milky Way is often referred to as the \textit{missing satellite problem} \citep{Klypin_1999, Moore_1994}. 

While a thorough understanding of baryonic physics feedback mechanisms \citep{Navarro_1996, Governato_2012, Brooks_2013, Chan_2015, Onorbe_2015, Read_2016, Morganti_2017} might help to alleviate some of these issues, they can be more easily disposed of assuming that DM is an ultralight ($m_\phi \sim 10^{-22}$eV) scalar/pseudoscalar or axion-like field, whose astrophysically large ($\mathcal{O}$(kpc)) de Broglie wavelength suppresses power on small scales.
Moreover, ultralight scalars are also generally present in string theory compactifications \citep{Green_1987, Svrcek_2006, Arvanitaki_2010}, which makes them interesting candidates for new physics as well.
CMB-based arguments are used to constrain $m_\phi \gtrsim 10^{-24}$eV \citep{Hlozek_2015}, while Lyman-$\alpha$ bounds push the limit up to $m_\phi \gtrsim 10^{-21}$eV, provided that the ultralight particles account for more than 30\% of the full dark matter budget \citep{Irsic_2017,Armengaud_2017,Kobayashi_2017,Nori_2018, Rogers_2021}. A lower limit of $m_\phi \sim 10^{-19}$eV is claimed by studies of ultra-faint dwarf (UFD) galaxies \citep{Hayashi_2021, Dalal_2022}, but a wide consensus is yet to be reached.  
In a seminal work, \cite{Khmelnitsky_2014} showed that the travel time of pulsar radio beams is affected by the gravitational potential induced by ULDM particles, making thus PTAs excellent facilities to investigate the existence of ULDM particles. Moreover, they represent complementary probes which do not suffer from the small-scale structure modelling uncertainties that affect non-CMB bounds \citep{Schive_2014, Zhang_2019}, as the aforementioned Lyman-$\alpha$ or UFD limits. 
In the following, we robustly assume that ULDM interacts only gravitationally, therefore giving rise to periodic oscillations in the TOAs of radio pulses as described in \cite{Khmelnitsky_2014}.

Being non-relativistic and with a very large characteristic occupation number, the ULDM scalar field can be described as a classical wave whose oscillation frequency is twice its mass $m_\phi$ \citep{Khmelnitsky_2014}. 
The periodic displacement that ULDM induces on the TOAs of signal from a pulsar $P$ can be written as follows \citep{Porayko_2018}:
\begin{equation}
    \delta t(t) = \frac{\Psi_c(\vec{x})}{2m_\phi} [\hat{\phi}^2_E\sin{(2m_\phi + \gamma_E )} - \hat{\phi}^2_P\sin{(2m_\phi + \gamma_P )} ],
    \label{eq:st}
\end{equation} 
where
\begin{equation}
    \Psi_c(\vec{x}) = \frac{\pi G \rho_\phi}{m^2_\phi} \approx 6.52\cdot 10^{-18} \left(\frac{10^{-22}\text{eV}}{m_\phi}\right)^2 \left(\frac{\rho_\phi}{0.4\text{GeV}/\text{cm}^3}\right),
\label{eq:psi_c}
\end{equation}
where $\gamma_P$ ($\gamma_E$) is a pulsar (Earth) dependent phase and   $\hat{\phi}^2_P $ ($\hat{\phi}^2_E$) takes into account the stochastic nature of the axion-like field near the pulsar (Earth). The parameters and their priors are summarised in Table \ref{tab:ULDM_priors}.
Considering a typical value of $v_\phi \sim 10^{-3}$ for the ULDM velocity, the region in which the scalar field oscillates coherently, i.e. with the same value of $\hat\phi$, is spanned by the \textit{coherence length}:
\begin{equation}
    l_c \approx \frac{2\pi}{m_\phi v_\phi} \approx 0.4\text{kpc} \left( \frac{10^{-22}\text{eV}}{m_\phi} \right).
    \label{eq:co_length}
\end{equation}
In particular, notice that $\hat{\phi}^2_E$ and $\hat{\phi}^2_{P}$ should be taken as:
\begin{itemize}
    \item \textit{different} parameters when the average pulsar-Earth and pulsar-pulsar distance is larger than the coherence length. 
    \item the \textit {same} parameter  when the average pulsar-Earth and pulsar-pulsar distance is smaller than the coherence length.
\end{itemize}
Following the procedure in \cite{Smarra_2023}, we analyze three separate cases, which we refer to as the \textit{uncorrelated}, the \textit{pulsar correlated} and the \textit{correlated} limit. 
As the average inter-pulsar and Earth-pulsar separation is $\sim$ kpc, 
the \textit{correlated} and \textit{uncorrelated} scenarios stand out as exact limits at the low mass and high mass end of the PTAs band, respectively. Instead, the \textit{pulsar correlated} limit holds when the coherence length of 
ULDM is smaller than the Galacto-centric radius probed by rotation curves (inner $\sim 20$ kpc), but larger than the average inter-pulsar and pulsar-Earth distance. More specifically, the \textit{correlated} regime holds for masses lower than $m_\phi \sim 2 \times 10^{-24}$ eV; the \textit{pulsar correlated} regime for $ 2 \times 10^{-24}\, \text{eV} \lesssim m_\phi \lesssim 5 \times 10^{-23}\, \text{eV} $ and the \textit{uncorrelated} limit for $m_\phi \gtrsim 5\times 10^{-23}~\text{eV}$.
We defer a more detailed study to future analysis.
\begin{table*}[ht]
\renewcommand{\arraystretch}{1.2}
\centering
\caption{Parameters used for the search and their respective priors. In the correlated limit, $\hat{\phi}^2_E = \hat{\phi}^2_P$ and can be reabsorbed in a redefinition of $\Psi_c$.}
\begin{tabular}{|c|c|c|c|}
\hline  \textbf{Parameter} & \textbf{Description} & \textbf{Prior} & \textbf{Occurrence} \\ \hline
\hline \multicolumn{4}{|c|}{ White Noise $\left(\sigma = E_f^2 \sigma^2_{ToA} + E_q^2\right)$} \\ \hline
\hline$E_f$ & EFAC per backend/receiver system & Uniform $[0,10]$ & 1 per pulsar \\
\hline$E_q$ & EQUAD per backend/receiver system & Log$_{10}$-Uniform $[-10,-5]$ & 1 per pulsar \\  \hline
\hline \multicolumn{4}{|c|}{ Red Noise } \\  \hline
\hline$A_{\text {red }}$ & Red noise power-law amplitude & Log$_{10}$-Uniform $[-20,-6]$ & 1 per pulsar \\
\hline$\gamma_{\text {red }}$ & Red noise power-law spectral index & Uniform $[0,10]$ & 1 per pulsar \\  \hline
\hline \multicolumn{4}{|c|}{ ULDM } \\  \hline
 \hline$\Psi_\text{c}$ & ULDM amplitude & Log$_{10}$-Uniform $[-20,-12]$ & 1 for PTA \\
\hline$m_\phi~[\mathrm{eV}]$ & ULDM mass & Log$_{10}$-Uniform $[-24,-22]$ & 1 for PTA \\
\hline$\hat{\phi}_E^2$ & Earth factor & $e^{-x}$ & 1 for PTA \\
\hline$\hat{\phi}_P^2$ & Pulsar factor & $e^{-x}$ & 1 per pulsar \\
\hline$\gamma_E$ & Earth signal phase & Uniform $[0,2 \pi]$ & 1 per PTA \\
\hline$\gamma_P$ & Pulsar signal phase & Uniform $[0,2 \pi]$ & 1 per pulsar \\  \hline
\hline \multicolumn{4}{|c|}{ Common spatially Uncorrelated Red Noise (CURN) } \\  \hline
\hline$A_{\mathrm{GWB}}$ & Common process strain amplitude & Log$_{10}$-Uniform $[-20,-6]$ & 1 for PTA \\
\hline
\end{tabular}
\label{tab:ULDM_priors}
\end{table*}

\begin{figure}[ht]
     \includegraphics[width=0.45\linewidth]{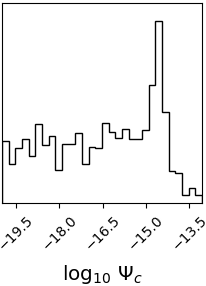}
     \includegraphics[width=0.45\linewidth]{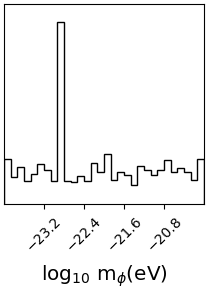}
     \\
     \includegraphics[width=0.45\linewidth]{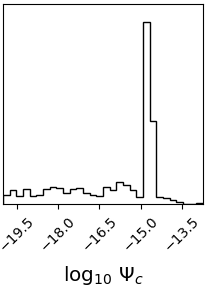}
     \includegraphics[width=0.45\linewidth]{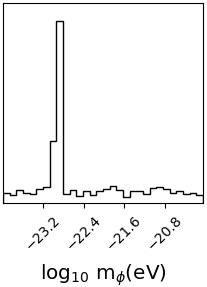}
  \caption{\footnotesize{Posterior probabilities for the ULDM amplitude $\Psi_c$ and mass $m_\phi$, from the correlated (top row) and uncorrelated (bottom row) analysis of the \texttt{DR2new} dataset. The pulsar correlated analysis is not shown, but displays the same features.}}     
    \label{fig:10yrULDM}
\end{figure}
Based on the above, we fit the model from Eq.~\eqref {eq:st} to the \texttt{DR2new} as in~\cite{Smarra_2023}. 
As an example, marginalised posterior distributions for $\Psi_c$ and $m_\phi$ are plotted in Fig.~\ref{fig:10yrULDM}. 
All the three limits peak at a ULDM mass $m_\phi \sim 10^{-23}\text{eV}$ with amplitude $\Psi_c \sim 10^{-15}$, which translates into a density $\rho_{\phi} \simeq 0.6\, \text{GeV}/\text{cm}^3$ of the scalar field. While this value is higher than the fiducial local DM density $\rho_{\text{DM}} \approx 0.4  \text{GeV}/\text{cm}^3$, it is well within the measurement uncertainties \citep{Bovy_2012,Read_2014,Sivertsson_2018, de_Salas_2020}.

Additionally, we search for a potential ULDM signature alongside the SMBHB gravitational-wave background. 
Thus, we introduce a new model that contains ULDM contributions alongside a common red signal to account for gravitational wave contributions.
We find no ULDM signal under this hypothesis, in agreement with the fact that the data support HD correlation, which naturally favours GWs over ULDM in a joint search. Thus we put 95\% upper limit on the amplitude $\Psi_c$ and the density $\rho_{\phi}$ of the scalar field.

\begin{figure}[h!]
    \centering
    \includegraphics[width=0.50\textwidth]{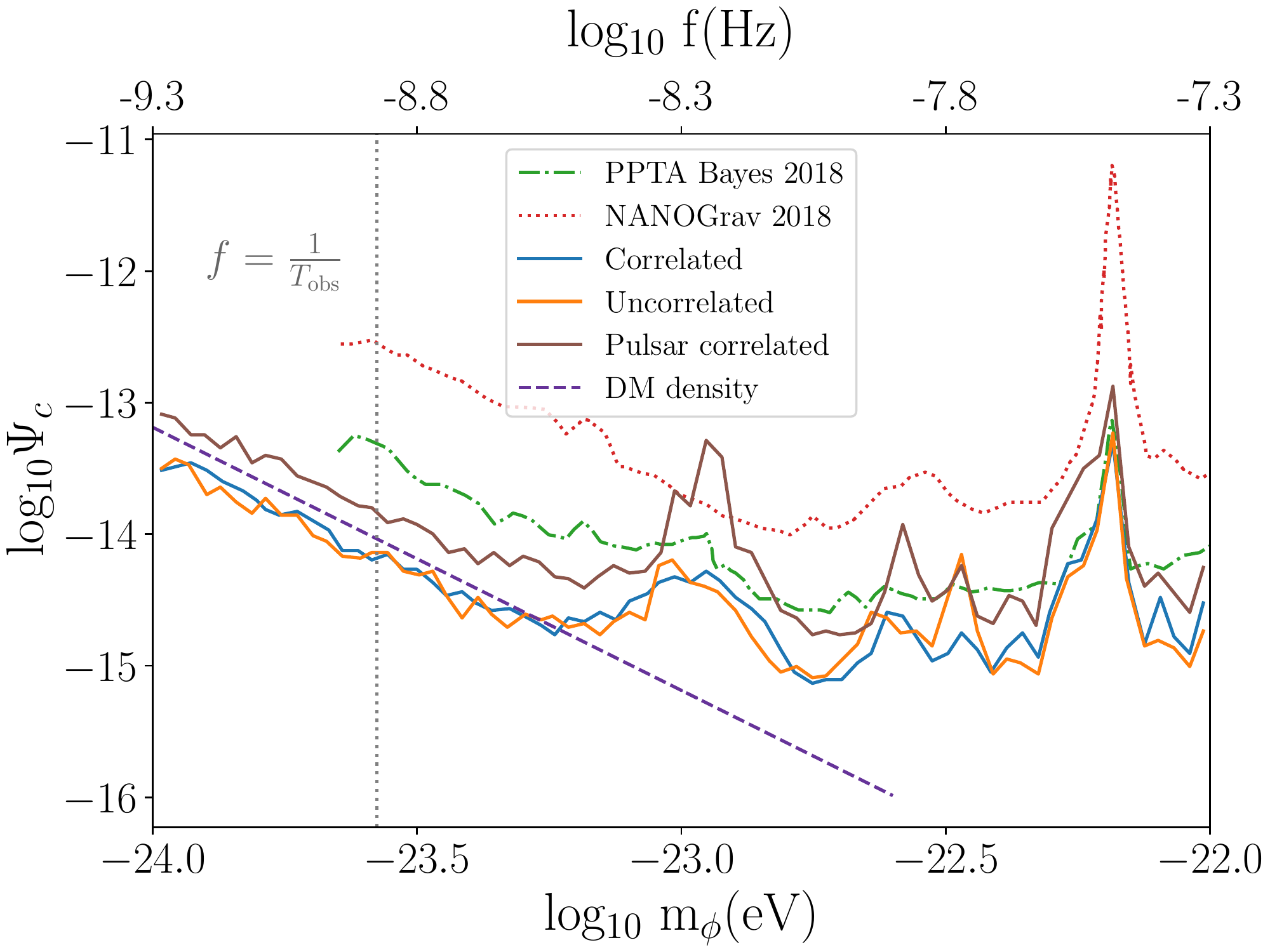}
    \caption{\footnotesize{Constraints on $\Psi_c$ as a function of $m_\phi$ using the EPTA \texttt{DR2new} dataset from PaperIII. Previous analyses are shown for comparison, cf. \cite{Porayko_2014, Porayko_2018} for further details. The blue, orange and brown lines represent the 95\% Bayesian upper limit on $\Psi_c$ obtained from the EPTA \texttt{DR2new} dataset with the correlated, uncorrelated and pulsar correlated analysis, respectively. The purple line shows the expected ULDM abundance computed from Eq.~\eqref{eq:psi_c}.}}
    \label{fig:24yrULDM}
\end{figure}

\begin{figure}[h!]
    \centering
    \includegraphics[width=0.5\textwidth]{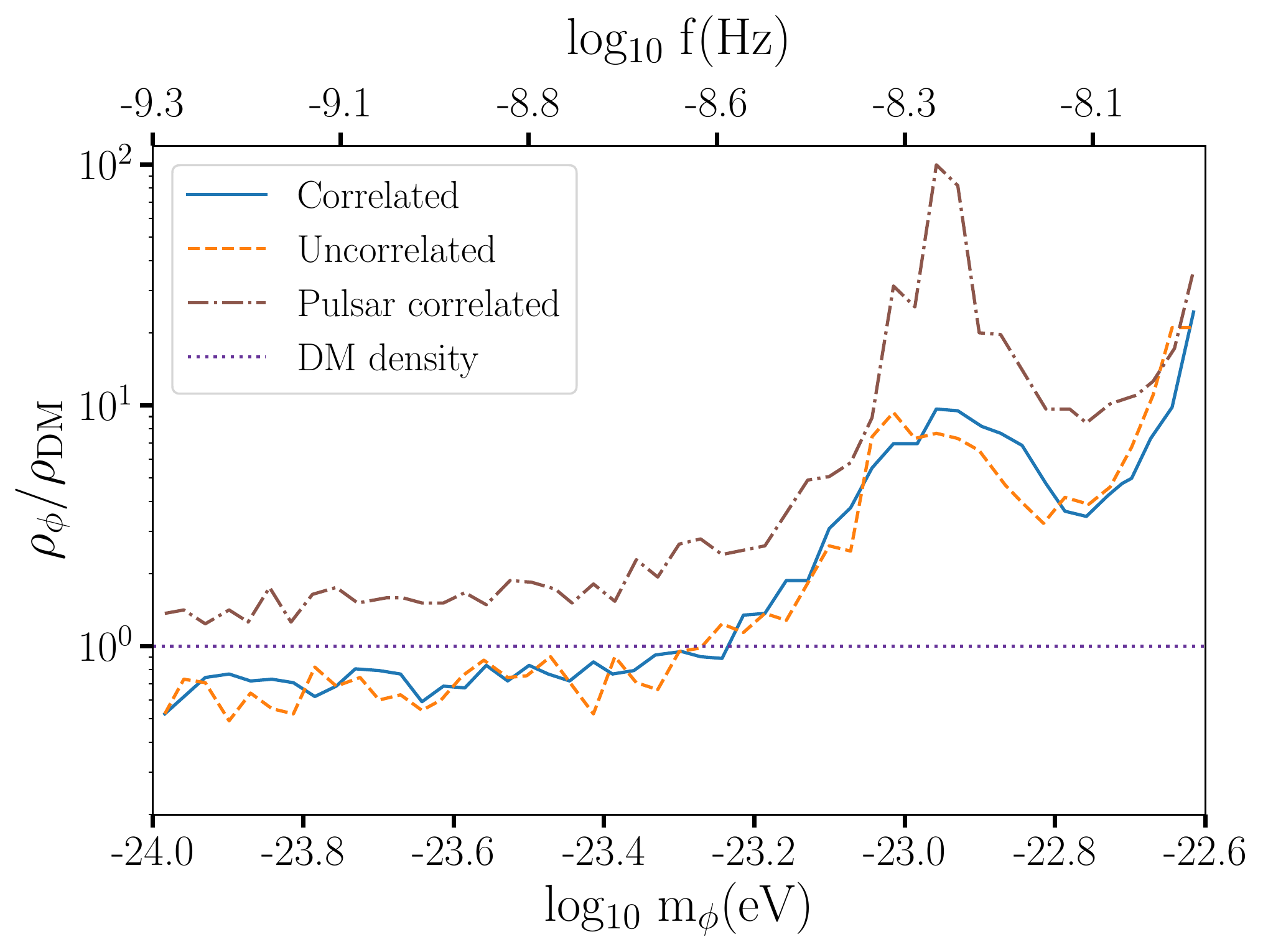}
    \caption{\footnotesize{Constraints on the ULDM density $\rho_{\phi}$ normalised to the DM background value $\rho_{\text{DM}} = 0.4 \text{GeV}/ \text{cm}^3$. The blue, orange and brown lines represent the 95\% Bayesian upper limit on $\rho_{\phi}$, obtained from the EPTA \texttt{DR2new} dataset with the correlated, uncorrelated and pulsar correlated analysis, respectively. The purple dotted line shows the fiducial local DM density value.}}
    \label{fig:24yrULDM_rho}
\end{figure}

Figs.~\ref{fig:24yrULDM} and \ref{fig:24yrULDM_rho} show that the EPTA at current sensitivity is able to constrain the presence of ULDM at the level of the expected DM abundance in the mass range $m_\phi \sim [10^{-24}\text{eV}, 10^{-23.2}\text{eV}]$. The \texttt{DR2Full} analysis performed in \cite{Smarra_2023} pushes these limits down, thus implying that such ULDM candidates cannot constitute the entire DM. 
We highlight that our bounds extend below the \texttt{DR2new} sampling frequency  $f = 1/T \approx 3\cdot10^{-9}\text{Hz}$, with $T\approx \text{10yr}$.
In fact, while it might naively be thought that the ULDM field needs to complete an oscillation in the timescale $\tau_{\text{PTA}}$ of the PTA experiment to produce a detectable effect, we point out that an ULDM wave with frequency $m_\phi < 1/\tau_{\text{PTA}}$ can still be approximated by an expansion in powers of  $m_\phi t$ \citep{Kaplan_2022}, though the sensitivity of PTAs will be reduced due to degeneracies with the timing model \citep{Ramani_2020}. Relying on the robust CMB bounds mentioned before, we fix the lower end of our search at $m_\phi = 10^{-24}\text{eV}$. 
Importantly, we remove the $\hat\phi_E = \hat\phi_P$ parameter from the search in the correlated limit, as it is fully degenerate with $\Psi_c$. In other words, building upon the observation that our Galaxy rotation curve measurements are performed within an ULDM coherence length in the correlated limit, we redefine a new variable, $\Psi_c^0 =  \Psi_c \phi^2$, which represents the instance of DM in the Milky Way. 
Finally, as shown by Fig. \ref{fig:24yrULDM}, we report a bump in upper limits at $m_\phi \sim 10^{-23}\text{eV}$, which is at around the maximum-a-posteriori boson mass in Fig.~\ref{fig:10yrULDM}. 
This mass further corresponds to the frequency of the CGW analyzed in \cite{wm4}. Looking at the posteriors in this mass range, we also find an additional contribution, on top of the CURN process.
A similar bump in upper limits is also present in the \texttt{DR2Full} dataset, as discussed in \cite{Smarra_2023}. 
However, the Bayes factor of ln$\mathcal{B} \sim 0.3$ we find in favour of the presence of ULDM signal is still inconclusive. 
We recommend following up on this bump in future work.
\section{Discussion and outlook}
\label{sec:conclusions}

In this paper, we have explored the implications of the common, correlated low-frequency signal observed in the latest data release of the EPTA+InPTA collaboration. Four different datasets were assembled, and the signal was more significant in those including only broadband, high-quality data taken with telescope backends of the new generation. Therefore, we took, as benchmark for our analysis, the signal measured in the \texttt{DR2new} dataset, for which the HD correlation is detected at high significance with a Bayes factor of $\approx 60$ or, equivalently, at a $p-$value of $\approx 0.001$, indicative of a $\gtrsim3\sigma$ confidence. The signal can be modelled by a single power law spectrum $S_h(f)$ as in Eq.~\eqref{eq:sh}, with best-fit parameters 
$A_{f=1{\rm yr}^{-1}}=-13.94_{-0.48}^{+0.23}$ and  $\gamma=2.71_{-0.71}^{+1.18}$ (PaperIII). 

We considered several physical processes separately, investigating the implications of the detected signal under the hypothesis that it is generated by that specific process. Our main findings can be summarised as follows.

{\bf SMBHBs}. The signal is consistent with a cosmic population of merging SMBHBs. Phenomenological models based on galaxy pairs observations can account for the power spectral distribution of the correlated signal. Those models can also be used to predict the chance to detect CGWs in the current data, for which the search has given inconclusive results so far \citep{wm4}. According to those models, there is roughly a 50\% chance to detect a CGW in \texttt{DR2new} at S/N$>3$. The relatively high amplitude of the signal can be used to place constraints on key properties of the cosmic SMBHB population. By exploiting the inference framework developed in \cite{2016MNRAS.455L..72M,csc2019}, we can infer that SMBHBs merge in less than 1Gyr following galaxy mergers, and that the SMBH-stellar bulge relation has a normalization ${\rm log}_{10}(M_*/{\rm M}_\odot)\approx8.4$ at a reference stellar mass bulge of $10^{11}$M$_\odot$, in line with recent compiled results \citep[e.g.][]{2013ARA&A..51..511K}. Finally, we investigated how the detected signal compares with predictions from state-of-the-art galaxy formation SAMs \citep{B12,2022MNRAS.509.3488I}. We found that the high amplitude of the signal imposes non-trivial constraints on galaxy and SMBH evolution models. In particular, boosted SMBH accretion and short SMBHB evolution timescales are needed to match the observed signal, which might lead to difficulties when trying to reproduce independent observables such as the quasar luminosity function. If the SMBHB origin of the observed signal is confirmed, it would be a major breakthrough for observational astrophysics and for our understanding of galaxy formation. This would be the first direct evidence that SMBHBs merge in nature, adding an important observational piece to the puzzle of structure formation and galaxy evolution.

{\bf Early Universe.} The measured signal would have different implications for each of the physical processes considered in this study. An inflationary origin requires non-standard inflationary scenarios breaking the slow-roll consistency relation, leading to a blue-tilted spectrum. In particular, the measured $\gamma\approx 2.7$ implies $n_T\approx 2.3$ for a radiation-dominated universe with $w=1/3$. 
A cosmic string origin would allow narrowing down the string tension to values of $-11\lesssim{\rm log}_{10}G\mu\lesssim-9.5$, depending on the specific distribution of loops in the string network. Conversely, the number of kinks cannot be constrained. 
A GWB induced by (M)HD turbulence at the QCD energy scale can also potentially explain the common red noise, but requires either high turbulent energy densities, of the same order of the radiation energy density, or a characteristic turbulent scale close to the horizon at the QCD epoch. 
Finally, the measured signal can be produced by the evolution of scalar perturbations at second order only if an excess of their primordial spectrum is present at large wavenumbers, compared to the level derived from CMB observations at small wave numbers. Notably, such an excess would lead to the production of PBHs which can non-negligibly contribute to the CDM density.

{\bf ULDM.} Finally, we searched for ULDM signatures in \texttt{DR2new}. The search returned a prominent peak in the posterior distribution of the ULDM particle mass around ${\rm log}_{10}(m_\phi/{\rm eV})\approx -23$. This corresponds to a field oscillation frequency of ${\rm log}_{10}(f/{\rm Hz})\approx -8.3$, which is consistent with the CGW candidate examined in \cite{wm4}. When a joint ULDM+GWB search is performed, however, the peak in the ULDM mass posterior distribution vanishes, as the data strongly prefer the presence of an HD correlation in the common power. We therefore conclude that an ULDM origin of the detected signal is disfavoured, placing a direct constraint on the abundance of ULDM in our Galaxy. The non-detection in \texttt{DR2new} implies that only about 80\% of the DM density in the solar neighbourhood can be attributed to ULDM with $-24<{\rm log}_{10}(m_\phi/{\rm eV})<-23.7$. More stringent constraints are obtained from \texttt{DR2full} and are presented in a separate paper \citep{Smarra_2023}.

It is interesting to remark that the best fit to the measured power-law slope is $\gamma=2.71$, well below the predicted $\gamma=13/3$ expected from a cosmic population of SMBHBs, often indicated as the primary candidate for generating nanohertz GWs. Before leaving the floor to speculations, we caution that this `inconsistency' might have multiple roots, as mentioned in Sec.~\ref{sec:considerations}. First, $\gamma=13/3$ assumes a population of circular GW-driven binaries. Eccentricity and environmental coupling can easily flatten the low-frequency spectrum. Even for circular binaries, a simple power-law of $\gamma=13/3$ is an ideal limit; small number statistics due to the sparseness of massive, nearby systems results in noisy spectra with a large variance, which can produce different spectral indices when fitted by a power-law. Second, the measured value can be subject to statistical and systematic biases. We have in fact shown that even when a power law with $\gamma=13/3$ is injected into the data, the measured values can be biased due to the statistical realization of the noise. Moreover, the mis-modeling of high-frequency noise can systematically bias the recovered $\gamma$, leading to flatter spectra if unaccounted high-frequency noise is present in the data. Therefore, caution should be taken when drawing conclusions from this measurement.

Conversely, a shallow spectral slope might indicate that the GW signal has a different origin, perhaps from some key physical process occurring in the early Universe, as investigated in Section \ref{sec:early_universe}. Another possibility, proposed by \cite{lackeos} is that GWs can lose energy while propagating through the intergalactic medium via the acceleration of charged particles. Owing to the much higher thermal velocity of electrons over protons and the presence of an intergalactic medium magnetic field, a significant fraction of the GWs can be converted to EM radiation. The ensuing electromagnetic intensity is inversely proportional to frequency. Partial GW loss via this mechanism could explain a flatter-than-expected GWB spectrum. The efficiency of such a mechanism depends on the strength of the intergalactic medium magnetic field.

It is also possible that the observed signal is coming from multiple overlapping GWBs. In our analyses, we have assumed a single origin for all the observed signal power in the EPTA+InPTA data, however, an overlap of GWBs of different origins can cause the spectral shape of the recovered signal to deviate from the expected value of any single GWB. Searching, disentangling and identifying the underlying physical processes will be part of the spectral characterisation moving forward \citep{2021NatAs...5.1268M,2022ApJ...938..115K}.

As time goes by and the PTA experiment improves, the low-frequency GWB will leave an increasingly distinctive signature in the data. Interpreting all of the details of this signature will be necessary to understand the nature of the signal and exploit the full potential of this new window into the Universe.
As mentioned in the introduction, a population of SMBHBs is expected to generate a highly non-Gaussian, partially anisotropic and perhaps non-stationary signal. As we increase the timespan of our data, improve our instrumentation and combine more pulsars, the detailed properties of the signal will eventually reveal themselves in the data 
\citep[e.g.][]{2013CQGra..30v4005C,2013PhRvD..88h4001T,2020PhRvD.102h4039T,2022ApJ...940..173P}. While many early Universe signals are expected to be isotropic and Gaussian, noticeable exceptions exist, such as GWBs from bursts of cosmic strings. Cross-correlating the power distribution of the nanohertz GW sky to the distribution of massive galaxies and large-scale structures in the low-$z$ universe can eventually provide the key to 
determining the true nature -- astrophysical vs. early Universe -- of this signal \citep{2014MNRAS.439.3986R,2017NatAs...1..886M}. In this respect, combining all of the available high-quality datasets within the IPTA framework is the next step towards the fulfilment of the promises of nanohertz GW science.




\begin{acknowledgements}
The European Pulsar Timing Array (EPTA) is a collaboration between
European and partner institutes, namely ASTRON (NL), INAF/Osservatorio
di Cagliari (IT), Max-Planck-Institut f\"{u}r Radioastronomie (GER),
Nan\c{c}ay/Paris Observatory (FRA), the University of Manchester (UK),
the University of Birmingham (UK), the University of East Anglia (UK),
the University of Bielefeld (GER), the University of Paris (FRA), the
University of Milan-Bicocca (IT), the Foundation for Research and 
Technology, Hellas (GR), and Peking University (CHN), with the
aim to provide high-precision pulsar timing to work towards the direct
detection of low-frequency gravitational waves. An Advanced Grant of
the European Research Council allowed to implement the Large European Array
for Pulsars (LEAP) under Grant Agreement Number 227947 (PI M. Kramer). 
The EPTA is part of the
International Pulsar Timing Array (IPTA); we thank our
IPTA colleagues for their support and help with this paper and the external Detection Committee members for their work on the Detection Checklist.

Part of this work is based on observations with the 100-m telescope of
the Max-Planck-Institut f\"{u}r Radioastronomie (MPIfR) at Effelsberg
in Germany. Pulsar research at the Jodrell Bank Centre for
Astrophysics and the observations using the Lovell Telescope are
supported by a Consolidated Grant (ST/T000414/1) from the UK's Science
and Technology Facilities Council (STFC). ICN is also supported by the
STFC doctoral training grant ST/T506291/1. The Nan{\c c}ay radio
Observatory is operated by the Paris Observatory, associated with the
French Centre National de la Recherche Scientifique (CNRS), and
partially supported by the Region Centre in France. We acknowledge
financial support from ``Programme National de Cosmologie and
Galaxies'' (PNCG), and ``Programme National Hautes Energies'' (PNHE)
funded by CNRS/INSU-IN2P3-INP, CEA and CNES, France. We acknowledge
financial support from Agence Nationale de la Recherche
(ANR-18-CE31-0015), France. The Westerbork Synthesis Radio Telescope
is operated by the Netherlands Institute for Radio Astronomy (ASTRON)
with support from the Netherlands Foundation for Scientific Research
(NWO). The Sardinia Radio Telescope (SRT) is funded by the Department
of University and Research (MIUR), the Italian Space Agency (ASI), and
the Autonomous Region of Sardinia (RAS) and is operated as a National
Facility by the National Institute for Astrophysics (INAF).

The work is supported by the National SKA programme of China
(2020SKA0120100), Max-Planck Partner Group, NSFC 11690024, CAS
Cultivation Project for FAST Scientific. This work is also supported
as part of the ``LEGACY'' MPG-CAS collaboration on low-frequency
gravitational wave astronomy. JA acknowledges support from the
European Commission (Grant Agreement number: 101094354). JA and SCha 
were partially supported by the Stavros
Niarchos Foundation (SNF) and the Hellenic Foundation for Research and
Innovation (H.F.R.I.) under the 2nd Call of the ``Science and Society --
Action Always strive for excellence -- Theodoros Papazoglou''
(Project Number: 01431). AC acknowledges support from the Paris
\^{I}le-de-France Region. AC, AF, ASe, ASa, EB, DI, GMS, MBo acknowledge
financial support provided under the European Union's H2020 ERC
Consolidator Grant ``Binary Massive Black Hole Astrophysics'' (B
Massive, Grant Agreement: 818691). GD, KLi, RK and MK acknowledge support
from European Research Council (ERC) Synergy Grant ``BlackHoleCam'', 
Grant Agreement Number 610058. This work is supported by the ERC 
Advanced Grant ``LEAP'', Grant Agreement Number 227947 (PI M. Kramer). 
AV and PRB are supported by the UK's Science
and Technology Facilities Council (STFC; grant ST/W000946/1). AV also acknowledges
the support of the Royal Society and Wolfson Foundation. JPWV acknowledges
support by the Deutsche Forschungsgemeinschaft (DFG) through thew
Heisenberg programme (Project No. 433075039) and by the NSF through
AccelNet award \#2114721. NKP is funded by the Deutsche
Forschungsgemeinschaft (DFG, German Research Foundation) --
Projektnummer PO 2758/1--1, through the Walter--Benjamin
programme. ASa thanks the Alexander von Humboldt foundation in
Germany for a Humboldt fellowship for postdoctoral researchers. APo, DP
and MBu acknowledge support from the research grant “iPeska”
(P.I. Andrea Possenti) funded under the INAF national call
Prin-SKA/CTA approved with the Presidential Decree 70/2016
(Italy). RNC acknowledges financial support from the Special Account
for Research Funds of the Hellenic Open University (ELKE-HOU) under
the research programme ``GRAVPUL'' (grant agreement 319/10-10-2022).
EvdW, CGB and GHJ acknowledge support from the Dutch National Science
Agenda, NWA Startimpuls – 400.17.608.
BG is supported by the Italian Ministry of Education, University and 
Research within the PRIN 2017 Research Program Framework, n. 2017SYRTCN. LS acknowledges the use of the HPC system Cobra at the Max Planck Computing and Data Facility.

\ifnum\wm>1 The Indian Pulsar Timing Array (InPTA) is an Indo-Japanese
collaboration that routinely employs TIFR's upgraded Giant Metrewave
Radio Telescope for monitoring a set of IPTA pulsars.  BCJ, YG, YM,
SD, AG and PR acknowledge the support of the Department of Atomic
Energy, Government of India, under Project Identification \# RTI 4002.
BCJ, YG and YM acknowledge support of the Department of Atomic Energy,
Government of India, under project No. 12-R\&D-TFR-5.02-0700 while SD,
AG and PR acknowledge support of the Department of Atomic Energy,
Government of India, under project no. 12-R\&D-TFR-5.02-0200.  KT is
partially supported by JSPS KAKENHI Grant Numbers 20H00180, 21H01130,
and 21H04467, Bilateral Joint Research Projects of JSPS, and the ISM
Cooperative Research Program (2021-ISMCRP-2017). AS is supported by
the NANOGrav NSF Physics Frontiers Center (awards \#1430284 and
2020265).  AKP is supported by CSIR fellowship Grant number
09/0079(15784)/2022-EMR-I.  SH is supported by JSPS KAKENHI Grant
Number 20J20509.  KN is supported by the Birla Institute of Technology
\& Science Institute fellowship.  AmS is supported by CSIR fellowship
Grant number 09/1001(12656)/2021-EMR-I and T-641 (DST-ICPS).  TK is
partially supported by the JSPS Overseas Challenge Program for Young
Researchers.  We acknowledge the National Supercomputing Mission (NSM)
for providing computing resources of ‘PARAM Ganga’ at the Indian
Institute of Technology Roorkee as well as `PARAM Seva' at IIT
Hyderabad, which is implemented by C-DAC and supported by the Ministry
of Electronics and Information Technology (MeitY) and Department of
Science and Technology (DST), Government of India. DD acknowledges the 
support from the Department of Atomic Energy, Government of India 
through Apex Project - Advance Research and Education in Mathematical 
Sciences at IMSc. \fi

The work presented here is a culmination of many years of data
analysis as well as software and instrument development. In particular,
we thank Drs. N.~D'Amico, P.~C.~C.~Freire, R.~van Haasteren, 
C.~Jordan, K.~Lazaridis, P.~Lazarus, L.~Lentati, O.~L\"{o}hmer and 
R.~Smits for their past contributions. We also
thank Dr. N. Wex for supporting the calculations of the
galactic acceleration as well as the related discussions.
\ifnum\wm=5 We would like to thank Prof. Drs. Alexey Starobinskiy, Sergei Blinnikov and Alexander Dolgov for discussions on the early Universe physics. \fi
\ifnum\wm=5 HM acknowledges the support of the UK Space Agency, Grant No. ST/V002813/1 and ST/X002071/1. Some of the computations described in this paper were performed using the University of Birmingham's BlueBEAR HPC service, which provides a High Performance Computing service to the University's research community. See~\url{http://www.birmingham.ac.uk/bear} for more details. \fi
The EPTA is also grateful
to staff at its observatories and telescopes who have made the
continued observations possible. 
\linebreak\linebreak\textit{Author contributions.}
The EPTA is a multi-decade effort and all authors have
contributed through conceptualisation, funding acquisition,
data-curation, methodology, software and hardware
 developments as well as (aspects of) the continued running of
the observational campaigns, which includes writing and
proofreading observing proposals, evaluating observations
and observing systems, mentoring students, developing
science cases. All authors also helped in (aspects of)
verification of the data, analysis and results as well as
in finalising the paper draft. Specific contributions from individual 
EPTA members are listed in the CRediT\footnote{\url{https://credit.niso.org/}} format below.

InPTA members contributed in uGMRT observations and data reduction to
create the InPTA data set which is employed while assembling the
\texttt{DR2full+} and \texttt{DR2new+} data sets. 

\ifnum\wm=1

JJan, KLi, GMS equally share the correspondence of the paper.

\linebreak\linebreak\textit{CRediT statement:}\newline
Conceptualisation: APa, APo, AV, BWS, CGB, CT, GHJ, GMS, GT, IC, JA, JJan, JPWV, JW, JWM, KJL, KLi, MK.\\
Methodology: APa, AV, DJC, GMS, IC, JA, JJan, JPWV, JWM, KJL, KLi, LG, MK.\\
Software: AC, AJ, APa, CGB, DJC, GMS, IC, JA, JJan, JJaw, JPWV, KJL, KLi, LG, MJK, RK.\\
Validation: AC, APa, CGB, CT, GMS, GT, IC, JA, JJan, JPWV, JWM, KLi, LG.\\
Formal Analysis: APa, CGB, DJC, DP, EvdW, GHJ, GMS, JA, JJan, JPWV, JWM, KLi.\\
Investigation: APa, APo, BWS, CGB, DJC, DP, GMS, GT, IC, JA, JJan, JPWV, JWM, KLi, LG, MBM, MBu, MJK, RK.\\
Resources: APa, APe, APo, BWS, GHJ, GMS, GT, HH, IC, JA, JJan, JPWV, JWM, KJL, KLi, LG, MJK, MK, RK.\\
Data Curation: AC, AJ, APa, BWS, CGB, DJC, DP, EG, EvdW, GHJ, GMS, GT, HH, IC, JA, JJan, JPWV, JWM, KLi, LG, MBM, MBu, MJK, MK, NKP, RK, SChe, YJG.\\
Writing – Original Draft: APa, GMS, JA, JJan, KLi, LG.\\
Writing – Review \& Editing: AC, AF, APa, APo, DJC, EB, EFK, GHJ, GMS, GT, JA, JJan, JPWV, JWM, KLi, MK, SChe, VVK.\\
Visualisation: APa, GMS, JA, JJan, KLi.\\
Supervision: APo, ASe, AV, BWS, CGB, DJC, EFK, GHJ, GMS, GT, IC, JA, JPWV, KJL, KLi, LG, MJK, MK, VVK.\\
Project Administration: APo, ASe, AV, BWS, CGB, CT, GHJ, GMS, GT, IC, JJan, JPWV, JWM, KLi, LG, MK.\\
Funding Acquisition: APe, APo, ASe, BWS, GHJ, GT, IC, JA, JJan, LG, MJK, MK.\\

\fi

\ifnum\wm=2
InPTA members contributed to the discussions that probed the impact of 
including InPTA data on single pulsar noise analysis. Furthermore, they 
provided quantitative comparisons of various noise models, wrote a brief 
description of the underlying \texttt{Tensiometer} package, and helped 
with the related interpretations.

APa, AC, MJK equally share the correspondence of the paper. 

\linebreak\linebreak\textit{CRediT statement:}\newline
Conceptualisation: AC, APa, APo, AV, BWS, CT, GMS, GT, JPWV, JWM, KJL, KLi, MJK, MK.\\
Methodology: AC, APa, AV, DJC, GMS, IC, JWM, KJL, KLi, LG, MJK, MK, SB, SChe, VVK.\\
Software: AC, AJ, APa, APe, GD, GMS, KJL, KLi, MJK, RK, SChe, VVK.\\
Validation: AC, APa, BG, GMS, IC, JPWV, JWM, KLi, LG, MJK.\\
Formal Analysis: AC, APa, BG, EvdW, GHJ, GMS, JWM, KLi, MJK.\\
Investigation: AC, APa, APo, BWS, CGB, DJC, DP, GMS, IC, JPWV, JWM, KLi, LG, MBM, MBu, MJK, RK, VVK.\\
Resources: AC, APa, APe, APo, BWS, GHJ, GMS, GT, IC, JPWV, JWM, KJL, KLi, LG, MJK, MK, RK.\\
Data Curation: AC, AJ, APa, BWS, CGB, DJC, DP, EvdW, GHJ, GMS, JA, JWM, KLi, MBM, MJK, MK, NKP, RK, SChe.\\
Writing – Original Draft: AC, APa, GMS, MJK.\\
Writing – Review \& Editing: AC, AF, APa, APo, BG, EB, EFK, GMS, GT, JA, JPWV, JWM, KLi, MJK, MK, SChe, VVK.\\
Visualisation: AC, APa, GMS, KLi, MJK.\\
Supervision: AC, APo, ASe, AV, BWS, CGB, DJC, EFK, GHJ, GT, JPWV, KJL, LG, MJK, MK, VVK.\\
Project Administration: AC, APo, ASe, AV, BWS, CGB, CT, GHJ, GMS, GT, JPWV, JWM, LG, MJK, MK.\\
Funding Acquisition: APe, APo, ASe, BWS, GHJ, GT, IC, LG, MJK, MK.\\
\fi

\ifnum\wm=3

Additionally, InPTA members contributed to GWB search efforts with 
\texttt{DR2full+} and \texttt{DR2new+} data sets and their interpretations. 
Further, they provided quantitative comparisons of GWB posteriors that 
arise from these data sets and multiple pipelines.

For this work specifically, SChen and YJG equally share the 
correspondence of the paper. 

\linebreak\linebreak\textit{CRediT statement:}\newline
Conceptualisation: AC, APa, APe, APo, ASe, AV, BG, CT, GMS, GT, IC, JA, JPWV, JWM, KJL, KLi, MK.\\
Methodology: AC, APa, ASe, AV, DJC, GMS, JWM, KJL, KLi, LS, MK, SChe.\\
Software: AC, AJ, APa, APe, GD, GMS, KJL, KLi, MJK, RK, SChe, VVK.\\
Validation: AC, APa, ASe, AV, BG, GMS, HQL, JPWV, JWM, KLi, LS, SChe, YJG.\\
Formal Analysis: AC, APa, ASe, AV, BG, EvdW, GMS, HQL, JWM, KLi, LS, MF, NKP, PRB, SChe, YJG.\\
Investigation: APa, APo, ASe, AV, BWS, CGB, DJC, DP, GMS, JWM, KLi, LS, MBM, MBu, MF, PRB, RK, SB, SChe, YJG.\\
Resources: AC, APa, APe, APo, ASe, AV, BWS, GHJ, GMS, GT, IC, JPWV, JWM, KJL, KLi, LG, LS, MJK, MK, RK.\\
Data Curation: AC, AJ, APa, BWS, CGB, DJC, DP, EvdW, GMS, JA, JWM, KLi, MBM, MJK, MK, RK, SChe.\\
Writing – Original Draft: AC, APa, BG, DJC, GMS, JA, KLi, SB, SChe, YJG.\\
Writing – Review \& Editing: AC, AF, APa, APo, ASe, AV, BG, DJC, EB, EFK, GMS, GT, JA, JPWV, JWM, KLi, LS, MBo, MK, NKP, PRB, SChe, VVK, YJG.\\
Visualisation: APa, BG, GMS, KLi, MF, PRB, SChe.\\
Supervision: APo, ASe, AV, BWS, CGB, DJC, EFK, GHJ, GMS, GT, JPWV, KJL, MK, SB.\\
Project Administration: APo, ASe, AV, BWS, CGB, CT, GHJ, GMS, GT, JPWV, JWM, LG, MK, SChe.\\
Funding Acquisition: APe, APo, ASe, AV, BWS, GHJ, GT, IC, JA, LG, MJK, MK, SB.\\
\fi

\end{acknowledgements}

%
%

\bibpunct{(}{)}{;}{a}{}{,}
\def\aap{A\&A}                
\def\aapr{A\&A~Rev.}          
\def\aaps{A\&AS}              
\def\aj{AJ}                   
\def\ajph{Australian J.~Phys.}
\def\alet{Astro.~Lett.}       
\def\ao{Applied Optics}       
\def\apj{ApJ}                 
\def\apjl{ApJ}                
\def\apjs{ApJS}              
\def\apss{Ap\&SS}             
\def\araa{ARA\&A}             
\def\asr{Av.~Space Res.}     
\def\azh{AZh}                 
\def\baas{BAAS}               
\def\cpc{Comput.~Phys.~Commun.} 
\def\gca{Geochim.~Cosmochim.~Acta} 
\def\iaucirc{IAU Circ.}       
\def\ibvs{IBVS}               
\def\icarus{Icarus}           
\def\jcomph{J.~Comput.~Phys.} 
\def\jcp{J.~Chem.~Phys.}      
\def\jgr{J.~Geophys.~R.}      
\def\jrasc{JRASC}             
\def\met{Meteoritics}         
\def\mmras{MmRAS}             
\def\mnras{MNRAS}             
\def\mps{Meteoritics and Planetary Science} 
\def\nast{New Astron.}        
\def\nat{Nature}              
\def\pasj{PASJ}               
\def\pasp{PASP}               
\def\phr{Phys.~Rev.}          
\def\pdra{Phys.~Rev.~A}       
\def\prb{Phys.~Rev.~B}       
\def\prc{Phys.~Rev.~C}       
\def\prd{Phys.~Rev.~D}       
\def\phrep{Phys.~Rep.}        
\def\phss{Phys.~Stat.~Sol.}        %
\def\procspie{Proc.~SPIE}     
\def\planss{Planet.~Space Sci.}  
\def\qjras{QJRAS}             
\def\rpph{Rep.~Prog.~Phys.}   
\def\rgsp{Rev.~Geophys.~Space Phys.~} 
\def\sal{Soviet Astron.~Lett.}
\def\sci{Science}             
\def\solph{Sol.~Phys.}        
\def\ssr{Space Sci.~Rev.}     
\def\zap{Z.~Astrophys.}       
\def\jasa{J.~Amer.~Stat.~Assoc.} 

\bibliographystyle{aa}
\bibliography{bibliography/references}

\begin{appendix}
\section{Supermassive black hole binaries - full corner plots}
\label{app:smbhb}

\begin{figure}[h]
    \includegraphics[width=\textwidth]{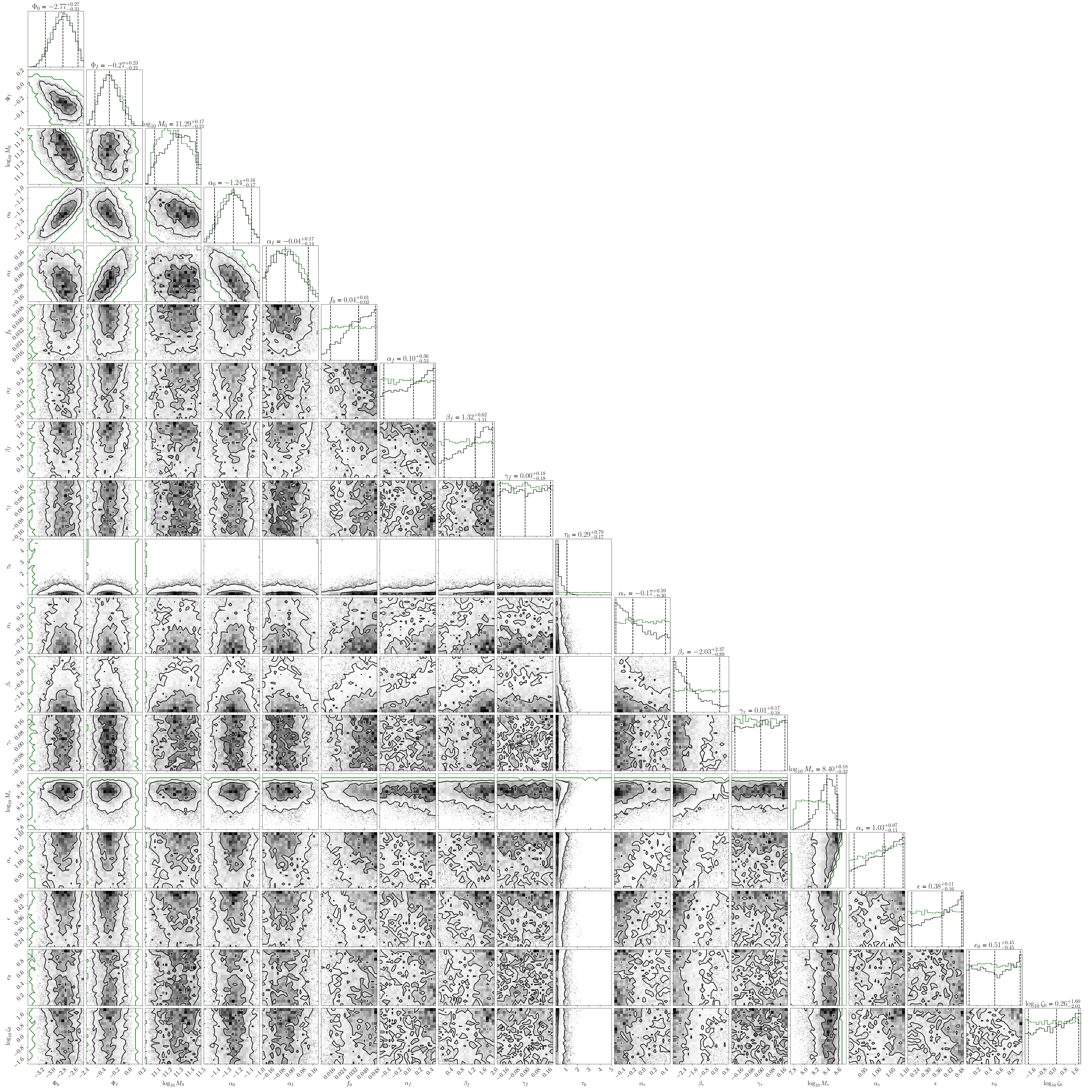}   \caption{\label{fig:astroInformedFullCorner_nFreq9}
    Marginalised posterior distributions for all $18$ parameters of the astrophysically-informed model. 
    The posterior and prior are shown in grey and green, respectively.
    }
\end{figure}

\begin{figure}
\begin{center}
\includegraphics[width=.5\textwidth]{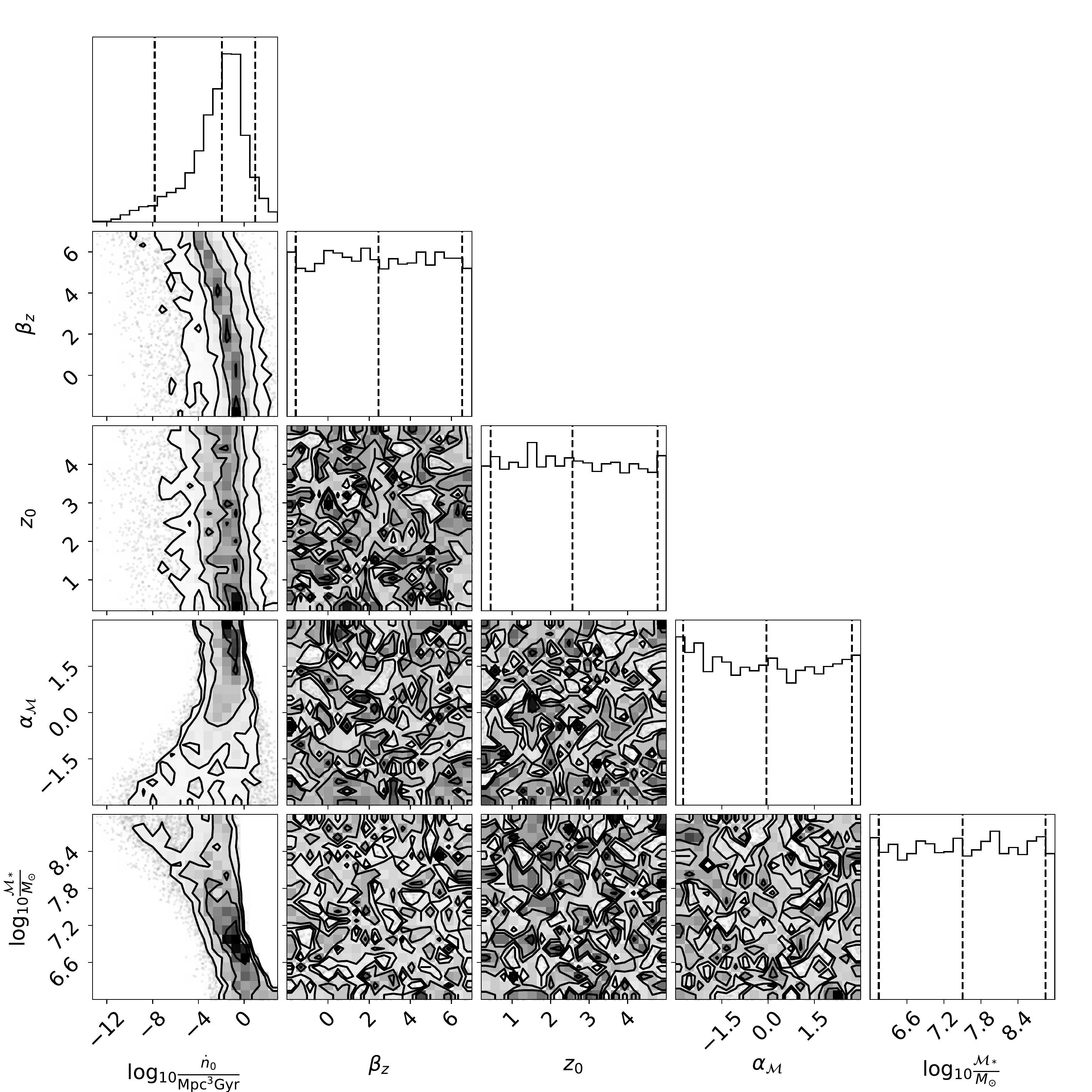}
\caption{\label{fig:agn_fullCorner}
Marginalised posteriors for all five parameters of the agnostic SMBHB model.
}    
\end{center}
\end{figure}

Figs.~\ref{fig:astroInformedFullCorner_nFreq9} and \ref{fig:agn_fullCorner} show the full posterior results for the astrophysically-informed and agnostic mode, respectively. Individual parameters are listed in the main text.

\end{appendix}

\end{document}